\documentclass[10pt,journal,compsoc]{IEEEtran}

\usepackage{pgfplots}
\pgfplotsset{compat=newest}
\usepackage{tikz}

\ifCLASSOPTIONcompsoc
  \usepackage[nocompress]{cite}
\else
  \usepackage{cite}
\fi

\usepackage{amsmath}
\usepackage{amssymb}
\usepackage{amsthm}

 \usepackage{caption}

\usepackage{comment}
\usepackage{amsmath,amssymb} 
\usepackage{color}
\usepackage{epsfig}
\usepackage{amsmath}
\usepackage{amssymb}
\usepackage{adjustbox} 
\usepackage{algorithm2e}
\usepackage{etoolbox}
\usepackage{multirow}
\usepackage{epigraph}
\usepackage{gensymb}
\usepackage{mathtools}
\usepackage{gensymb}
\usepackage{arydshln}
\usepackage{pifont}
\newcommand{\cmark}{\ding{51}}%
\usepackage[capbesideposition={inside,right},capbesidesep=quad]{floatrow}

\usepackage[utf8x]{inputenc}
\usepackage{pgfplots}
\usepackage{pgfplotstable}
\usepackage{tikz}
\usetikzlibrary{backgrounds}
\usepackage{color}
\usepackage{booktabs,siunitx} 
\usepackage{makecell}
\usepackage{enumitem}
\usepackage[caption=false]{subfig}
\usepackage{array,multirow,graphicx}
\usepackage{rotating}

 \usepackage{hyperref}
\hypersetup{
    colorlinks=true,
    linkcolor=blue,
    filecolor=magenta,      
    urlcolor=cyan,
}

\newcommand{\etal}{\textit{et al}.}
\newcommand{\ie}{\textit{i}.\textit{e}. }

\newfloatcommand{capbtabbox}{table}[][\FBwidth]
\definecolor{car}                 {RGB}{  0, 0,100}
\definecolor{train}               {RGB}{  0, 80,100}
\definecolor{motorcycle}          {RGB}{  0,  0,230}

\begin{document}

\title{Binaural SoundNet: Predicting Semantics, Depth and Motion with Binaural Sounds}


%
\author{Dengxin~Dai,
        Arun Balajee Vasudevan,
        Jiri Matas, 
        and~Luc~Van~Gool
\IEEEcompsocitemizethanks{
\IEEEcompsocthanksitem D.~Dai is with the Vision for Autonomous Systems Group, MPI for Informatics. 
\protect\\  E-mail: ddai@mpi-inf.mpg.de
\IEEEcompsocthanksitem D.~Dai, A.~Balajee Vasudevan and L.~Van Gool are with the Computer Vision Lab, ETH Z\"urich.
\IEEEcompsocthanksitem J. Matas is with the Center for Machine Perception, Czech Technical University.
\IEEEcompsocthanksitem L.~Van Gool is also with the Department of Electrical Engineering, KU Leuven.}
}
\maketitle

\begin{abstract}
Humans can robustly recognize and localize objects by using visual and/or auditory cues. While machines are able to do the same with visual data already, less work has been done with sounds. This work develops an approach for scene understanding purely based on binaural sounds. 
The considered tasks include predicting the semantic masks of sound-making objects, the motion of sound-making objects, and the depth map of the scene. To this aim, we propose a novel sensor setup and record a new audio-visual dataset of street scenes with eight professional binaural microphones and a 360\degree camera. 
The co-existence of visual and audio cues is leveraged for supervision transfer. In particular, we employ a cross-modal distillation framework that consists of multiple vision `teacher' methods and a sound `student' method -- the student method is trained to generate the same results as the teacher methods do. This way, the auditory system can be trained without using human annotations. To further boost the performance, we propose another novel auxiliary task, coined Spatial Sound Super-Resolution, to increase the directional resolution of sounds. We then formulate the four tasks into one end-to-end trainable multi-tasking network aiming to boost the overall performance.
Experimental results show that 1) our method achieves good results for all four tasks, 2) the four tasks are mutually beneficial -- training them together achieves the best performance, 3) the number and orientation of microphones are both important, and 4) features learned from the standard spectrogram and features obtained by the classic signal processing pipeline are complementary for auditory perception tasks.
The data and code are released on the project page: \url{https://www.trace.ethz.ch/publications/2020/sound_perception/index.html}.

\end{abstract}

\section{Introduction}

Autonomous vehicles and other intelligent robots will have a substantial impact on people’s daily lives. While great progress has been made in the past years with visual perception systems~\cite{everingham2010pascal,Cordts2016Cityscapes,SDV21_arXiv}, we argue that auditory perception and sound processing also play a crucial role in this context~\cite{soundscape:icra17,hear:around:corners:21,vehicle:tracking:sound:iccv19}. 
As known, animals such as bats, dolphins, and some birds have specialized on ``hearing" their environment. To some extent, humans are able to do the same -- to ``hear" the shape, distance, and density of objects around us~\cite{echolocating:listeners:00,gaver1993world}. 
In fact, humans surely need to use this capability for many daily activities such as for driving -- certain alerting stimuli, such as horns of cars and sirens of ambulances, police cars, fire trucks and human speech are meant to be heard, i.e. are primarily acoustic~\cite{soundscape:icra17,ORSpoken,deruyttere2019talk2car}. 
Auditory perception can be used to localize  common objects like a running car, which is especially useful when visual perception fails due to adverse visual conditions or occlusions. 
Future intelligent robots are expected to have the same perception capability to be robust and to be able to interact with humans naturally.

Numerous interesting tasks have been defined at the intersection of visual and auditory sensing like sound source localization~\cite{sound:pixels:eccv18}, scene-aware audio generation~\cite{scene-aware-audio}, geometry estimation for rooms using sound echos~\cite{room:geometry:acoustic:response:12}, sound source separation using videos~\cite{separate:sound:watching:video:18}, and scene~\cite{aytar2016soundnet} and object~\cite{vehicle:tracking:sound:iccv19} recognition using audio cues. There are also works to learn the correlation of visual objects and sounds~\cite{sound:pixels:eccv18,sounds:motion:iccv19}. While great achievements have been made, previous methods mostly focus on specific objects, e.g. musical instruments or noise-free rooms, or on individual tasks only, e.g., sound localization or geometry estimation. This work aims to learn auditory perception of multiple tasks for general, unconstrained environments.  

\begin{figure*}
    \centering
    \subfloat[super-resolved sounds]{\includegraphics[trim=150 350 150 300,clip,width=0.24\textwidth]{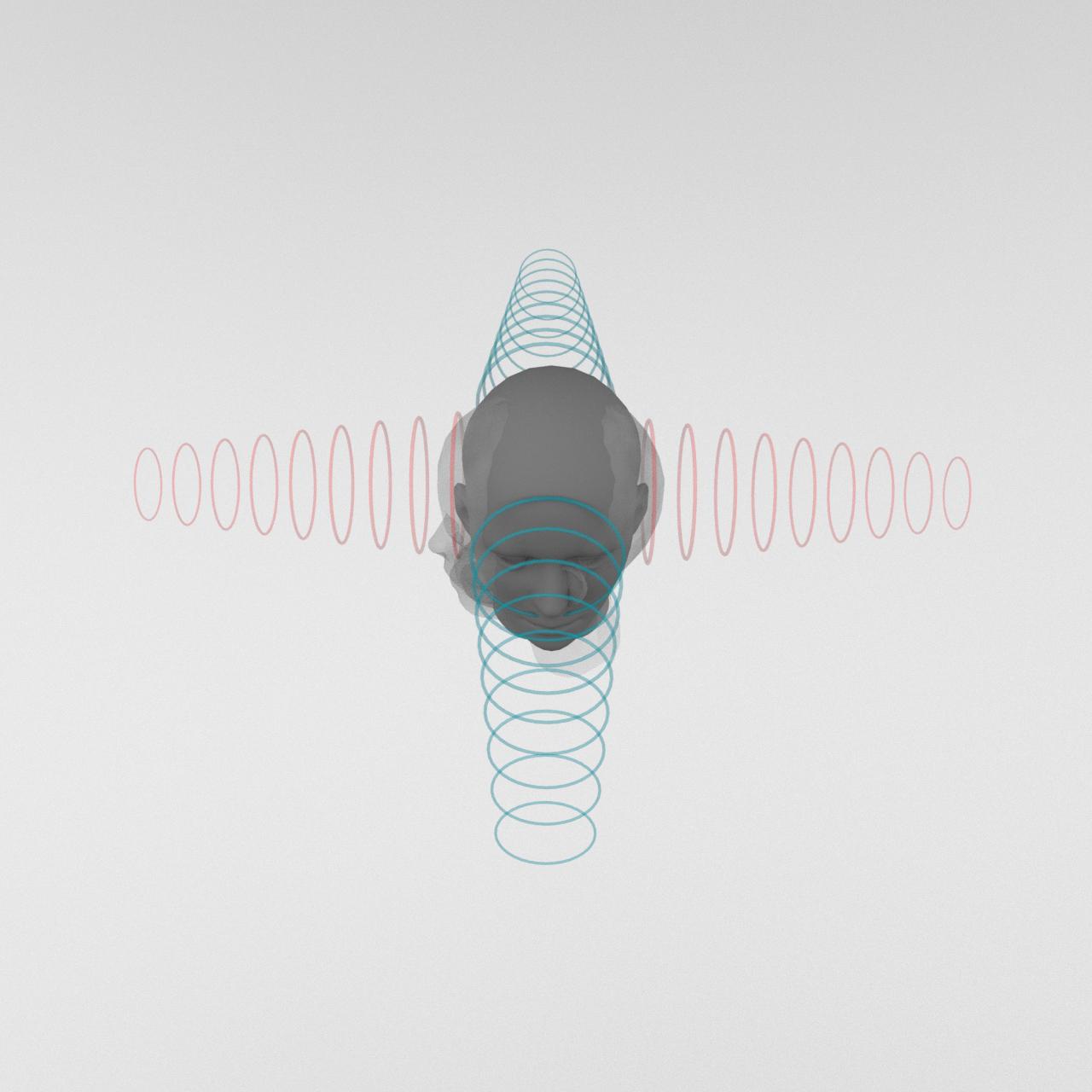}}
    \hfil
    \subfloat[semantic labels]{\includegraphics[width=0.24\textwidth,height=0.155\textwidth]{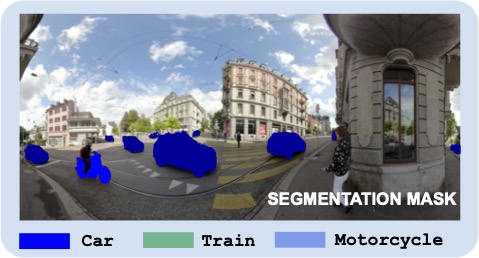}}
    \hfil
    \subfloat[depth values]{\includegraphics[width=0.24\textwidth,height=0.155\textwidth]{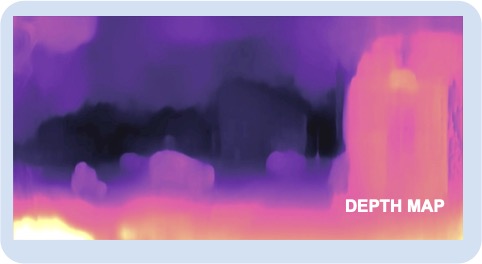}}
    \hfil 
    \subfloat[flow values]{\includegraphics[width=0.24\textwidth,height=0.155\textwidth]{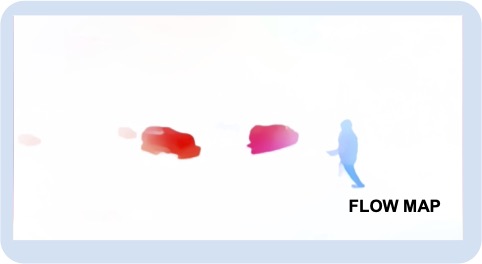}}
    \hfil
    \caption{An illustration of our four tasks:  (a) super-resolved binaural sounds -- from azimuth angle $0^\circ$ (red) to other angles such as $90^\circ$ (blue), (b) auditory semantic prediction for three sound-making object classes,  (c) auditory depth prediction and (d) auditory motion prediction. }
  \label{fig:teaser}
\end{figure*}

This work investigates three scene understanding tasks: semantic prediction, depth prediction, and motion prediction based on binaural sounds, aiming to replicate human auditory capabilities. 
To further enhance the performance, this work proposes a novel auxiliary task called spatial sound super-resolution (S$^{3}$R).
S$^{3}$R is a novel task aiming to increase the directional resolution of audio signals, e.g. from Stereo Audio to Surround Audio. S$^3$R, as an auxiliary task, is motivated from the studies~\cite{sound:localization:head:movement:14,sound:localization:head:movement:67} showing that humans are able to better recognize and localize the sounding sources by changing their head orientation.  S$^{3}$R is also a standalone contribution and has its own applications. For instance, spatially resolved sounds can improve the spatial hearing effects for AR/VR applications.
It offers better environmental perception for users and reduces the ambiguities of sound source localization~\cite{huang2012human}. The four tasks are demonstrated in Fig.~\ref{fig:teaser}.

In order to tackle these tasks, we propose a sensor setup containing eight professional binaural microphones and a $360^\circ$ camera. We used it to record the new `Omni Auditory Perception Dataset' on public streets. The semantic labels and motion labels of the sound-making objects, and the depth of the scene are inferred from the video frames using well-established vision systems. The co-existence of visual and audio cues is leveraged for supervision transfer to train the auditory perception systems without using human annotations. In particular, we employ a cross-modal framework that consists of  vision teacher methods and sound student methods to let the students imitate the performance of the teachers. For evaluation, we have manually annotated a test set. The task S$^{3}$R has accurate ground truth to train and evaluate, thanks to our multi-microphone rig.

We then formulate the three primary prediction tasks and the auxiliary task S$^{3}$R into a multi-tasking network which can be trained in an end-to-end fashion. The multi-task learning is inspired by many recent works~\cite{mousavian2016joint,multitasking:review:20} showing that semantic prediction, depth estimation, and motion estimation are correlated. The goal of S$^{3}$R -- enhancing the spatial sense of sounds -- is also aligned with the goals of the primary tasks. Thus, we believe that learning them together can boost the overall performance. This conjecture is confirmed by the experimental results.

Finally, in order to study what feature representation performs better for the considered auditory perception tasks, we also compare features learned from the standard spectrograms and features obtained by the classic (differentiable) digital signal process method \cite{engel2019ddsp} extensively.   

We evaluate our method on our new Omni Auditory Perception Dataset. Extensive experimental results reveal that 1) our method achieves good results for auditory semantic perception, auditory depth prediction, auditory motion prediction, and spatial sound super-resolution;  2) the four tasks are mutually beneficial -- training  them  together  achieves  the best results;  3) both the number of microphones and the orientations of microphones are important for auditory perception; and 4) features learned from the standard spectrogram and features obtained by the classic signal processing pipeline are complementary for auditory perception tasks.

This work makes multiple contributions: 1) a novel approach to three major scene understanding tasks based on binaural sounds; 2) a method for spatial sound super-resolution, which is novel both as a standalone task and as an auxiliary task for other auditory perception tasks; and 3) a new Omni Auditory Perception Dataset with four pairs of binaural sounds ($360^\circ$ coverage) accompanied by synchronized $360^\circ$ videos. 
  
A preliminary version of this work has been published at ECCV 2020 \cite{huang2012human}. Compared to it, this journal version has made the following additional contributions: 
\begin{itemize}
    \item A new task called auditory motion prediction is included into the multi-task learning framework. This leads to improved performance for all other tasks. 
    \item In addition to the spectrogram representation used in the conference paper, this journal version also employs the recent differentiable digital signal processing (DDSP) method \cite{engel2019ddsp} to represent audio signals. The two methods are thoroughly compared and are also combined. The combined representation yields the best performance for all the four considered tasks. 
    \item More results and ablation studies are added. 
    \item Other sections such as Related Works and Approach are enhanced. 
\end{itemize}

\section{Related Works}
\label{sec:related}


\noindent
\textbf{Audio-Visual Learning}. 
The interplay among senses is basic to the sensory organization in human brains \cite{merging:of:sense} and is the key
to understand the complex interaction of the physical world. Fortunately, most videos like those available in the Internet contain both visual information and audio information, which provide a bridge linking those two domains and enable many interesting learning approaches. Aytar et al. propose an audio-based scene recognition method by cross-domain distillation to transfer supervision from the visual domain to the audio domain~\cite{aytar2016soundnet}. A similar system was proposed for emotion recognition~\cite{emotion:cross-model:transfer:18}. It has shown that ambient sounds can also provide supervision for visual learning~\cite{ambient:sound:eccv16}. 

There is also a surge of interest in learning audio-visual feature representations in a self-supervised manner \cite{Owens_2018_ECCV,spatial:alignment:nips20,left:right:flipping:audio:cvpr20,VisualEchoes:eccv20,audio-visual:instance:cvpr21,distill:audio:visual:cvpr21}. 
The supervision can be generated by predicting whether video frames and audio are temporally aligned~\cite{Owens_2018_ECCV} or spatially aligned \cite{left:right:flipping:audio:cvpr20,spatial:alignment:nips20}, by predicting the correct camera orientation at which the agent would receive sound echoes \cite{VisualEchoes:eccv20}, or by using cross-modal prediction for co-clustering \cite{alwassel_2020_xdc}. 
The learned representations have been successfully applied to multiple down-streaming tasks such as action recognition, sound localization, sound separation, depth estimation and robot navigation. 
Our spatial sound super-resolution can be understood as another novel self-supervised audio feature representation method. It aims to learn features useful for spatial prediction tasks which is well-aligned with our targeted  tasks, \ie semantic mask prediction, depth prediction, and motion  prediction. 

\noindent
\textbf{Sound Source Localization}. 
One of the most important capability of human auditory system is to localize sound by using three different cues: interaural time difference, interaural level difference, and head-related transfer function \cite{blauert1997spatial}.  Sound localization has been well studied with applications such as localizing sniper fire on the battle field, cataloging wildlife in rural areas, and localizing noise pollution sources in an urban environment. It also enriches human–robot interaction by complementing the robot’s perceptual capabilities~\cite{survey:sound:localization:15,localization:sound:source:review:17}. The task is often tackled by the beamforming technique with a microphone array~\cite{SoundCompass:2014}, with a notable exception that relies on a single microphone only~\cite{sound:location:09}. 
Another classical technique uses kalman filter that processes auto-correlations from multiple channels~\cite{klee2006kalman}. 
The recent advance of deep learning enables acoustic camera systems for real-time reconstruction of acoustic camera spherical maps~\cite{deepwave:nips19}.

Recently, there is a large body of work that localize the sources of sounds in visual scenes~\cite{localize:sound:source:scene:cvpr18,separate:sound:watching:video:18,look:listen:learn:iccv17,objects:sound:eccv18,Harmony:in:Motion:cvpr07,sound:pixels:eccv18,cosegmenting:sounds:objects:iccv19,sounds:motion:iccv19,tian2018audio}. The localization is mostly done by analyzing the consistency between visual motion cues and audio motion cues over a large collection of video data. These methods generally learn to locate image regions which produce sounds and separate the input sounds into a set of components that represent the sound for each pixel. Our work uses binaural sounds rather than monaural sounds. Our method also localizes and recognizes sounds at the pixel level in image frames, but it performs the task with sounds as the only input.



\noindent
\textbf{Auditory Perception for Robotic Applications}. 
Auditory scene analysis has been also widely applied to robotic applications. For instance, auditory cues are used to determine the
occurrence of abnormal events in driving scenarios~\cite{soundscape:icra17} or an alarm in the city \cite{marchegiani2018listening}. An acoustic safety emergency system has also been proposed~\cite{acoustic:based:safety:emergency:09}. Salamon et al. have presented a taxonomy of urban sounds and a new dataset, UrbanSound, for automatic urban sound classification~\cite{dataset:taxonomy:14}. In order to navigate safely in unknown environments, Zürn \etal \cite{zuren2019tro} have proposed a novel terrain classification framework leveraging an unsupervised classifier that learns from vehicle-terrain interaction sounds. 
Schulz \etal \cite{hear:around:corners:21} have developed a method to recognize the approaching directions of cars at intersections before they appear in line of sight to gain more reaction time.  Audio signals could also let a vehicle better understand the nature of the road surface, for example how wet it is \cite{wet:audio}.   

The closest to our work is the  recent work of car detection/tracking with stereo sounds~\cite{vehicle:tracking:sound:iccv19,more:than:meets:eye} in which a 2D bounding box is proposed for each moving car in an image frame and the proposed bounding boxes over frames are merged into tracklets depending on their overlaps. While being similar in spirit, our semantic prediction is dense and for multiple classes instead of a 2D bounding box prediction for a single class. Our method also includes dense depth prediction, motion prediction, and spatial sound super-resolution. Furthermore, Binaural sounds are different from general stereo sounds. 
Another similar work to ours is done by Irie \etal~\cite{irie2019seeing}. While their goal is also to predict semantic segmentation results from multichannel audio signals, our method differs significantly: 1) our method considers multiple semantic classes, 2) our method works in unconstrained real outdoor environment instead of a controlled lab setting, and 3) our method addresses multiple tasks. 

Audio has also been used for estimating geometric scene layouts~\cite{3d:room:geometry:audio-visual:sensors:17,glass:reconstruction:acoustic:15,room:geometry:acoustic:response:12,accoustic:echoes:reveal:room:shape:13}. The general idea is that the temporal relationships between the arrival times of echoes allows us to estimate the geometry of the environment. 
The systems are complementary to vision systems, especially when smooth surfaces and glasses are prevalent. Recently, deep neural networks have been applied to this task as well. For instance, Christensen \etal \cite{BatVision:icra:20} used an encoder-decoder type of network to predict scene layouts with echos from two artificial human ears. This setting is further extended by integrating visual cues from a paired monocular camera \cite{depthwithechoes:cvpr21}.      Specifically, Parida \etal \cite{depthwithechoes:cvpr21}
 propose a novel network with a multimodal fusion module to incorporates material properties of the scene and to fuse the two modalities with spatial attention maps.
 
 Recently, multiple works have introduced an exciting new task audio-visual navigation \cite{Gan2020LookLA,soundspace:eccv20,audio-visual:nav:cvpr21,waypoint:audio-visual:nav}, for which the agent needs to learn to navigate to a sounding object by both seeing and hearing in an acoustically and visually realistic 3D environment. These works show that audio can greatly benefit visual navigation in 3D spaces.
 Eliakim \etal \cite{eliakim2018fully} take a step further and present a fully autonomous bat-like terrestrial robot that relies on echolocation to move through a novel environment while mapping it solely based on sound. 
 

\noindent
\textbf{Sound Generation with/for Visual Scenes}. 
There is increasing effort in characterizing the soundscapes around us for more compelling and immersive experiences \cite{3d:sound,spatial:soundscapes:2020}. To that aim, multiple methods have been proposed to generate sounds for a given visual scene or enrich the spatial sense of sounds with the guidance of a given visual scene. For example, the notable work by Owens \etal~\cite{Owens_2016_CVPR} has shown that the sounds of striking an object can be learned based on the visual appearance of the object.
There are also works to add scene-aware spatial audio to 360$^{\circ}$ videos in typical indoor scenes~\cite{scene-aware-audio,360:sound:generation:nips18} by using a conventional mono-channel microphone and a speaker and by analyzing the structure of the scene.  Gao and Grauman \cite{gao20192} proposed the Mono2bianural pipeline to convert common monaural audio into binaural audio conditioned on a video frame embedding. Zhou \etal \cite{sep-stereo:eccv20} extended the idea by tackling stereophonic audio generation and source separation at the same by using mono data.  Xu \etal \cite{visually:informed:binaural:audio:21}  developed PseudoBinaural, an effective pipeline to generate visually coherent binaural audios by training with synthetic data. Very recently, Richard \etal \cite{binauralspeechsynthesis} presented a neural sound spatialization method that can generate spatially accurate waveform outputs in realtime. 
Our S$^3$R task enriches this repository by generating binaural sounds of multiple heading orientations given the binaural sound of one orientation. Our data features visual and acoustic scenes of real-world, outdoor environments. This complements existing works.     




\section{Approach}
\label{sec:approach} 

\begin{figure*}[!tb]
  \centering
  \includegraphics[width=0.97\textwidth,height=0.35\textwidth]{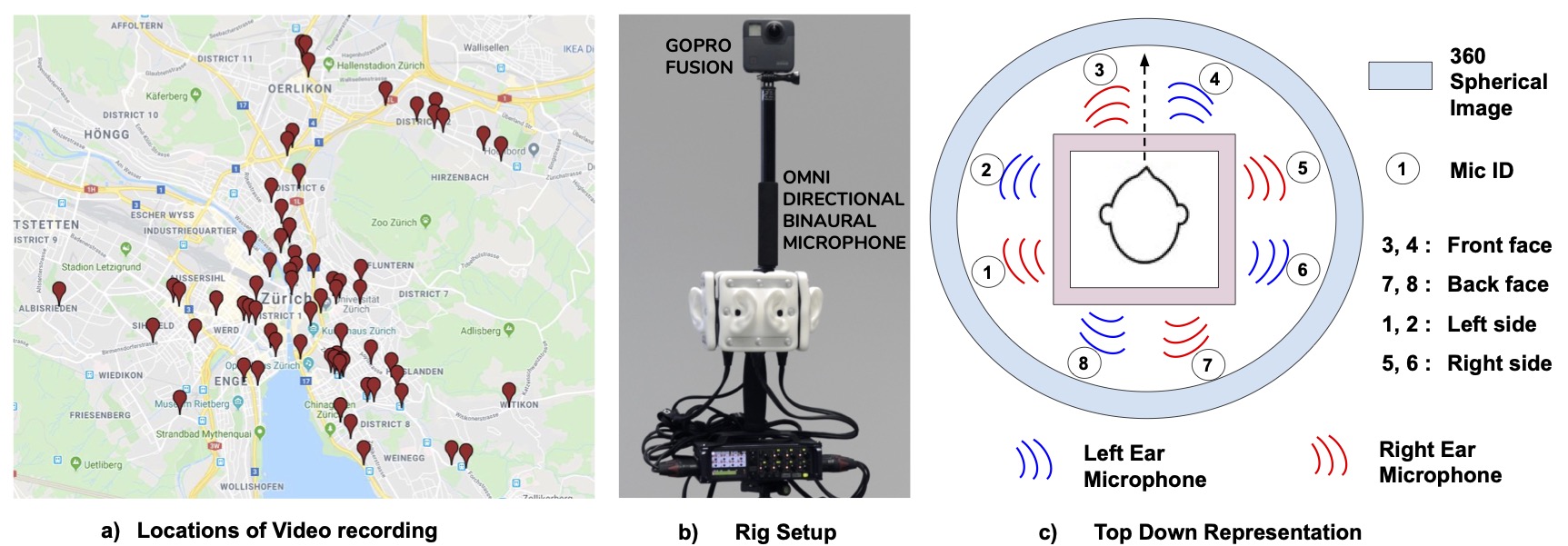}
  \caption{Sensor and dataset: a) data capture locations, b) our custom rig and c) abstract depiction of our recording setup with sensor orientations and microphone ids.}
  \label{fig:setup}
\end{figure*}

\subsection{Omni Auditory Perception Dataset} 
\label{sec:dataset} 


Training our method requires a large collection of omni-directional binaural audios and accompanying $360^{\circ}$ videos.  Since no public video dataset fulfills this requirement, we collect a new dataset with a custom rig. As shown in Fig.~\ref{fig:setup}, we assembled a rig consisting of a 3Dio Omni Binaural Microphone, a $360^{\circ}$ GoPro Fusion camera and a Zoom F8 MultiTrack Field Recorder, all attached to a tripod. 
We mounted the GoPro camera on top of the 3Dio Omni Binaural Microphone by a telescopic pole to capture all of the sights with minimum occlusion from the devices underneath. This custom rig enables omni seeing and omni binaural hearing with 4 pairs of human ears. 
The 8 microphones are connected to 8 tracks of the MultiTrack Recorder. The recorder has 8 high-quality, super low-noise mic preamps to amplify the sounds, provides accurate control of microphone sensitivity and allows for accurate synchronization among all microphones. The GoPro Fusion camera captures 4K videos using a pair of $180^\circ$ cameras housed on the front and back faces. The two cameras perform synchronized image capture and are later fused together to render $360^\circ$ videos using the GoPro Fusion Software. We further use the clap sound to synchronize the camera and 3Dio microphones. 
After the sensors are turned on, we do hand clapping near them. This clapping sounds recorded by both the binaural microphones and the built-in microphones of the camera are used to synchronize the binaural sounds and the video. The video and audio signals recorded by the camera are synchronized by default. Videos are recorded at $30$ fps. Audios are recorded at $96$ kHz.

We recorded videos on the streets of Zurich covering 165 locations within an area of 5km $\times$ 5km as shown in Fig.~\ref{fig:setup}(a). We choose the locations next to road junctions, where we kept the rig stationary for the data recording of the traffic scenes. For each location, we recorded data for around 5-7 minutes. 
Our dataset consists of 165 city traffic videos and audios with an average length of 6.49 minutes, totalling 15 hours. We post-process the raw video-audio data into 2 second segments, resulting in $64,250$ video clips. The videos contain numerous sound-making objects such as cars, trams, motorcycles, buses and trucks. 
We perform a careful selection of training samples such that they are not in silent phase of the traffic scene. A sample is used only if its audio energy is beyond a chosen threshold. 
Also, we only use samples of which the images have at least $5$\% different semantic labels than the corresponding background images (see Sec. \ref{sec:semantic:obj:loc} for how to obtain them). 


It is worth noticing that the professional 3Dio binaural mics simulate how human ears receive sound, which is different from the general stereo microphones or monaural microphones. Humans localize sound sources by using three primary cues \cite{localization:sound:source:review:17}: interaural time difference (ITD), interaural level difference (ILD), and head-related transfer function (HRTF). ITD is caused by the difference between the times sounds reach the two ears. ILD is caused by the difference in sound pressure level reaching the two ears due to the acoustic shadow casted by the listener's head. HRTF is caused because the pinna and head affect the intensities of sound frequencies. 
All these cues are missing in monaural audio, thus in this work, we focus on learning to localize semantic objects with binaural audios. This difference makes our dataset more suitable for cognitive applications.

\subsection{Auditory Semantic Prediction} 
\label{sec:semantic:obj:loc}
Since it is costly to create a large collection of human annotations, we follow the teacher-student learning strategy to transfer knowledge from vision to audio~\cite{aytar2016soundnet,vehicle:tracking:sound:iccv19}. Thus, our auditory semantic object prediction system is composed of two key components: a teacher vision network and a student audio network. The difference to previous methods is that we learn to transfer precise semantic segmentation results instead of scene labels \cite{aytar2016soundnet} or bounding boxes \cite{vehicle:tracking:sound:iccv19}.

\noindent
\textbf{Vision network}. We employ the DeepLabv3+ model \cite{chen2018encoder}. We pick the middle frame of our 2-second video clip as the target frame and feed it to the teacher network to generate the semantic map. During training, each target frame is fed into a Cityscapes \cite{Cordts2016Cityscapes} pre-trained DeepLabv3+ to assign a semantic label to each pixel. Since objects in many classes such as \emph{sky}, \emph{road} and \emph{parked cars} are not sound-making, it is very challenging to predict their semantic masks. Therefore, an object selection policy needs to be designed in order to collect the semantic masks of major sound-making objects. 

\noindent
\textbf{Sound-making Object Collection}.
Our dataset contains numerous sound making objects such as cars, trams, motorcycles, pedestrians, buses and bicycles. The target objects must be constrained to make sure that the task is challenging but still achievable by current sensing systems and learning methods. In this work, we focus on \emph{car}, \emph{tram}, \emph{motorcycle} due to their high occurrences in the datasets and because they produce sufficient noise when they move. 

As to the motion status, we employ background subtraction to remove the background classes, such as road, building and sky, and the stationary foreground classes such as parked cars. In the end, only the semantic masks of moving trams, moving cars and moving motorcycles are taken as the prediction targets. This selection guides the attention of the learning method to major sound-making objects and avoids localizing `rare' sound-making objects and sound-irrelevant objects such as \emph{parked car} and \emph{sky}.  

There is a rich body of studies for background estimation~\cite{background:substraction:11}, in this work we employ a simple method based on majority voting. The method works surprisingly well. Specifically, the background image is computed as 
\begin{equation}
I_{bg}(h,w) = \text{Mode}\{I_{1}(h,w), I_{2}(h,w), ..., I_T(h,w) \},
\label{eq:modeOp} 
\end{equation}
where $T$ is the length of the complete video sequence, $(h,w)$ are pixel indexes, and $\text{Mode}\{.\}$ computes the number which appears most often in a set of numbers.  Since the complete video sequence is quite long (about 5-7 mins), the background estimation is accurate and reliable. 

The sound-making objects are detected by the following procedure: given an video frame $I_t$ and its corresponding background image $I_{bg}$, we use DeepLabv3+ to get their semantic segmentation results $Y_t$, and $Y_{bg}$. Fig.~\ref{fig:pipeline} gives an illustration. The detection is done as
\begin{equation}
    \label{eq:sounding:obj} 
    S(h,w) = \left\{ 
    \begin{array}{rl}
    1 & \text{if  } Y_t(h,w) \in \{car, train, motorcycle\} \\ 
    & \text{ and } Y_t(h,w) \neq Y_{bg}(h,w),\\
    0 & \text{otherwise},
   \end{array} 
   \right.
\end{equation}
where $1$ indicates pixel locations of sound-making objects and $0$ otherwise.  
Fig.~\ref{fig:result1} shows examples of the detected background and the detected sound-making target objects (i.e. ground truth).

\noindent
\textbf{Audio network}.
We treat auditory semantic prediction from the binaural sounds as a dense label prediction task. We take the semantic labels produced by the teacher vision network and filter by Eq.~\ref{eq:sounding:obj} as 
pseudo-labels, and then train a student audio network (BinauralSematicNet) to predict the pseudo semantic labels directly from the audio signals. A cross-entropy loss is used. The description of the network architecture can be found in Sec.~\ref{sec:network:architecture}.  

\subsection{Auditory Depth Prediction} 
\label{sec:depth:prediction} 
Similar to the auditory semantic prediction, our auditory depth prediction method is composed of a teacher vision network and a student audio network. 

\noindent
\textbf{Vision network}. We employ the MonoDepth2 model \cite{monodepth2:iccv19} due to its good performance. We again pick the middle frame of our 2-second video clip as the target frame and feed it to the teacher network to generate the depth map. The model is pre-trained on KITTI~\cite{Geiger2013IJRR}. We estimate the depth for the whole scene instead of for the sound-making objects only. This design is similar to previous methods for holistic scene layout estimation with sounds \cite{room:geometry:acoustic:response:12,3d:room:geometry:audio-visual:sensors:17,depthwithechoes:cvpr21}. 

\noindent
\textbf{Audio network}.
We treat depth prediction from the binaural sounds as a dense regression task. We take the depth values produced by the teacher vision network as 
pseudo-labels, and then train a student audio network
(BinauralDepthNet) to regress the pseudo depth labels directly from the audio signals. The L2 loss is used. The description of the network architecture can be found in Sec.~\ref{sec:network:architecture}.

\subsection{Auditory Motion Estimation} 
\label{sec:flow:estimation} 
In line with auditory semantic prediction and auditory depth prediction, here again we use the same setup of a teacher vision network and a student audio network for our auditory motion estimation method. We represent motion with optical flow maps. 

\noindent
\textbf{Vision network}.
For the teacher vision network, we use the FlowNet2 model \cite{ilg2017flownet}. We again pick the middle frame (\ie at 1 second) of our 2-second video clip as the source frame and the frame at 1.25 seconds as the target frame. We then feed the two frames to the teacher network to generate the optical flow estimate. The model is pre-trained on KITTI~\cite{Geiger2013IJRR}. 

\noindent
\textbf{Audio network}.
We treat optical flow estimation from the binaural sounds as a regression task. We take the flow estimates produced by the teacher vision network as pseudo-labels. We then train a student audio network (BinauralFlowNet) to regress the pseudo labels directly from the audio signals. L2 loss is used. The network architecture is described in Sec.~\ref{sec:network:architecture}.

\begin{figure*}[!tb]
  \centering
  \includegraphics[width=0.9\textwidth]{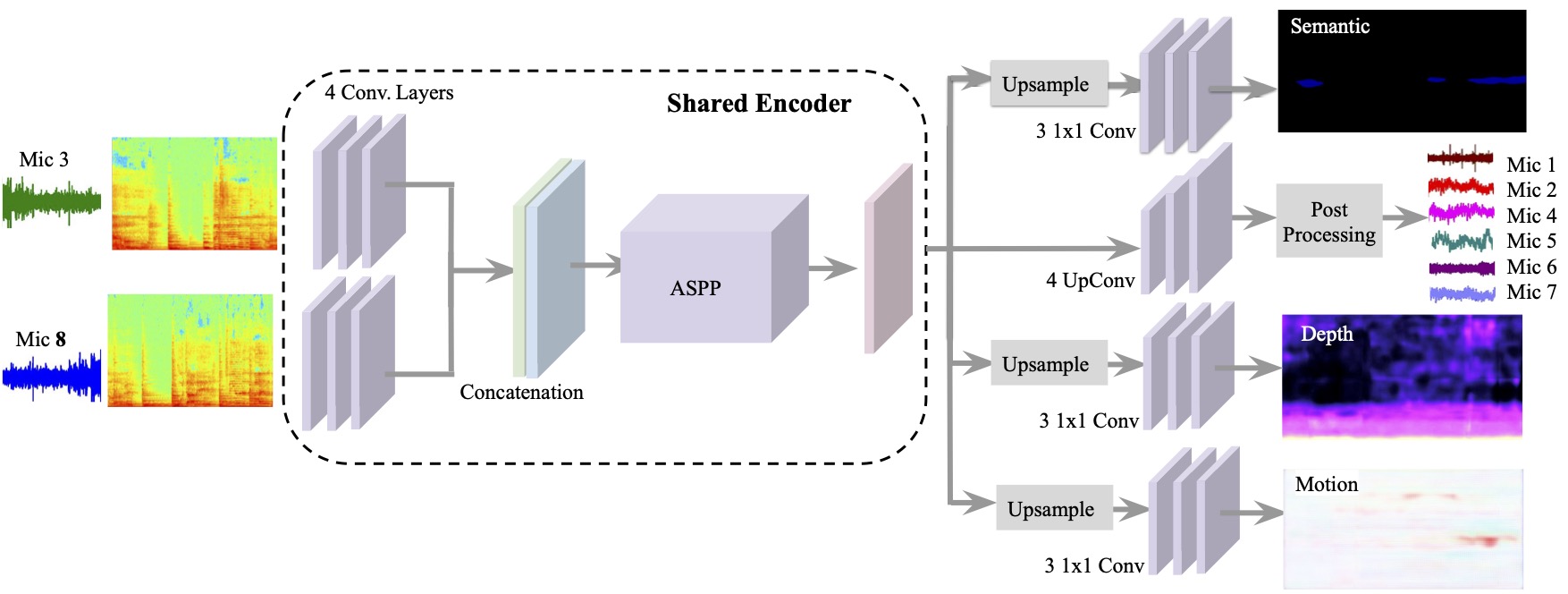}
  \caption{The diagram of our method for the four considered tasks. The encoder is shared by all tasks and each task has its own decoder.}
  \label{fig:pipeline}
\end{figure*}

\subsection{Spatial Sound Super-resolution (S$^3$R)}
\label{sec:sasr} 
We leverage our omni-directional binaural microphones to design a novel task of spatial sound super-resolution (S$^3$R), to provide auxiliary supervision for our three primary tasks.
The S$^3$R task is motivated by the well-established studies~\cite{sound:localization:head:movement:14,role:head:movements:sound:localization:40} about the effects of head movement in improving the accuracy of sound localization.  Previous studies also found that rotational movements of the head occur most frequently during source localization~\cite{sound:localization:head:movement:67}. 
Inspired by these findings, we study the effect of head rotation on auditory semantic and depth perception. Our omni-directional binaural microphone set is an ideal device to simulate head rotations at four discrete angles, i.e. to an azimuth angle of $0^{\circ}$ $90^{\circ}$, $180^{\circ}$ and $270^{\circ}$, respectively.  This method also shares some similarity with the inspiring work VisualEchoes \cite{VisualEchoes:eccv20}, in which feature representation is learned by training a network to predict the correct camera orientation at which the agent receives certain echoes. 

Specifically, the S$^3$R task is to train a neural network to predict the binaural audio signals at other azimuth angles given the signals at the azimuth angle of $0^\circ$.
We denote the received signal by the left and right ears at azimuth $0^{\circ}$ by $x^{L_0}(t)$ and $x^{R_0}(t)$, respectively. We then feed those two signals into a deep network to predict the binaural audio signals $x^{L_\alpha}(t)$ and $x^{R_\alpha}(t)$ at azimuth $\alpha^{\circ}$. Inspired by~\cite{gao20192}, we predict the difference of the target signals to the input signals, instead of directly predicting the absolute values of the targets. This way, the network is forced to learn the subtle difference. Specifically, we predict the difference signals:
\begin{equation}
\begin{split}
   & x^{DL_{\alpha}}(t) =  x^{L_0}(t) -  x^{L_\alpha}(t)   \\
   & x^{DR_{\alpha}}(t) =  x^{R_0}(t) -  x^{R_\alpha}(t), 
   \end{split}
\end{equation}
where $\alpha \in \{ 90^{\circ}, 180^{\circ}, 270^{\circ}\}$. 
In order to leverage the image processing power of convolutional neural network, we follow the literature and choose to work with the spectrogram representation. Following ~\cite{gao20192}, real and imaginary components of complex masks are predicted. The masks are multiplied with input spectrograms to get the spectrograms of the difference signals; the raw waveforms of the difference signals are then produced by applying Inverse Short-time Fourier Transform (ISTFT) \cite{invert:fourier:transform:83}; and finally the target signals are reconstructed by adding back the reference raw waveform.

\subsection{Network Architecture} 
\label{sec:network:architecture} 

Here, we present our multi-tasking audio network for all the four tasks. The network is composed of one shared encoder and four task-specific decoders.
The pipeline of the method is shown in Fig.~\ref{fig:pipeline}.
 As to the encoder, we convert the two channels of binaural sounds to log-spectrogram representations. Each spectrogram is passed through 4 strided convolutional (conv) layers with shared weights before they are concatenated. Each conv layer performs a $4\times4$ convolution with a stride of $2$. Each conv layer is followed by a BN layer and a ReLU activation. The concatenated feature map is further passed to a Atrous Spatial Pyramid Pooling (ASPP) module~\cite{chen2017deeplab}. ASPP has one $1\times1$ convolution and three $3\times3$ convolutions with dilation rates of $6$, $12$, and $18$. 
Each of the convolutions has $64$ filters and a BN layer. ASPP concatenates all the features and passes them through a $1\times1$ conv layer to generate binaural sound features. This feature map is taken as the input to our decoders.

Below, we present the three task-specific decoders.
For the semantic prediction task, we employ a decoder to predict the dense semantic labels from the above feature map given by the shared encoder. The decoder comprises of an upsampling layer and three $1\times1$ conv layers. For the first two conv layers, each is followed by a BN and a ReLU activation; for the last one, it is followed by a softmax activation. We use the same decoder architecture for the depth prediction task and the motion prediction task.
For the S$^3$R task, we perform a series of $5$ up-convolutions for the binaural feature map, each convolution layer is followed by a BN and a ReLU activation. The last layer is followed by a sigmoid layer which predicts a complex valued mask. We perform a few post processing steps to convert this mask to binaural sounds at other azimuth angles as mentioned in Sec.~\ref{sec:sasr}.

\subsubsection{Alternative Encoders} 
While the spectrogram representation is sufficient to express audio signals, it does not utilize strong domain knowledge of auditory signal processing and perception. The recent work Differentiable Digital Signal Processing (DDSP)  \cite{engel2019ddsp} is proposed to enable direct integration of classic signal processing elements with deep learning methods. 

The key idea of DDSP~\cite{engel2019ddsp} is to make digital signal processing (DSP) units interpretable and differentiable. The parameter of these DSP elements can be carefully controlled to create complex realistic signals. In the work of~\cite{engel2019ddsp}, the authors show that differentiable oscillators, filters, and reverberation from the DDSP library allows to train for high-quality audio synthesis models and can do so with fewer parameters and lesser data.  In previous sections, we discussed how we extracted features from sounds by converting them to time-frequency representations followed by convolutional feature encoding. Also, sound generation in S$^{3}$R task is elaborately discussed which comprises of multiple post-processing steps. The potential of DDSP has been showcased by the application of audio synthesis~\cite{engel2019ddsp}. Therefore, it stands to ask how DDSP performs for our tasks and whether it can be combined with spectrogram to further improve the performance. To answer these two questions, we design and evaluate two alternative encoders -- one with DDSP only and the other with DDSP and spectrogram combined.

\begin{figure}[!bt]
\begin{tabular}{cccc}
  \centering 
  \includegraphics[width=1\textwidth]{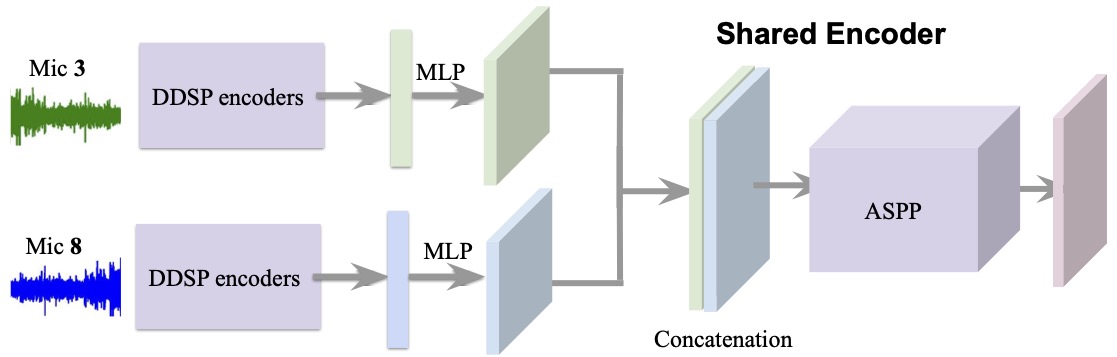} & \\ \text{(a) the encoder with DDSP} & \\
\includegraphics[width=1\textwidth]{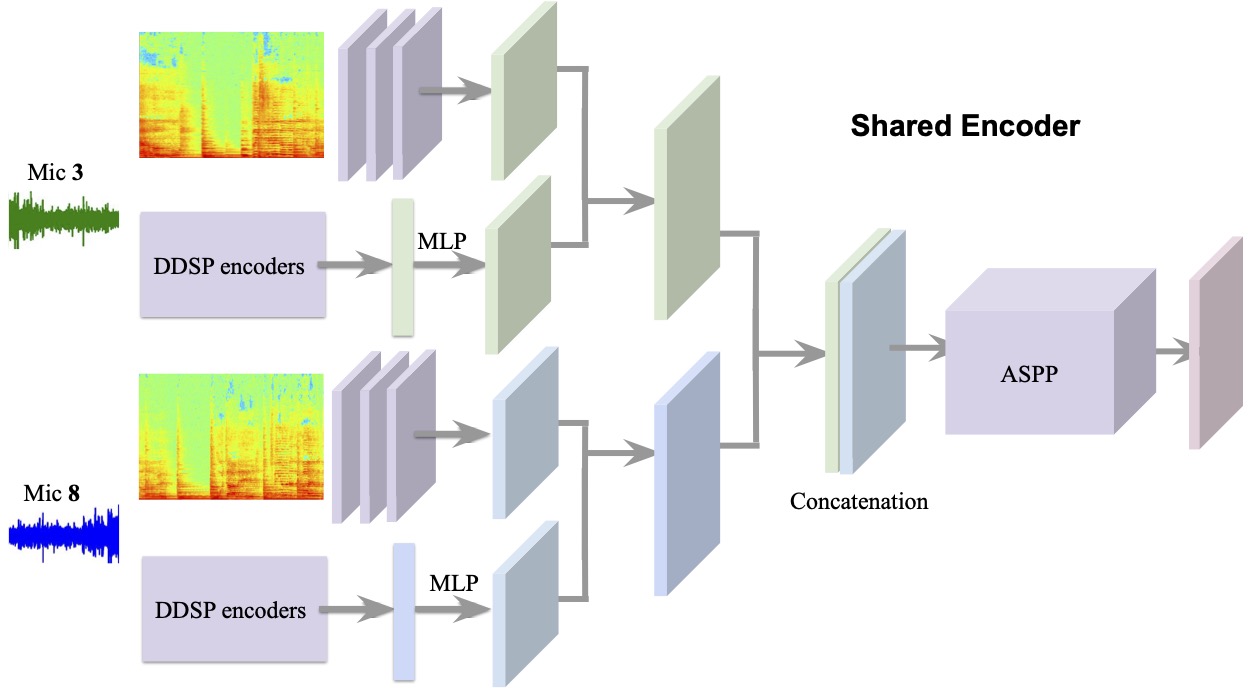} &\\ 
 \text{(b) the encoder with both DDSP and spectrogram}  &
  \end{tabular}
  \caption{The diagram of our two alternative encoders.}
  \label{fig:encoders}
  \end{figure}
  
Following the work of~\cite{engel2019ddsp}, we have an encoder that maps the input raw waveforms to $(f(t), l(t), z(t))$, which are fundamental frequencies, loudness and latent encoding, respectively. Specifically, we use the  CREPE model~\cite{kim2018crepe} pretrained on a large collection music database to extract the fundamental frequency $f(t)$. The loudness $l(t)$ is extracted directly from the audio. We utilize an A-weighting of the power spectrum which computes tensor of amplitude attenuation weights corresponding to the frequencies tensor. This gives greater emphasis on higher frequencies.  Weights in decibel scale (log scaling) are then computed. Finally, loudness vector is normalized over all samples in our dataset. For latent encoding $z(t)$, we first extract MFCCs (Mel Frequency Cepstrum Coefficients) from the sounds, which is computed from the log-mel-spectrogram of the sound with a FFT. 
The MFCCs are then passed through a normalization layer and a 512-unit GRU which is followed by a 512-unit linear layer to get $z(t)$. The computed $(f(t), l(t), z(t))$ are concatenated and passed to an multi-layer perceptron (MLP) to obtain a latent vector which is then reshaped to a 3d tensor. The two tensors from the two audio signals are concatenated and passed to the ASPP for context encoding. This DDSP-based encoder is shown in Fig.~\ref{fig:encoders}(a).

In order to test whether DDSP and spectrogram are complementary, we also  design another encoder to combine the features from ddsp-based method and spectrogram-based method. Basically, the features from DDSP and spectrogram are simply concatenated before being fed to the ASPP module. The structure of the combined encoder is shown in Fig.~\ref{fig:encoders}(b).

\subsubsection{Loss Function}  We train the complete model shown in Fig.~\ref{fig:pipeline} in an end-to-end fashion. We use a) cross-entropy loss for the semantic prediction task which is formulated as dense pixel labeling to $3$ classes, b) L2 loss for the depth prediction task; c) L2 loss for  the motion prediction task to minimize the distance between the predicted values and the ground-truth  values, and d) L2 loss for the S$^{3}$R task to minimize the distance between the predicted complex spectrogram and the ground truths. Hence, the total loss $L$ for our multi-tasking learning is 
\begin{equation}
   L = L_{semantic} + \lambda_{1}L_{depth} + \lambda_{2}L_{motion} + \lambda_{3}L_{s^3r}\\
   \label{eqn:loss}
\end{equation}
where $\lambda_{1}$, $\lambda_{2}$ and $\lambda_{3}$ are weights to balance the losses.

For the sound generation in S$^{3}$R task under ddsp-based models, we convert the generated and ground truth sounds to spectrograms $\hat{S}_{i}$ and $S_{i}$, respectively, with a FFT of size $i$. A combined L1 loss is computed between their magnitudes as
\begin{equation}
   L_{s^3r} = \sum_{i} {\|\hat{S}_{i} - S_{i}\|}_{1} + \alpha {\|\log \hat{S}_{i} - \log S_{i}\|}_{1} \\
   \label{eqn:ddsploss}
\end{equation}
where $\alpha$ is set to $1$ in our experiments and $i$ comprises of  (256, 128, 64).
 
When the shared encoder gets input from both ddsp models and  spectrograms as shown in Fig.~\ref{fig:encoders}(b), we use the loss in Eqn.~\ref{eqn:ddsploss} for S$^{3}$R  and the loss in Eqn.~\ref{eqn:loss} for the overall loss.

\section{Experiments}
\label{sec:expts}

\noindent
\textbf{Data Preparation}.
Our dataset comprises of $64,250$ video segments, each of 2 seconds long. We split the samples into three parts: $51,400$ for training, $6,208$ for validation and $6,492$ for testing. We use 2-seconds segments following ~\cite{vehicle:tracking:sound:iccv19}, which shows that performances are stable for $\geq$1 second segments. For each scene, a background image is also precomputed according to Eqn.~\ref{eq:modeOp}. 
For the middle frame of each segment, we generate the ground truth for semantic segmentation task by using  Deeplabv3+~\cite{chen2018encoder} pretrained on Cityscapes dataset~\cite{Cordts2016Cityscapes}  the depth map by using   Monodepth2~\cite{monodepth2:iccv19} pretrained on KITTI dataset~\cite{Geiger2013IJRR}, and the motion results by FlowNet2~\cite{ilg2017flownet}.  
We use \emph{AuditoryTestPseudo} to refer a large test set generated this way. In order to more reliably evaluate the method, we manually annotate the middle frame of $80$ test video segments for the three considered classes, namely car, train and motorcycle. We carefully select the video segments such that they cover diverse scenarios such as daylight, night, foggy and rainy.
We use LabelMeToolbox~\cite{russell2008labelme} for the annotation and follow the annotation procedure of Cityscapes~\cite{Cordts2016Cityscapes}. We call this test set \emph{AuditoryTestManual}.

For all the experiments, a training or a testing sample consists of a 2-second video segment and eight 2-second audio channels. We preprocess audio samples following techniques from ~\cite{sound:pixels:eccv18,gao20192}. We keep the audio samples at 96kHz and their amplitude is normalized to a desired RMS level, which we set to $0.1$ for all the audio channels. For normalization, we compute mean RMS values of amplitude over the entire dataset separately for each channel. An STFT is applied to the normalized waveform, with a window size of $512$ ($5.3$ms), hop length of $160$ ($1.6$ms) resulting a Time-Frequency representation of size of $257\times601$ pixels. Video frames are resized to $960\times1920$ pixels to fit to the GPU.

\smallskip
\noindent
\textbf{Implementation Details}. We train our complete model using Adam solver~\cite{kingma2014adam} with a learning rate of 0.00001 and we set a batch size of $2$. We train our models on GeForce GTX 1080 Ti GPUs for 20 epochs. For joint training of all the four tasks, we keep $\lambda_{1}=0.2$, $\lambda_{2}=0.2$ and $\lambda_{3}=0.2$ in Eq.~\ref{eqn:loss} for spectrogram inputs while we keep $\lambda_{1}=0.2$, $\lambda_{2}=0.2$ and $\lambda_{3}=0.02$ in Eq.~\ref{eqn:loss} for ddsp model and the model combining ddsp and spectrogram inputs. The parameters are set to let all losses contribute at a similar magnitude. We find that this simple method works very well for our tasks.  A grid search approach can be employed if one wants to find even better values for those parameters.

\smallskip
\noindent
\textbf{Evaluation metrics}. We use the standard mean IoU for the semantic prediction task.
For audio super resolution, we use MSE error for the  spectrograms and the envelope error for the waveforms as used  in~\cite{gao20192}. For depth prediction, we employ RMSE, MSE, Abs Rel and Sq Rel by following~\cite{monodepth2:iccv19}. For motion result, we use the standard metric EPE used for optical flow evaluation~\cite{ilg2017flownet}.

\begin{table*}[!tb]
  \centering
  \begin{adjustbox}{max width=\textwidth,max totalheight=\textheight}
  \begin{tabular}{lccccccccccccccc}
\toprule
& \multirow{ 2}{*}{Methods}  & \multicolumn{2} {c} {Microphone} & \multicolumn{3} {c} {Auxiliary Tasks} & \multicolumn{4} {c} {AuditoryTestPseudo} & \multicolumn{4} {c} {AuditoryTestManual} \\
 & &  Mono & Binaural & S$^3$R & Depth & Motion  & Car & MC & Train & All & Car & MC & Train & All \\ \midrule
& Visual  & & & & & & - & - & - & - & 79.01 & 39.07 & 77.34 & 65.35 \\ \cdashline{1-15}
\parbox[t]{2mm}{\multirow{9}{*}{\rotatebox[origin=c]{90}{Spectrogram}}} &  Mono   &  \cmark & & & & & 33.53 & 7.86 & 24.99 & 22.12 & 30.13 & 9.21 & 24.1 & 21.14 \\ 
& Ours(B)   &   & \cmark & & & & 35.80 & 19.51 & 40.71 & 32.01 & 35.30 & 13.28& 35.48& 28.02\\
& Ours(B:S)  &   & \cmark &  \cmark & & & 35.62 & 36.81 & 56.49  & 42.64 & 38.12 & 26.5 & 49.02 & 37.80 \\ 
& Ours(B:D)  &   & \cmark &   &  \cmark & & 33.53 & 28.01 &  55.32 & 38.95  & 32.42 & 25.8 & 50.12 & 36.11 \\ 
& Ours(B:M)  &   & \cmark &   &  & \cmark & 33.86 & 25.87 &  53.24 & 37.66  & 32.51 & 23.73 & 49.13 & 35.12 \\ 
& Ours(B:SD)  &   & \cmark &  \cmark &  \cmark & & 35.81 & 38.14 & 56.25  & 43.40  & 35.51 & 28.51 & 50.32 & 38.01 \\ 
& Ours(B:SM)  &   & \cmark &  \cmark &   & \cmark & 35.58 & 37.68 & 56.28  & 43.12  & 37.59 & 26.92 & 49.92 & 38.14 \\
& Ours(B:DM)  &   & \cmark &   &  \cmark &  \cmark & 35.23 & 37.34 & 56.17  & 42.91  & 36.26 & 26.56 & 49.95 & 37.59 \\
& Ours(B:SDM)  &   & \cmark & \cmark & \cmark & \cmark & 36.34 & 37.18 &  58.23 & \textbf{43.92}  & 37.81 & 26.34 & 50.67 & \textbf{38.27} \\ 
\midrule
\parbox[t]{2mm}{\multirow{9}{*}{\rotatebox[origin=c]{90}{DDSP}}} & Mono  & \cmark & &   &  & & 23.30 & 2.05 & 15.98  & 13.77 & 21.47 & 2.56 & 13.07 & 12.33 \\ 
& Ours(B)  &   & \cmark &   &  & & 32.01 & 7.57 & 29.44  & 23.01 & 31.11 & 5.25 & 25.67 & 20.67 \\  
& Ours(B:S)  &   & \cmark &  \cmark &  & & 34.47 & 10.32 & 31.31  & 25.40 & 34.22 & 9.78 & 31.16 & 25.05 \\ 
& Ours(B:D)  &   & \cmark &   & \cmark & & 31.22 & 8.55 & 30.23  & 23.33  & 30.64 & 7.87 & 27.66 & 22.05 \\
& Ours(B:M)  &   & \cmark &   &  &\cmark & 31.75 & 9.02 &  31.11 & 23.96  & 30.55 & 8.24 & 28.78 & 22.52 \\
& Ours(B:SD)  &   & \cmark &  \cmark & \cmark & & 34.12 & 9.89 & 31.65 & 25.22 & 32.84 & 10.03 & 31.55 & 24.80 \\ 
& Ours(B:SM)  &   & \cmark &  \cmark &   & \cmark & 34.21 & 10.31 & 32.56  & 25.69  & 33.03 & 10.11 & 32.43 & 25.19 \\
& Ours(B:DM)  &   & \cmark &   &  \cmark & \cmark & 33.11 & 9.52 & 31.26  & 24.63 & 32.06 & 9.13 & 30.97 & 24.05 \\
& Ours(B:SDM)  &   & \cmark & \cmark  & \cmark & \cmark & 34.93 & 9.45 & 33.46 & \textbf{25.95}  & 34.18 & 9.24 & 32.91 & \textbf{25.44} \\ 
\midrule
\parbox[t]{2mm}{\multirow{9}{*}{\rotatebox[origin=c]{90}{Spectrogram + DDSP}}} & Mono  & \cmark & &  & & & 29.78 & 12.67 & 31.10  & 24.51  & 27.63 & 12.58 & 23.45 & 21.22 \\ 
& Ours(B)  &   & \cmark &   &  & & 36.02 & 25.77 & 46.60  & 36.13 & 34.58 & 21.37 & 40.88 & 32.27 \\
& Ours(B:S)  &   & \cmark & \cmark  & & & 40.22 & 33.85 & 59.19 & 44.42 & 38.55 & 30.32 & 52.36 & 40.41 \\ 
& Ours(B:D)  &   & \cmark &   & \cmark & & 36.70 & 32.79 & 54.20  & 41.23  & 34.07 & 28.11 & 50.35 & 37.51 \\
& Ours(B:M)  &   & \cmark &   &  &\cmark & 37.35 & 28.68 &  55.25 & 40.43  & 34.62 & 26.48 & 51.23 & 37.44 \\ 
& Ours(B:SD)  &   & \cmark & \cmark  & \cmark & & 40.36 & 37.31& 58.97 & 45.54  & 39.78 & 32.45 & 51.83 & 41.35 \\
& Ours(B:SM)  &   & \cmark &  \cmark &   & \cmark & 40.23 & 37.02 & 59.23  & 45.49  & 40.02 & 32.12 & 52.09 & 41.41 \\
& Ours(B:DM)  &   & \cmark &   &  \cmark & \cmark & 40.10 & 36.92 & 58.76  & 45.26 & 39.47 & 31.84 & 51.51 & 40.94 \\
& Ours(B:SDM)  &   & \cmark & \cmark & \cmark & \cmark & 41.07 & 36.82 &  59.80 & \textbf{45.89}  & 40.72 & 32.32 & 53.45 & \textbf{42.16} \\ 
\bottomrule 
\end{tabular} 
\end{adjustbox}
\caption{Results of auditory semantic prediction. The results of DeepLabv3+ on the target middle frame (Visual) are reported for reference purpose.  mIoU (\%) is used. MC denotes Motorcycle.} 
  \label{tab:maintable}
\end{table*} 

\subsection{Auditory Semantic Prediction}
We compare the performance of different methods and report the results in Tab.~\ref{tab:maintable}. The table shows that our method learning with sounds can generate promising results for  dense semantic object prediction.  We also find that using binaural sounds \emph{Ours(B)} generates significant better results than using \emph{Mono} sound. The conclusion is true for all three considered encoders. This is because the major cues for sound localization such as ILD, ITD, and HRTF are missing in \emph{Mono} sound.  
Another very exciting observation is that joint training with the depth prediction task, the motion estimation task, and the S$^3$R task is beneficial to the semantic prediction task. This is because all the four tasks benefit from the same common goal -- extracting spatial information from the input binaural sounds. A joint training of the four tasks   allows to share supervision across tasks, and that can regularize network training. Below, we present our ablation studies.

\noindent 
\textbf{S$^{3}$R, depth prediction and motion prediction all give a boost}. As can be seen in Tab.~\ref{tab:maintable}, by adding any one of the three tasks as an auxiliary task, indicated by \emph{Ours(B:S)}, \emph{Ours(B:D)} or \emph{Ours(B:M)} respectively, improves the performance of our baseline \emph{Ours(B)} clearly. 
It can be also observed that using two of the auxiliary tasks, indicated by \emph{Ours(B:SD)}, \emph{Ours(B:SM)} or \emph{Ours(B:DM)} yields better performance than using one, and using all the three tasks together, indicated by \emph{Ours(B:SDM)} achieves the best performance.  This shows that the four tasks are mutually beneficial and they should be trained together.

\begin{figure}[t]
    \centering
    \subfloat[Visual scene]{\includegraphics[width=0.4\textwidth]{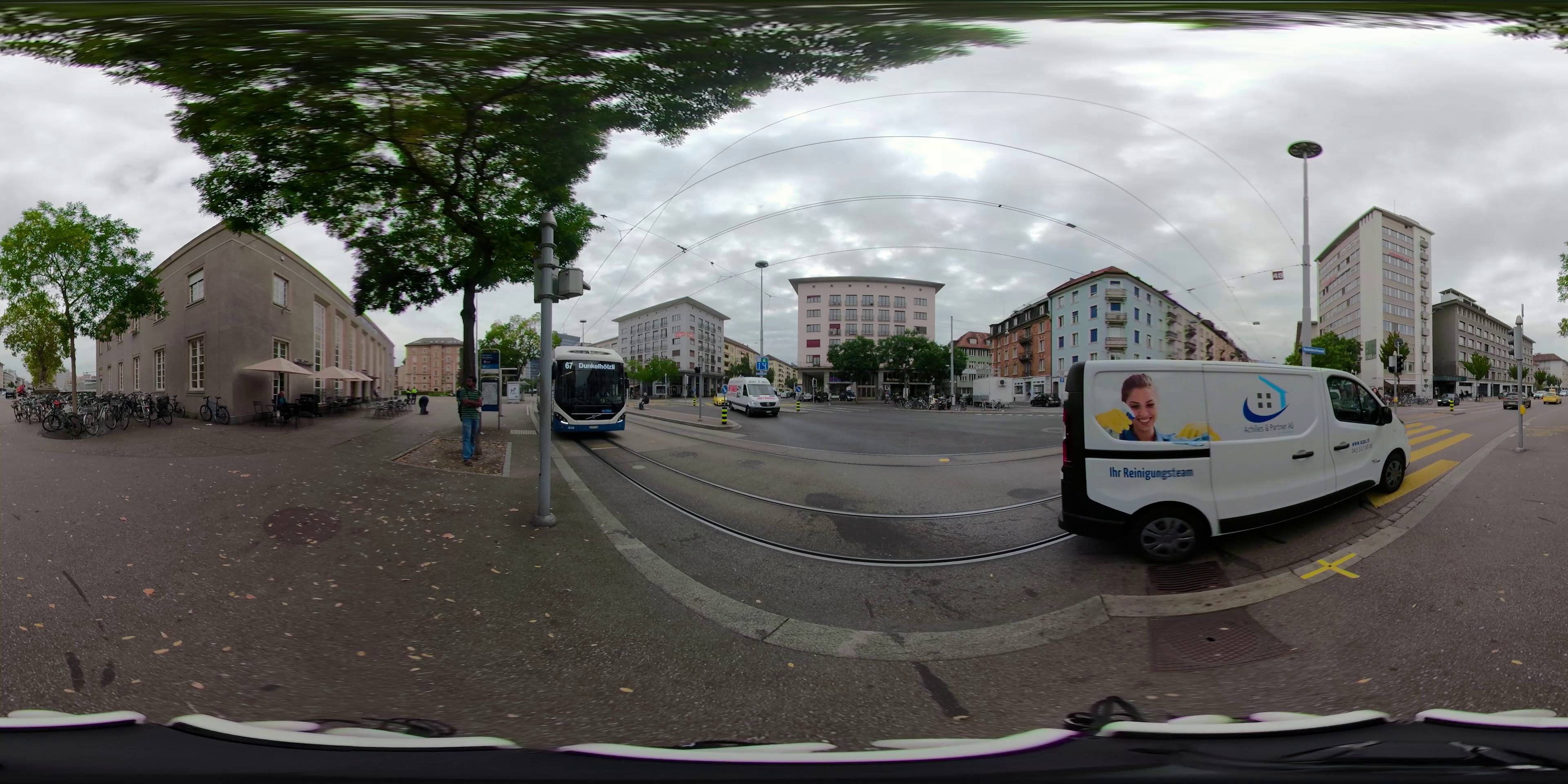}}
    \hfil
    \subfloat[Semantic prediction]{\includegraphics[width=0.4\textwidth]{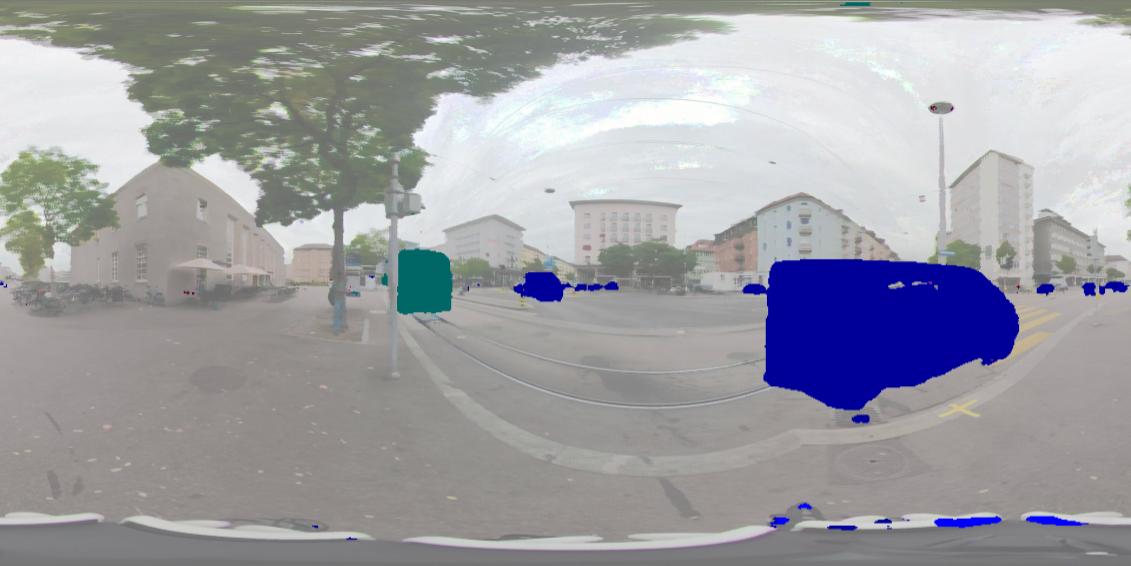}}
    \\ 
    \subfloat[Spec: 0]{\adjustbox{trim={.12\width} {.2\height} {0.11\width} {.150\height},clip}{\includegraphics[width=0.3\textwidth,height=0.22\textwidth]{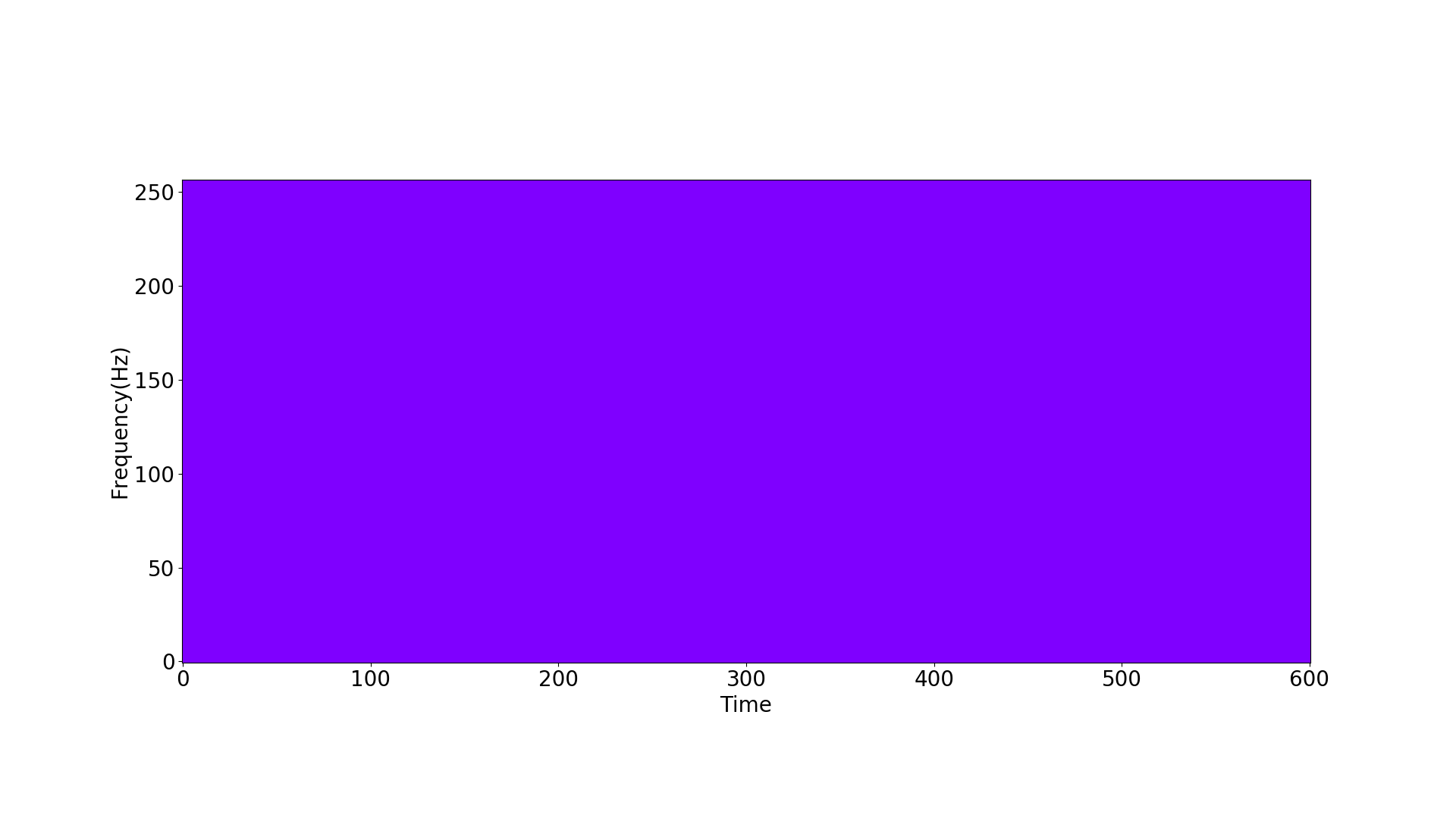}}}
    \hfil
    \subfloat[Spec: 0.5]{\adjustbox{trim={.12\width} {.2\height} {0.11\width} {.150\height},clip}{\includegraphics[width=0.3\textwidth,height=0.22\textwidth]{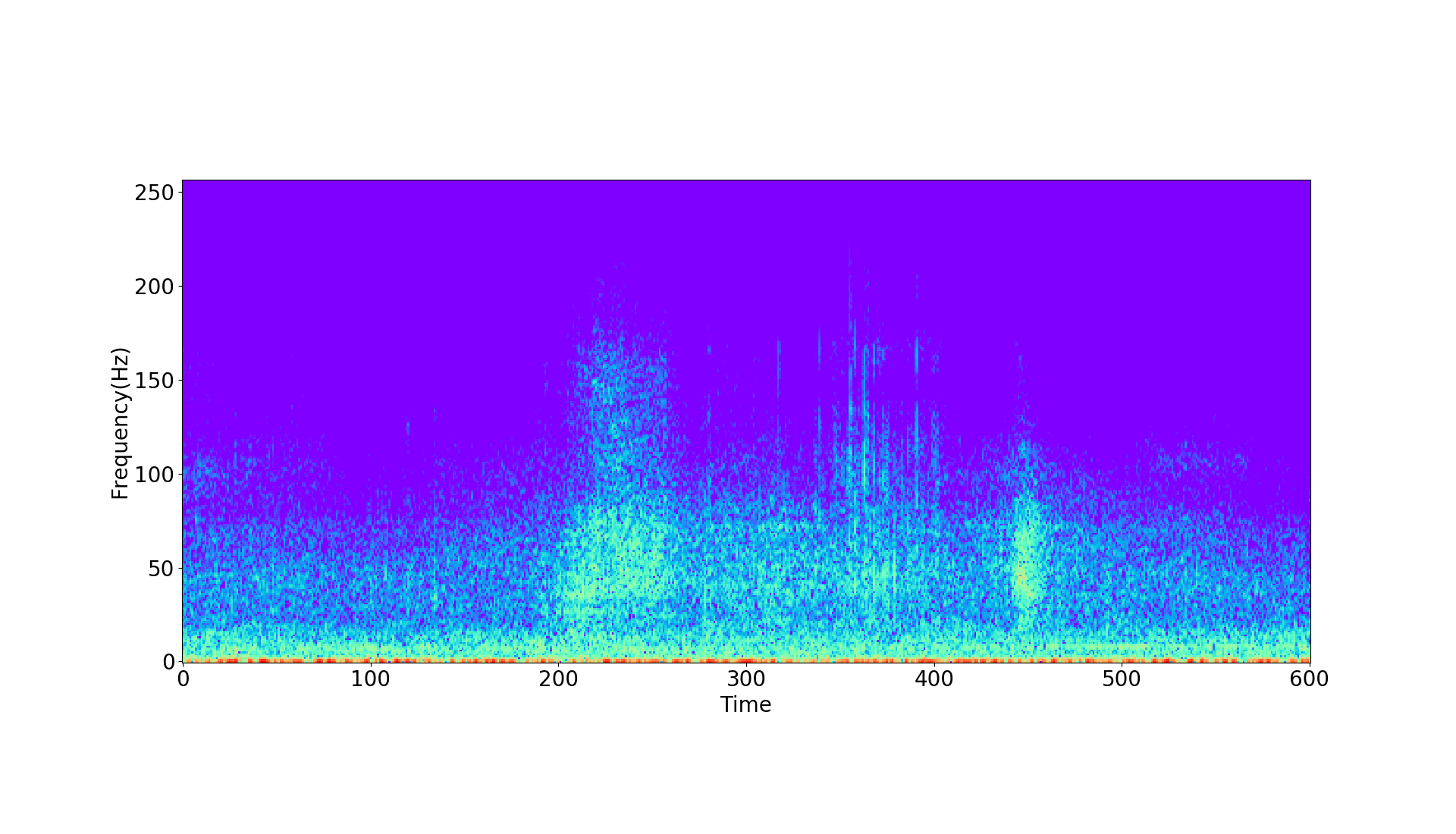}}}
    \hfil
    \subfloat[Spec: 1]{\adjustbox{trim={.12\width} {.2\height} {0.11\width} {.150\height},clip}{\includegraphics[width=0.3\textwidth,height=0.22\textwidth]{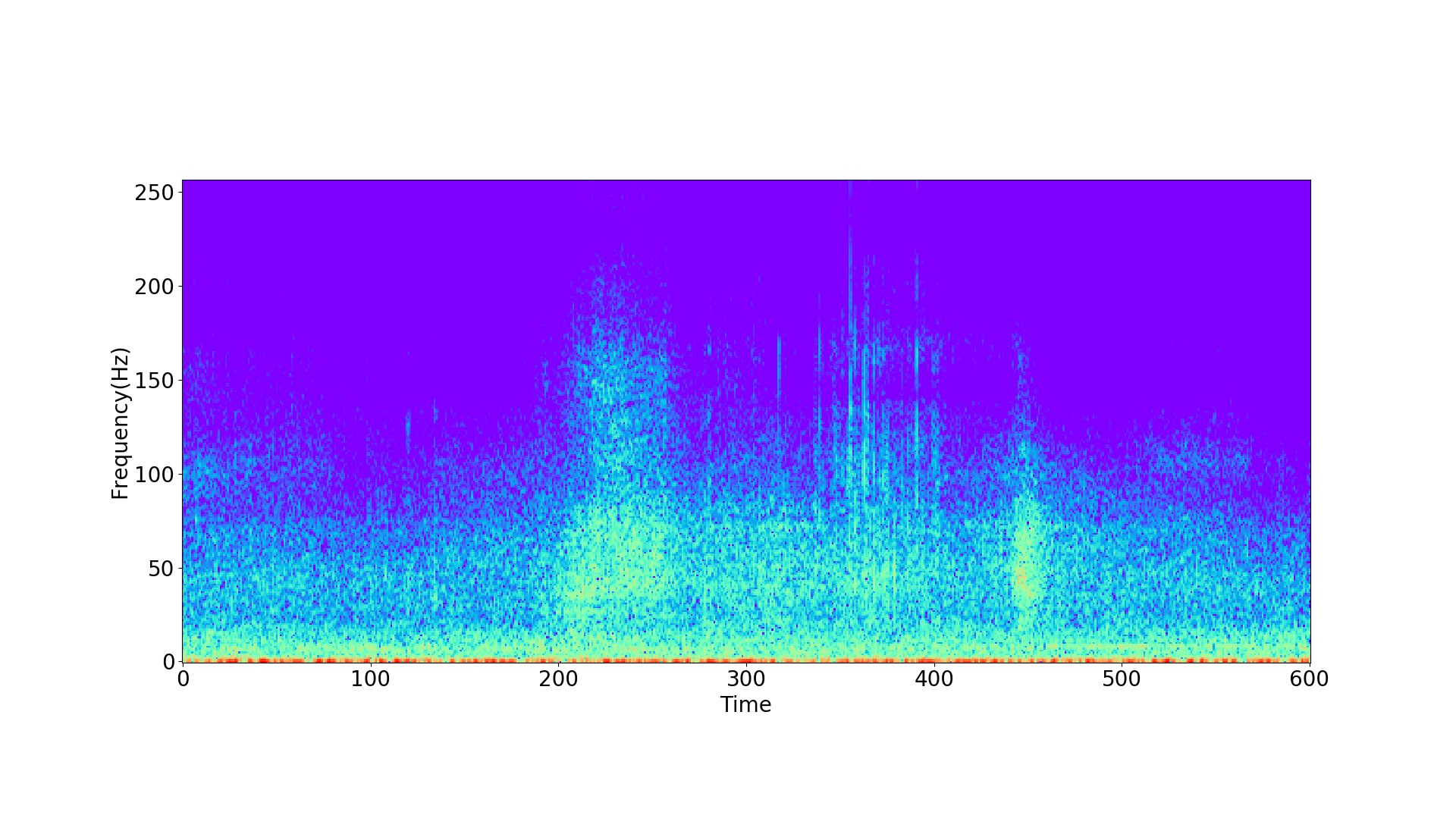}}}
    \hfil
    \subfloat[Spec: 2]{\adjustbox{trim={.12\width} {.2\height} {0.11\width} {.150\height},clip}{\includegraphics[width=0.3\textwidth,height=0.22\textwidth]{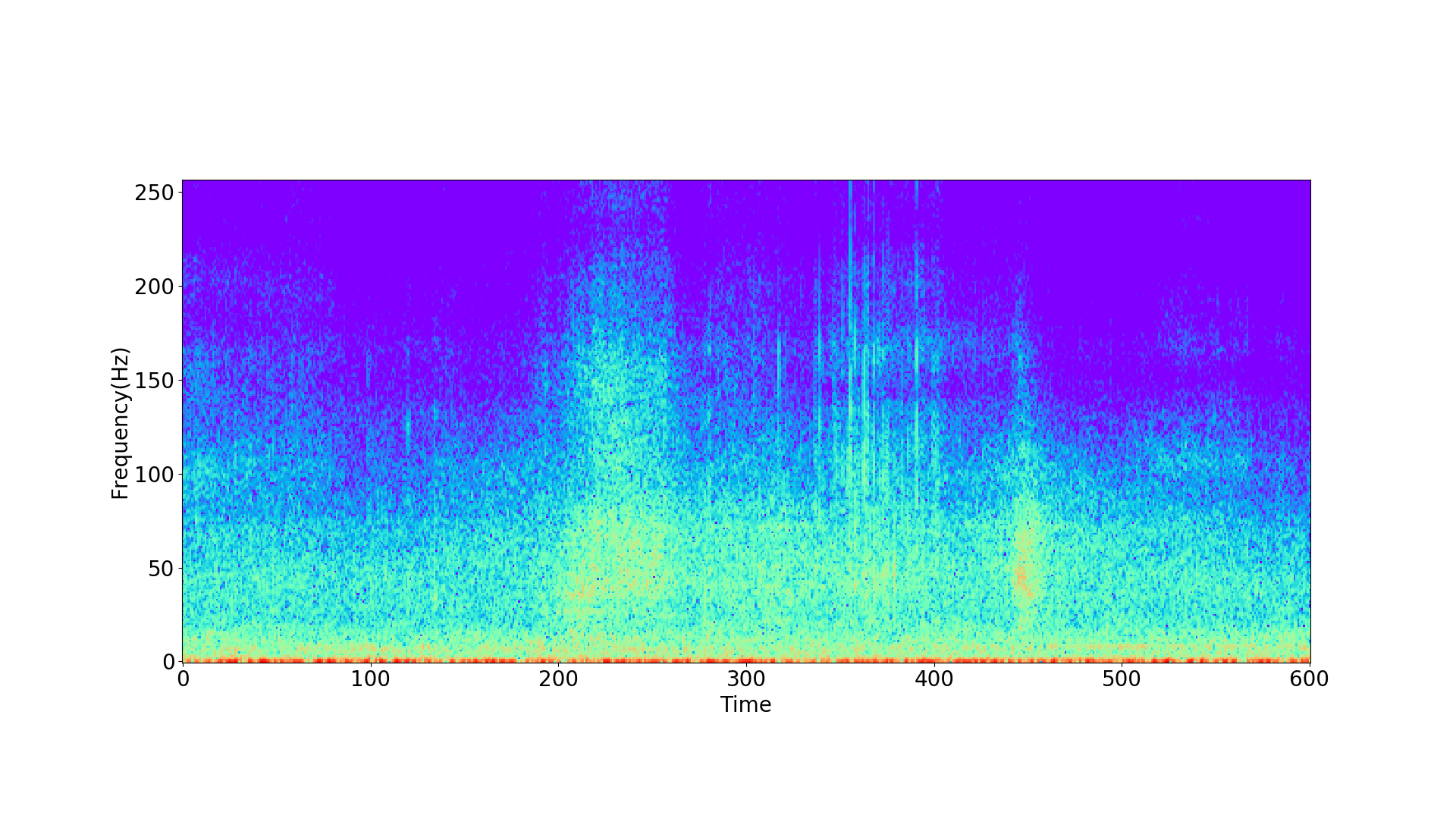}}}
    \\
    \subfloat[Result: 0]{\includegraphics[width=0.23\textwidth,height=0.13\textwidth]{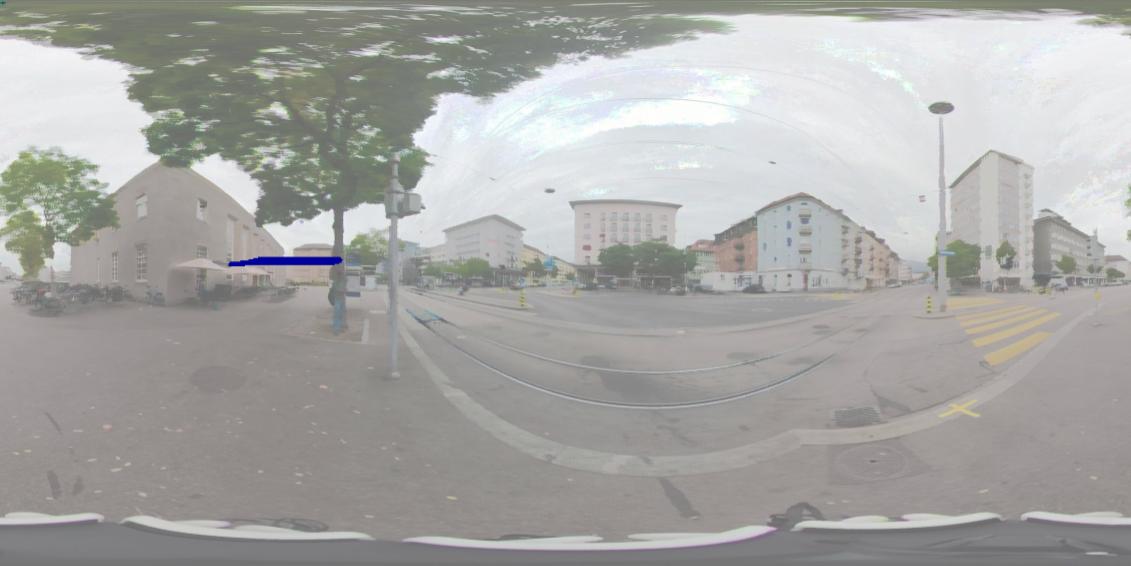}}
    \hfil
    \subfloat[Result: 0.5]{\includegraphics[width=0.23\textwidth,height=0.13\textwidth]{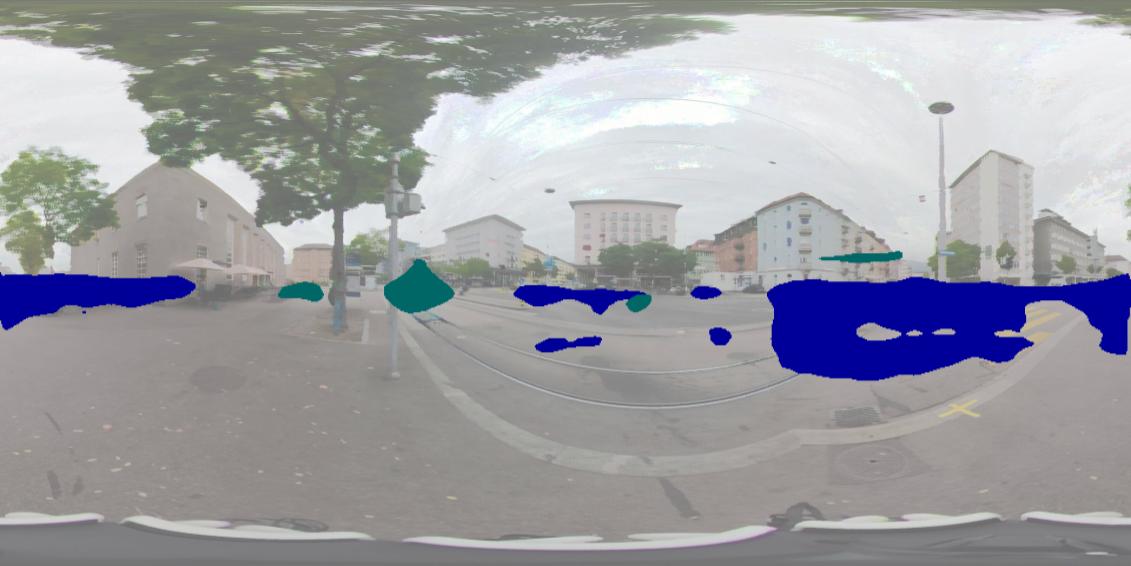}}
    \hfil
    \subfloat[Result: 1]{\includegraphics[width=0.23\textwidth,height=0.13\textwidth]{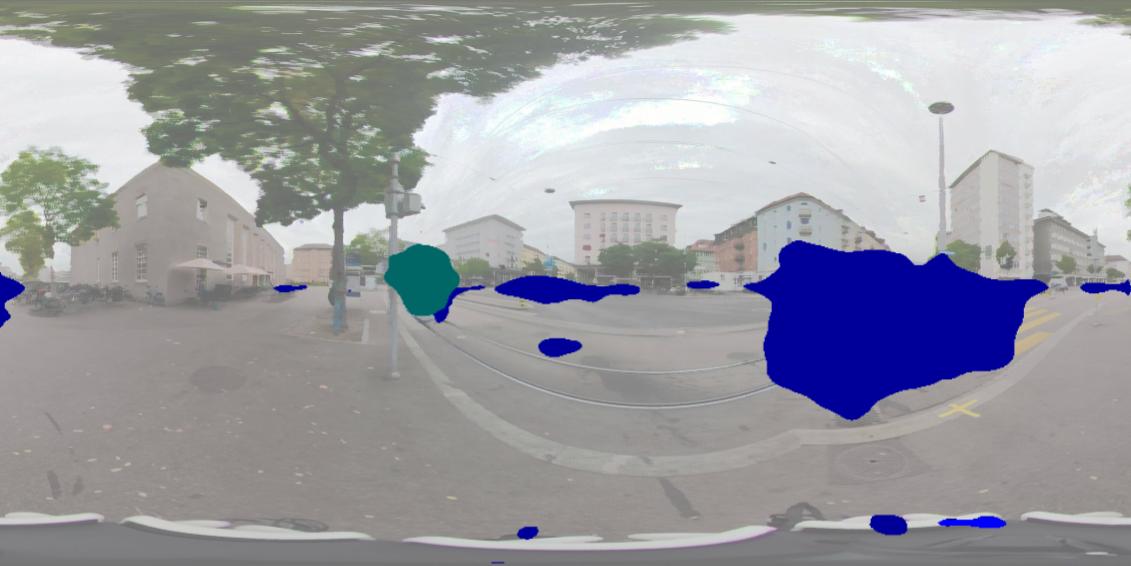}}
    \hfil
    \subfloat[Result: 2]{\includegraphics[width=0.23\textwidth,height=0.13\textwidth]{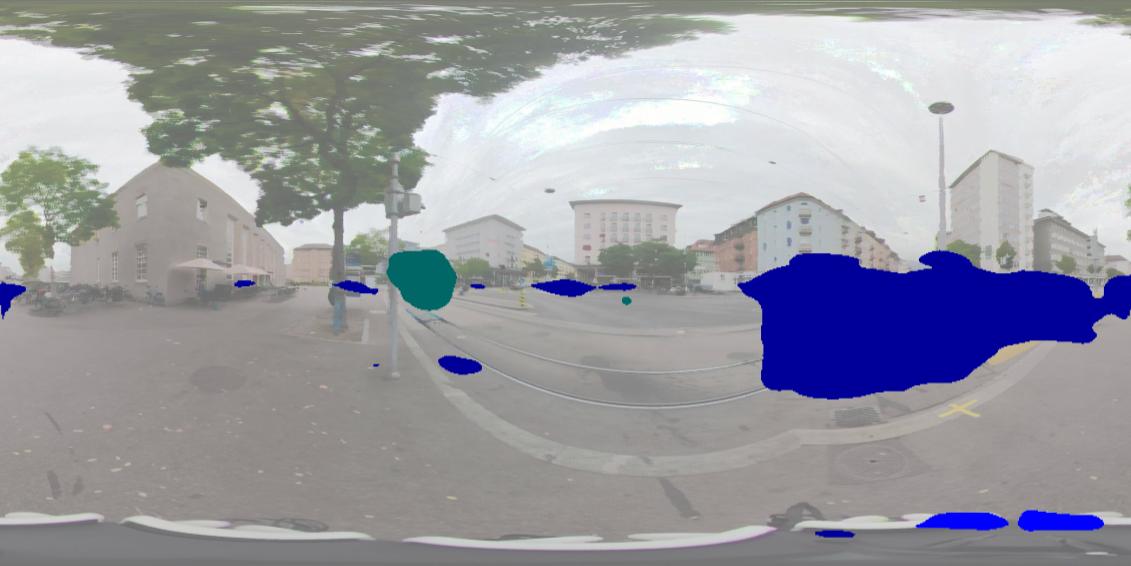}}
    \\
    
    \resizebox{0.7\linewidth}{!}{
\begin{tikzpicture}[tight background, scale=0.75, every node/.style={font=\small}]
	\draw[white, fill=car, draw=white] (0,0) rectangle (1* 4, 1) node[pos=0.5] {Car};
	\draw[white, fill=train, draw=white] (1 * 4,0) rectangle (2 * 4, 1) node[pos=0.5] {Train};
	\draw[white, fill=motorcycle, draw=white] (2 * 4,0) rectangle (3 * 4, 1) node[pos=0.5] {Motorcycle};
\end{tikzpicture}}
    \caption{Qualitative results of auditory semantic prediction by our approach with  input audio signal of different amplification ratios (\ie $0, 0.5, 1,$ and $2$) at inference stage. We show the input spectrograms (Spec) and the multiplier of amplitude for the signal from (c) to (f) and corresponding semantic prediction in (g) to (j). We show the visual scene and ground truth semantic prediction in (a) and (b) respectively.}
    \label{fig:varyamplitude}
\end{figure}
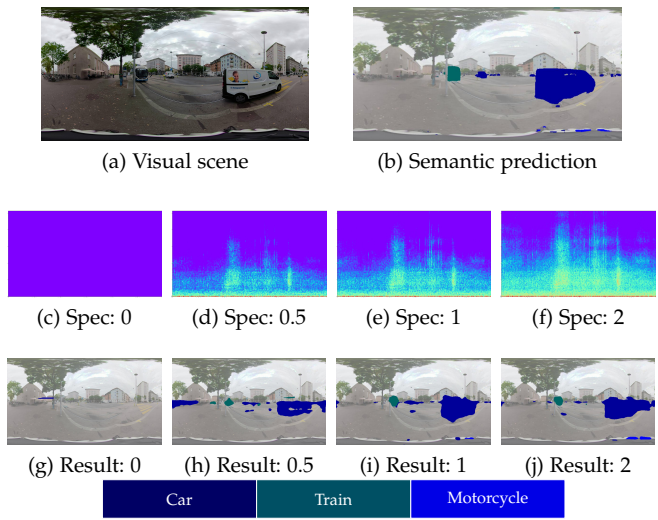

\begin{figure*}[!tb]
\begin{tabular}{cccc}
  \centering \hspace{-2mm}
  \includegraphics[trim=160 20 140 65,clip,width=0.35\textwidth,height=0.19\textwidth]{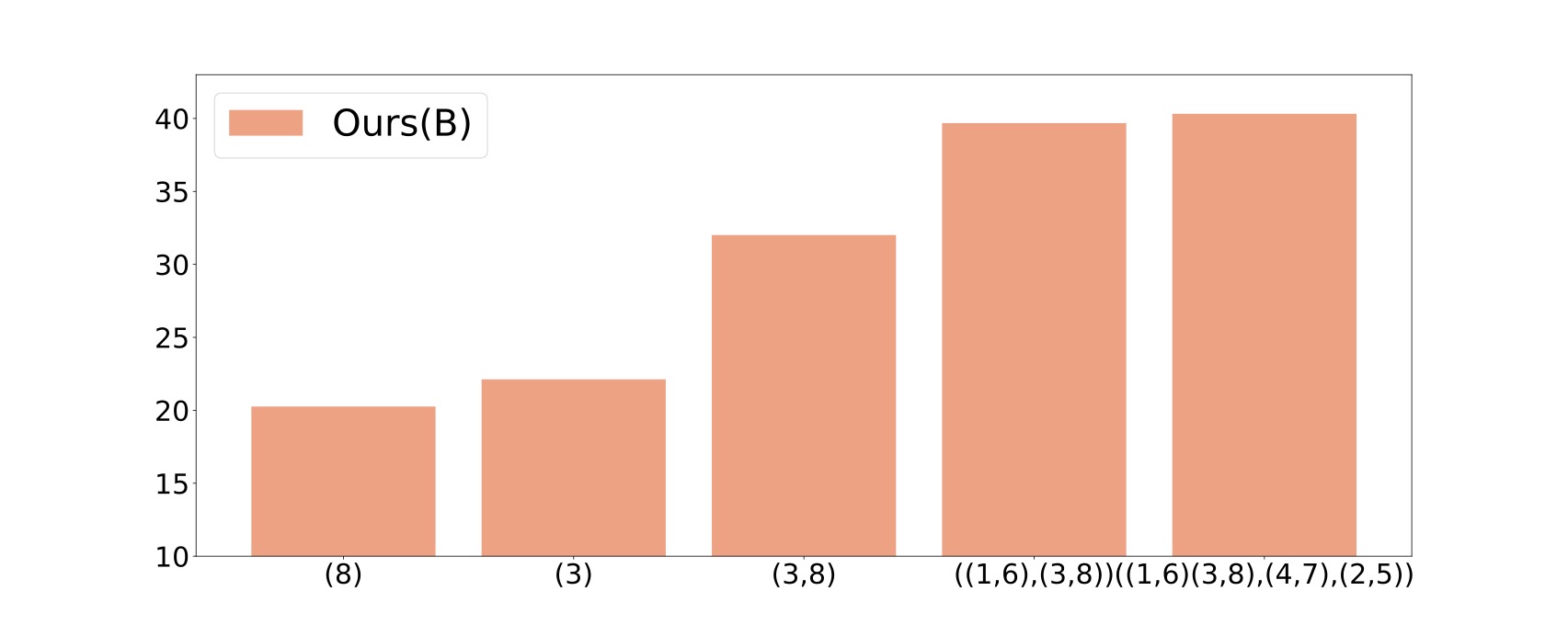} & \hspace{-4mm}
  \includegraphics[trim=45 10 50 45,clip,width=0.22\textwidth,height=0.19\textwidth]{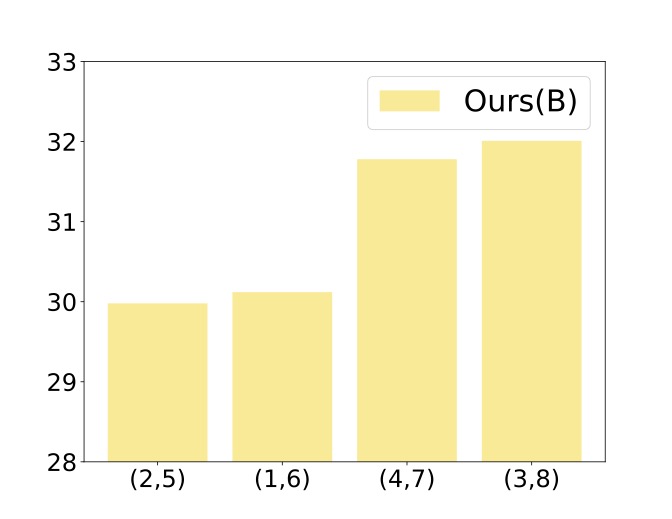} & \hspace{-4mm}
  \includegraphics[trim=150 20 100 65,clip,width=0.42\textwidth,height=0.19\textwidth]{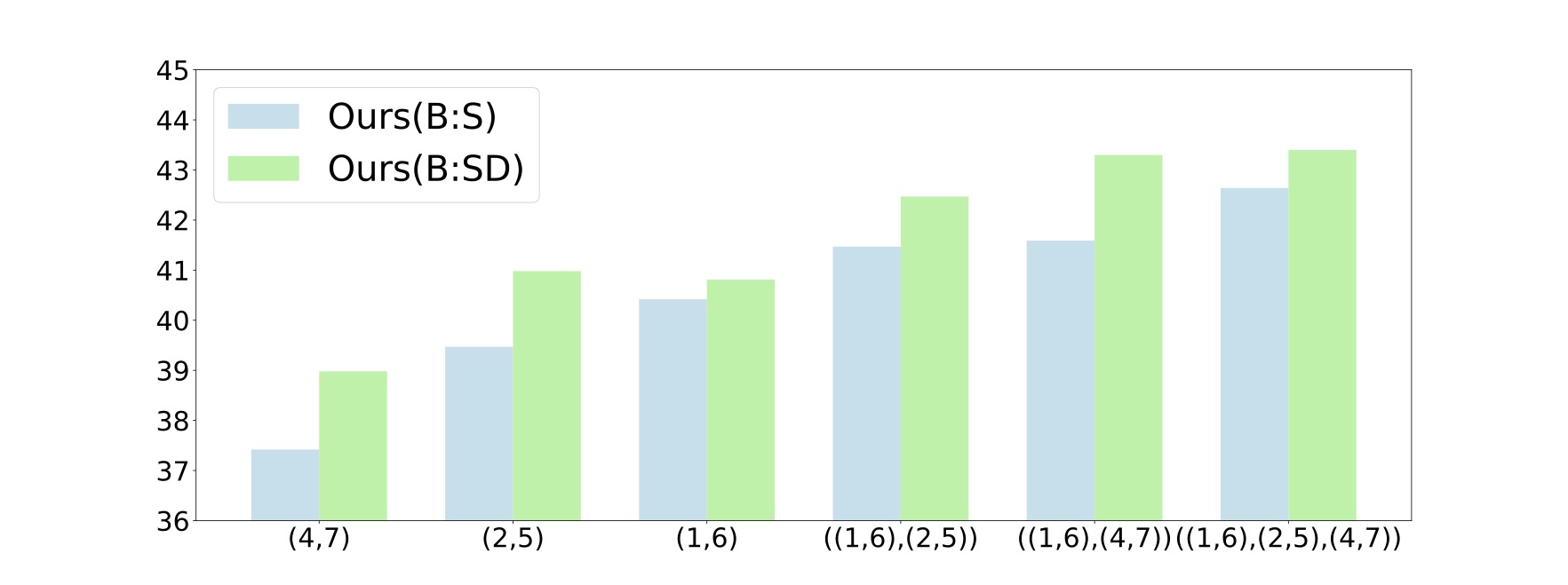}  
  \\
  \text{(a) number of \textbf{input} microphones}  &
  \text{(b) \textbf{input} microphone pairs} &\text{(c) number of of \textbf{output} microphone} 
  \\ 
  \caption{Results (mIoU) of semantic prediction: different sets of microphones used as inputs under \emph{Ours(B)} in (a) and (b) and varying number of output microphones for the auxiliary task S$^3$R, under \emph{Ours(B:S)} and \emph{Ours(B:SD)}, in (c).}
  \label{fig:mic_ablation}
  \end{tabular} 
\end{figure*}

\noindent
\textbf{Adding input channels increases the performance}. We compare the auditory semantic prediction (without using auxiliary tasks) accuracies in Fig.~\ref{fig:mic_ablation}(a) for different set of input microphones. Here, we experiment with a) Mono sound from 3 (front) or 8 (back) microphone, b) binaural sounds from the pair (3,8), c) 2 pairs of binaural sound channels ((1,6),(3,8)) which faces in four orthogonal directions, and d) 4 pairs of binaural sound channels. We see that semantic prediction accuracy increases from $22.12$\% when using \emph{Mono} sound to $40.32$\% when using all 8 channels. This shows that the performance of semantic prediction improves with the number of sound channels.
 
\smallskip
\noindent
\textbf{Orientation of microphones matter}. 
Fig.~\ref{fig:mic_ablation}(b) shows the auditory semantic prediction results (without using auxiliary tasks) from different orientations of the input binaural pairs for the same scene. The pair (3,8) is aligned in parallel with the front facing direction 
of the camera as can be seen in Fig.~\ref{fig:setup}. We define this as orientation of 0\degree. 
Then, we have other pairs (1,6), (4,7) and (2,5) orientating at azimuth angles of 90\degree, 180\degree and 270\degree respectively. 
We observe that (3,8) and (4,7) outperform the other two pairs. Intuitively, the left-right pair should be better for auditory perception.  The reason that a front-back pair performs better could be that for semantic prediction, especially for long-range objects, the microphone facing straight to the objects has the advantage to detect  weak audio signals. Also, at road intersections, many sounding objects move in the left-right direction, so the front-rear mic pair is practically under the left-right configuration relative to those sounding objects. The optimal  orientations of microphones for auditory perception is indeed an interesting problem that deserves more investigation. Our dataset provides a great testbed to study this problem.

\begin{table*}
  \setlength\tabcolsep{1pt}
    \begin{minipage}{.6\textwidth}
        \centering
    \begin{tabular}{cccccccccccccc}
\toprule
 \multirow{2}{*}{} &  \multicolumn{2} {c} {Microphone} & \multicolumn{3}{c}{Joint Tasks} &  \multicolumn{4}{c}{Metrics} \\
 & Mono & Binaural & Semantic & S$^3$R & Motion  & Abs Rel & Sq Rel & RMSE & MSE \\  \midrule
\parbox[t]{2mm}{\multirow{9}{*}{\rotatebox[origin=c]{90}{Spectrogram}}} & \cmark& & & & & 118.88 & 622.58 & 5.413& 0.365 \\
 & &\cmark & & & & 108.59 & 459.69 & 5.263& 0.331 \\
 & &\cmark&\cmark & & & 87.96 & 290.06 & 5.136 & 0.315 \\
 & &\cmark& & \cmark & & 90.43 & 400.53 & 5.193 & 0.318 \\ 
 & &\cmark& &  & \cmark & 92.34 & 412.11 & 5.213 & 0.322 \\ 
  & &\cmark&\cmark & \cmark & & 84.24 &  222.41 &  5.117 &  0.310 \\
 & &\cmark&\cmark & & \cmark & 86.45 & 261.54 & 5.127 & 0.313 \\ 
 & &\cmark& &\cmark & \cmark& 89.71 & 375.66 & 5.185 & 0.318 \\ 
 & &\cmark& \cmark &\cmark & \cmark& \textbf{83.22} & \textbf{210.13} & \textbf{5.108} & \textbf{0.308} \\
 \midrule
\parbox[t]{2mm}{\multirow{9}{*}{\rotatebox[origin=c]{90}{DDSP}}} & \cmark& & & & & 134,56 & 798.22 & 5.572 & 0.397 \\
& &\cmark & & & & 122.16 &  628.39 & 5.452 & 0.381 \\
& &\cmark&\cmark & &  & 112.45 &  532.78 & 5.357 & 0.361 \\
& &\cmark& & \cmark & & 110.61 &  501.07  & 5.334 & 0.347 \\
& &\cmark& &  & \cmark & 115.22 & 563.25 & 5.388 & 0.363 \\ 
& &\cmark&\cmark & \cmark & & 108.83 &  467.54 & 5.301 &  0.341 \\
& &\cmark&\cmark & & \cmark & 111.34 & 519.35 & 5.343 & 0.354 \\ 
& &\cmark& &\cmark & \cmark& 109.46 & 485.40 & 5.314 & 0.344 \\ 
& &\cmark& \cmark &\cmark & \cmark& \textbf{108.11} & \textbf{451.64} & \textbf{5.292} & \textbf{0.339} \\
 \midrule
\parbox[t]{2mm}{\multirow{9}{*}{\rotatebox[origin=c]{90}{Spectrogram + DDSP}}} & \cmark& & & & & 117.21 & 605.38 & 5.393 & 0.364 \\
& &\cmark & & & & 105.71 & 447.82 & 5.244 & 0.327 \\
& &\cmark&\cmark & & & 87.33 & 274.43 & 5.154 & 0.316 \\ 
& &\cmark& & \cmark & & 86.82 & 266.66 & 5.131 & 0.315 \\ 
& &\cmark& &  & \cmark & 90.34 & 391.22 & 5.191 & 0.318 \\
& &\cmark&\cmark & \cmark & & 83.49 &  215.22 & 5.110 &  0.308 \\
& &\cmark& \cmark & & \cmark& 86.10 & 256.75 & 5.125 & 0.313 \\
& &\cmark&  &\cmark & \cmark& 85.51 & 244.86 & 5.120 & 0.310 \\
& &\cmark& \cmark &\cmark & \cmark& \textbf{82.36} & \textbf{201.57} & \textbf{5.106} & \textbf{0.302} \\
\bottomrule
\end{tabular} 
\text{(a) auditory depth prediction}
    \end{minipage}%
    \begin{minipage}{.4\textwidth}
        \centering
        \begin{tabular}{ccccccccccc}
\toprule
\multirow{2}{*}{} &  \multicolumn{2} {c} {Microphone} & \multicolumn{3}{c}{Joint Tasks} &  \multicolumn{1}{c}{Metric} \\
& Mono & Binaural & Semantic & S$^3$R  & Depth & EPE  \\  \midrule
\parbox[t]{2mm}{\multirow{9}{*}{\rotatebox[origin=c]{90}{Spectrogram}}} & \cmark& & & &  & 22.11  \\
&  &\cmark & & &  & 16.22 \\
&  &\cmark&\cmark & &  & 15.01  \\
&  &\cmark& & \cmark &  & 15.11  \\ 
&  &\cmark& &  & \cmark & 15.48  \\ 
&  &\cmark& \cmark & \cmark &  &  14.29 \\
&  &\cmark&\cmark &  &  \cmark &  14.67 \\
&  &\cmark& & \cmark &  \cmark &  14.91 \\
&  &\cmark&\cmark & \cmark &  \cmark &  \textbf{14.04} \\
 \midrule
\parbox[t]{2mm}{\multirow{9}{*}{\rotatebox[origin=c]{90}{DDSP}}} & \cmark& & & &  & 25.43  \\
& &\cmark & & &  & 21.40 \\
& &\cmark&\cmark & &  & 20.51  \\
& &\cmark& & \cmark &  & 20.77  \\ 
& &\cmark& &  & \cmark & 21.01  \\ 
& &\cmark& \cmark & \cmark &  &  20.11 \\
& &\cmark&\cmark &  &  \cmark &  20.43 \\
& &\cmark& & \cmark &  \cmark &  20.62 \\
& &\cmark&\cmark & \cmark &  \cmark &  \textbf{19.92} \\
 \midrule
\parbox[t]{2mm}{\multirow{9}{*}{\rotatebox[origin=c]{90}{Spectrogram + DDSP}}} & \cmark& & & &  & 20.28  \\
& &\cmark & & &  & 15.24 \\
& &\cmark&\cmark & &  & 14.34  \\
& &\cmark& & \cmark &  & 14.51  \\ 
& &\cmark& &  & \cmark & 14.83  \\ 
& &\cmark& \cmark & \cmark &  &  13.98 \\
& &\cmark&\cmark &  &  \cmark &  14.29 \\
& &\cmark& & \cmark &  \cmark &  14.21 \\
& &\cmark&\cmark & \cmark &  \cmark &  \textbf{13.77} \\
 \bottomrule
\end{tabular} 
\text{(b) auditory motion prediction}
    \end{minipage}
\caption{Results of auditory depth prediction and motion prediction. For all values, lower is better.}
\label{tab:depth:motion}
\end{table*}

\begin{table}[!bt]
  \centering 
  \small
  \setlength\tabcolsep{2pt}
  \begin{adjustbox}{max width=\textwidth,max totalheight=\textheight}
  \begin{tabular}{cccccccccccccc}
\cmidrule[\heavyrulewidth]{1-10}
  & \multicolumn{3}{c}{Joint Tasks} &  \multicolumn{2}{c}{Mic Ids} &  \multicolumn{4}{c}{Metrics} \\ 
 & Semantic & Depth & Motion & In & Out & MSE-1   & ENV-1  & MSE-2   & ENV-2 \\   \cmidrule(lr){1-10}
\parbox[t]{2mm}{\multirow{9}{*}{\rotatebox[origin=c]{90}{Spectrogram}}} & & & & (3,8) & (1,6) &0.1228 & 0.0298 &0.1591 & 0.0324\\ 
& \cmark & & & (3,8) & (1,6) &  0.0956 & 0.0214 & 0.1001 & 0.0243 \\
&  & \cmark & & (3,8) & (1,6) & 0.0984 & 0.0221 & 0.1044 & 0.0267 \\
&  &  & \cmark & (3,8) & (1,6) & 0.0980 & 0.0220 & 0.1045 & 0.0270 \\
&    \cmark  & \cmark & & (3,8) & (1,6) & 0.0978 & 0.0218 & 0.1040 & 0.0264\\ 
&   \cmark  &  & \cmark & (3,8) & (1,6) & 0.0943 & 0.0212 & 0.0977 & 0.0230 \\ 
&  & \cmark & \cmark & (3,8) & (1,6) & 0.0979 & 0.0219 & 0.1041 & 0.0266 \\ 
 &   \cmark  & \cmark & \cmark & (3,8) & (1,6) & \textbf{0.0940} & \textbf{0.0211} & \textbf{0.0971} & \textbf{0.0224} \\ 
 \cmidrule(lr){1-10}
\parbox[t]{2mm}{\multirow{9}{*}{\rotatebox[origin=c]{90}{DDSP}}} & & & & (3,8) & (1,6) & 0.1285 & 0.0301 & 0.1617 & 0.0344 \\
& \cmark & & & (3,8) & (1,6) &  0.0987 & 0.0233 & 0.1123 & 0.0262\\
 & & \cmark & & (3,8) & (1,6) & 0.1180 & 0.0271 & 0.1303 & 0.0295 \\
&  & & \cmark & (3,8) & (1,6) & 0.1151 & 0.0265 & 0.1282 & 0.0287 \\
&  \cmark  & \cmark &  & (3,8) & (1,6) & 0.1003 & 0.0259 & 0.1191 & 0.0263. \\ 
&    \cmark  &  & \cmark & (3,8) & (1,6) & 0.0981 & 0.0225 & 0.1101 & 0.0260\\ 
&  & \cmark & \cmark & (3,8) & (1,6) & 0.1104 & 0.0252 & 0.1243 & 0.0275 \\ 
&  \cmark  & \cmark & \cmark & (3,8) & (1,6) & \textbf{0.0979} & \textbf{0.0219} & \textbf{0.1092} & \textbf{0.2590}\\ 
 \cmidrule(lr){1-10}
\parbox[t]{2mm}{\multirow{9}{*}{\rotatebox[origin=c]{90}{Spectro. + DDSP}}} & & & & (3,8) & (1,6) & 0.1202 & 0.0288 & 0.1514 & 0.0311 \\
 & \cmark & & & (3,8) & (1,6) &  0.0921 & 0.0200 & 0.0986 & 0.0233 \\
 &  & \cmark & & (3,8) & (1,6) & 0.0987 & 0.0226 & 0.1088 & 0.0282 \\
 & & &\cmark & (3,8) & (1,6) & 0.0983 & 0.0221 & 0.1076 & 0.0260 &\\
  &   \cmark  & \cmark & & (3,8) & (1,6) & 0.0973 & 0.0217 & 0.1056 & 0.0251 \\
 &  \cmark  &  & \cmark & (3,8) & (1,6) & 0.0912 & 0.0196 & 0.0971 & 0.0225 \\ 
  &  & \cmark & \cmark & (3,8) & (1,6) & 0.0980 & 0.0219 & 0.1069 & 0.0258 \\ 
   &  \cmark  & \cmark & \cmark & (3,8) & (1,6) & \textbf{0.0910} & \textbf{0.0195} & \textbf{0.0965} & \textbf{0.0212} \\ 
\cmidrule[\heavyrulewidth]{1-10} \\
\end{tabular} \end{adjustbox} 
\caption{The results of our S$^3$R task. MSE1 and MSE2 represent mean squared error while ENV1 and ENV represent envelope error for the 2 output channels of binaural sounds. For all metrics, lower score is better.}
  \label{tab:spatialresolutionresults}
\end{table} 

\smallskip
\noindent 
\textbf{Removing output channels degrades the performance}.
We vary the number of output microphone pairs for S$^3$R under the two multi-tasking models \emph{Ours(B:S)} and \emph{Ours(B:SD)}. We fix the input to pair (3,8) and experiment with different number of output pairs, ranging from 1 to 3. The results are presented in Fig.~\ref{fig:mic_ablation}(c) under these settings.
We see that spatial sound resolution to 3 binaural pairs performs better than to  1 or 2 binaural pairs. The more output channels we have, the better the semantic prediction results are.

\noindent
\textbf{ASPP is a powerful audio encoder}. 
We have found in our experiments that ASPP is a powerful encoder for audio as well. 
We compare our audio encoder with and without the ASPP module. For instance, Mono sound with ASPP clearly outperforms itself without ASPP -- adding ASPP improves the performance from 13.21 to 22.12  for mean IoU. The same trend is observed for other cases. 

\noindent
\textbf{Changing sound volumes}.
We examine the robustness of our model against the volume of the audio signals. We experiment by scaling the input binaural sounds with different multipliers: $0$, $0.5$, $1$ and $2$. We find that the semantic prediction remains robust with different scales of the input sounds.  Fig.~\ref{fig:varyamplitude} shows visual examples.

\subsection{Auditory depth prediction}
We report the depth prediction results in Tab.~\ref{tab:depth:motion}(a). 
The first row represents the depth prediction from Mono sounds alone while the second row represents our baseline method where the depth map is predicted from binaural sounds. As a simple baseline, we also compute the mean depth over the training dataset and evaluate over the test set. The RMSE and MSE scores are $15.024$ and $0.864$, respectively, which are 2.5 times worse than our binaural sound baseline. The multi-task learning with S$^{3}$R, semantic prediction and motion prediction under shared audio encoder also provides pronounced improvements for depth prediction. Joint training of the four tasks also yields the best performance for depth prediction. The conclusion holds for all the three considered encoders.

\subsection{Auditory motion prediction}
We report the motion prediction results in Tab.~\ref{tab:depth:motion}(b). 
The first row represents the motion prediction from Mono sounds alone while the second row represents our baseline method where the motion map is predicted from binaural sounds. The multi-task learning with S$^{3}$R, semantic prediction and depth prediction under shared audio encoder also provides pronounced improvements for motion prediction. It can be found that joint training  of the four tasks yields the best performance for motion prediction as well. Again, this conclusion holds for all the three considered encoders.

\begin{figure}
    \centering
 \includegraphics[trim=35 0 0 55,clip,width=0.8\textwidth]{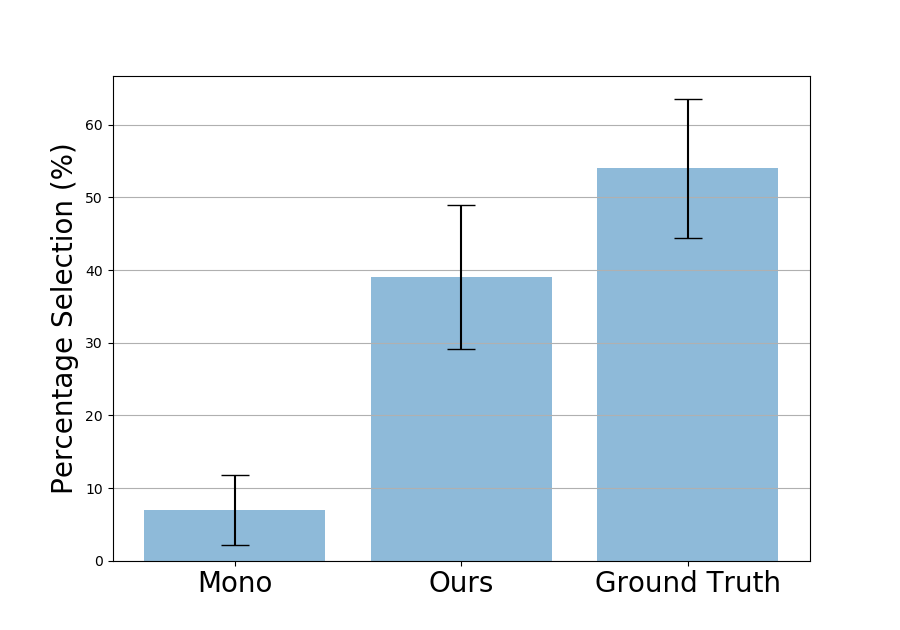} 
    \caption{Subjective assessment of the generated binaural sounds. The results are the percentages of times each result is chosen by users as the preferred one. }
    \label{fig:subjective:sounds}
\end{figure}

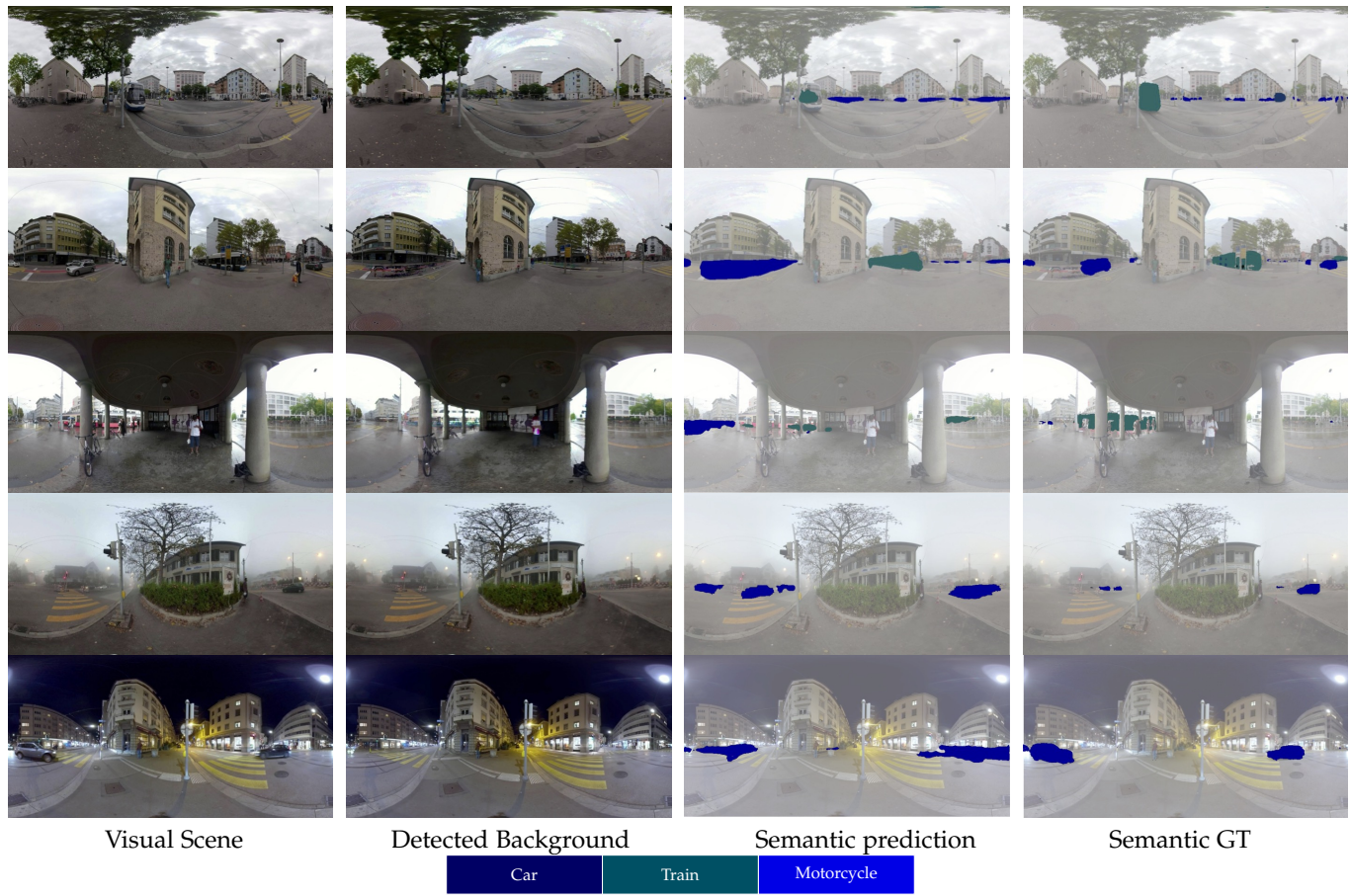
\begin{figure*}[tb]
    \centering
    \includegraphics[width=0.24\textwidth]{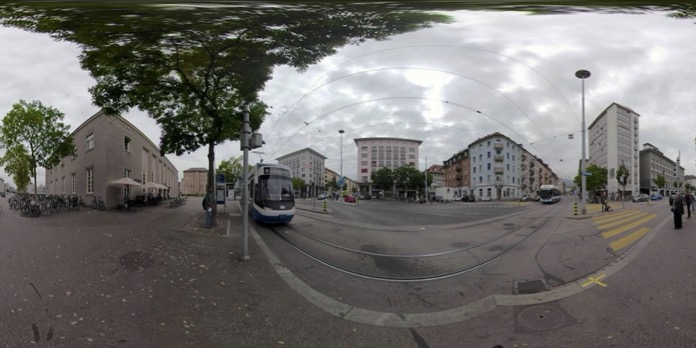}
    \hfil
    \includegraphics[width=0.24\textwidth]{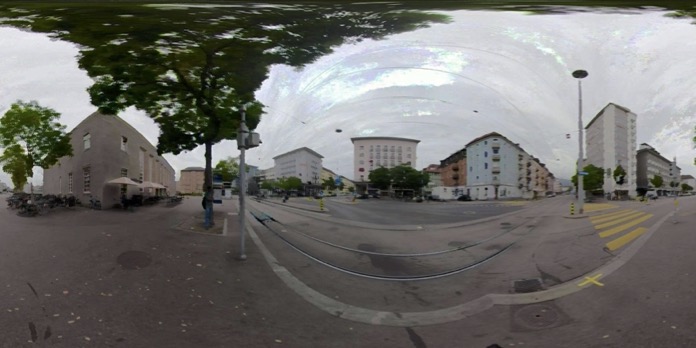}
    \hfil
    \includegraphics[width=0.24\textwidth]{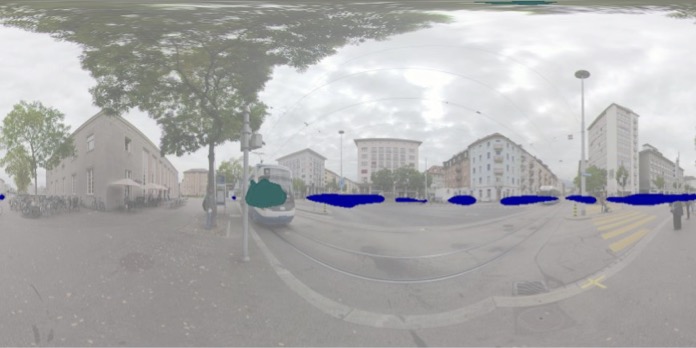}
    \hfil
    \includegraphics[width=0.24\textwidth]{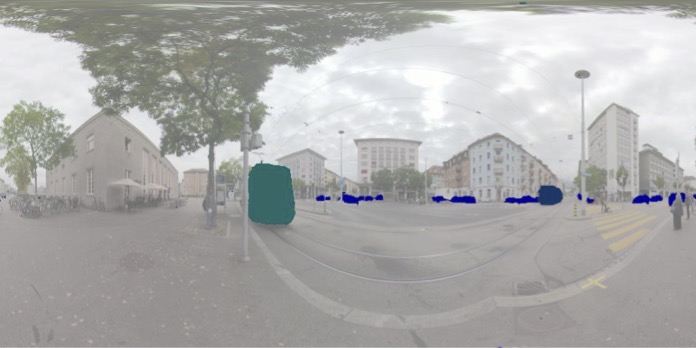}
    \\ 
    \includegraphics[trim=0 40 0 0,clip,width=0.24\textwidth,height=0.12\textwidth]{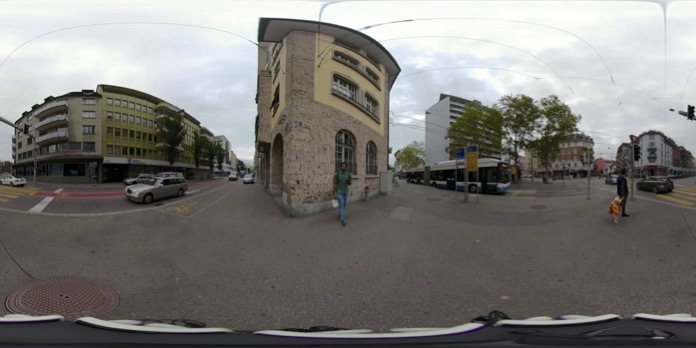}
    \hfil
    \includegraphics[width=0.24\textwidth]{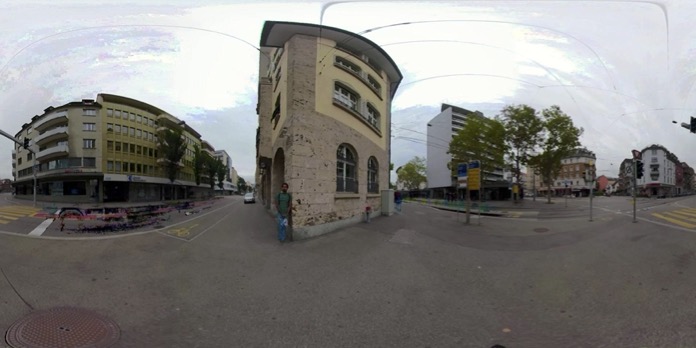}
    \hfil
    \includegraphics[width=0.24\textwidth]{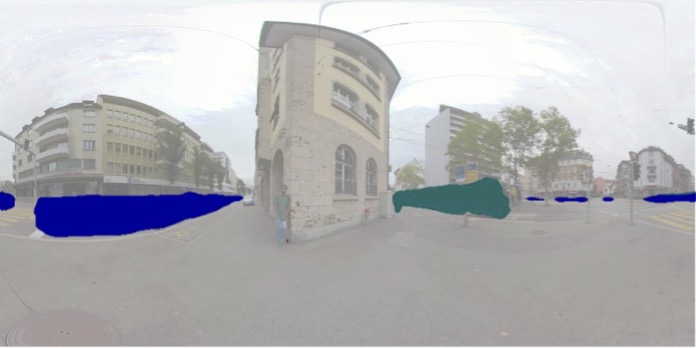}
    \hfil
    \includegraphics[width=0.24\textwidth]{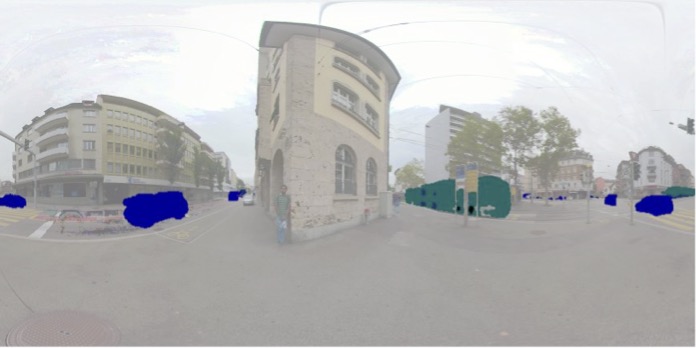}
    \\ 
    \includegraphics[trim=0 30 0 0,clip,width=0.24\textwidth,height=0.12\textwidth]{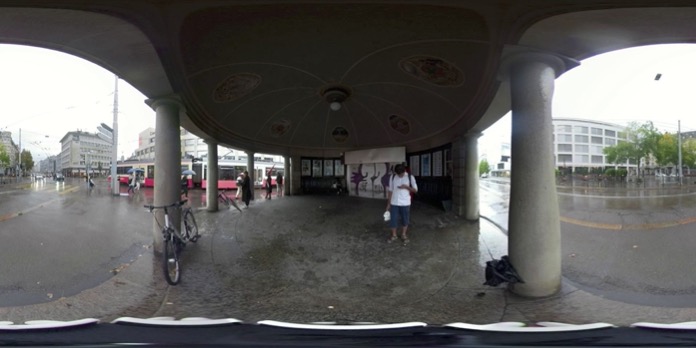}
    \hfil
    \includegraphics[trim=0 30 0 0,clip,width=0.24\textwidth,height=0.12\textwidth]{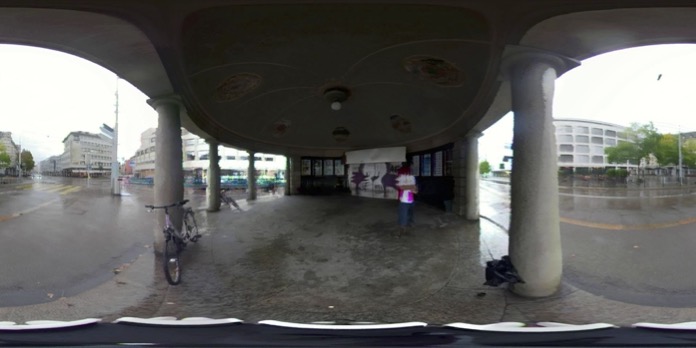}
    \hfil
    \includegraphics[width=0.24\textwidth]{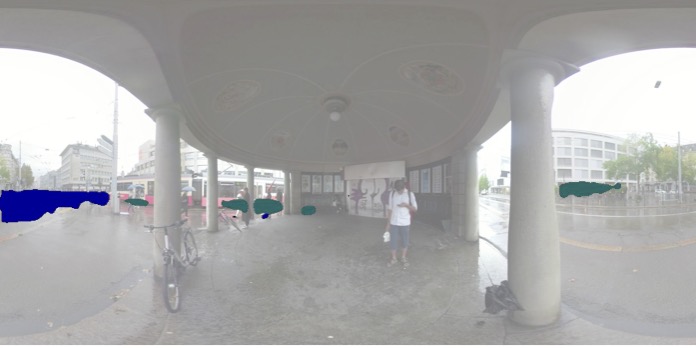}
    \hfil
    \includegraphics[width=0.24\textwidth]{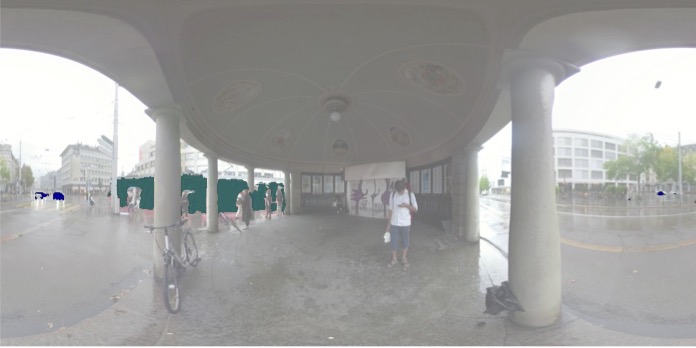}
    \\ 
    \includegraphics[trim=0 40 0 0,clip,width=0.24\textwidth,height=0.12\textwidth]{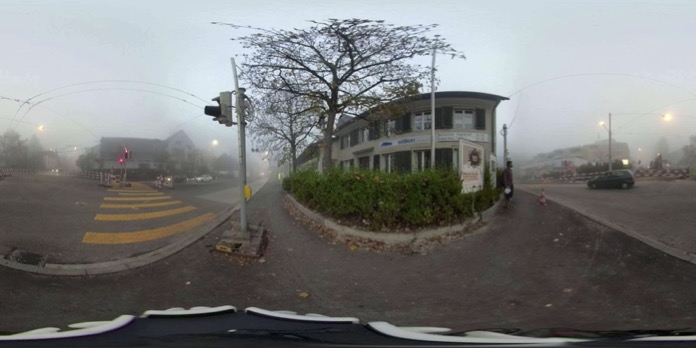}
    \hfil
    \includegraphics[trim=0 40 0 0,clip,width=0.24\textwidth,height=0.12\textwidth]{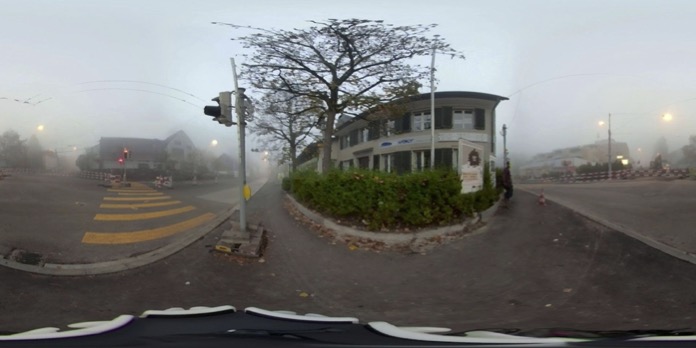}
    \hfil
    \includegraphics[width=0.24\textwidth]{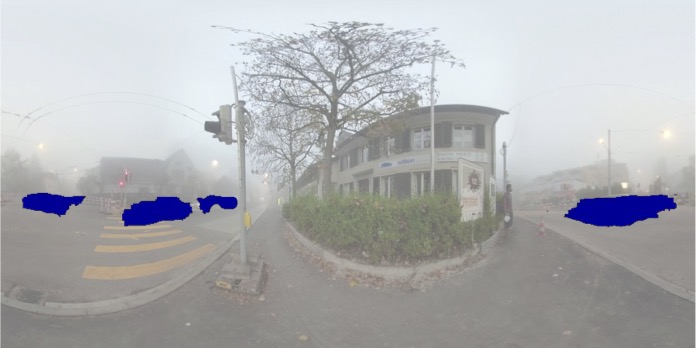}
    \hfil
    \includegraphics[width=0.24\textwidth]{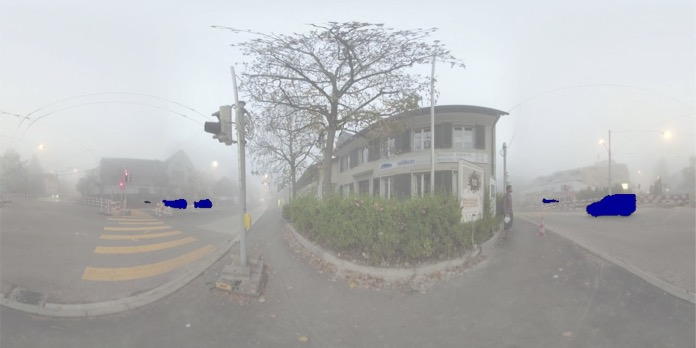}
    \\ 
    \includegraphics[trim=0 35 0 0,clip,width=0.24\textwidth,height=0.12\textwidth]{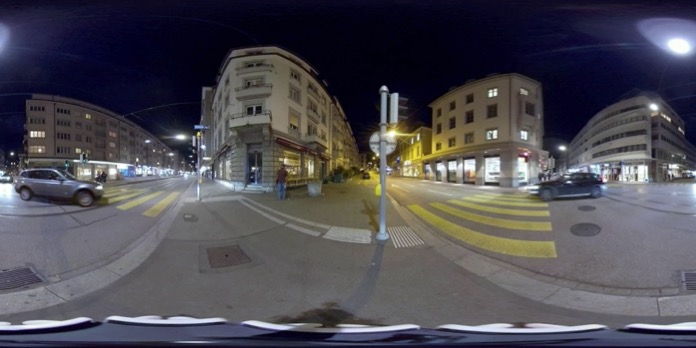}
    \hfil
    \includegraphics[trim=0 35 0 0,clip,width=0.24\textwidth,height=0.12\textwidth]{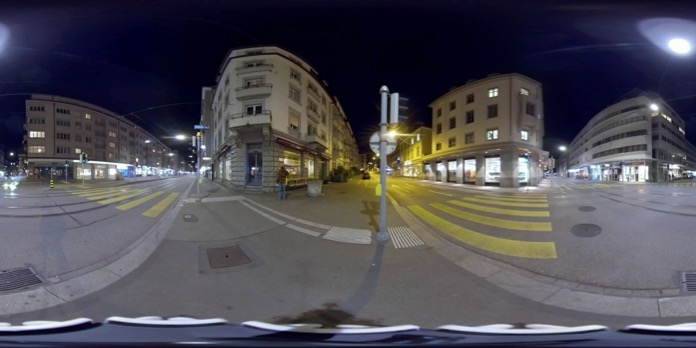}
    \hfil
    \includegraphics[width=0.24\textwidth]{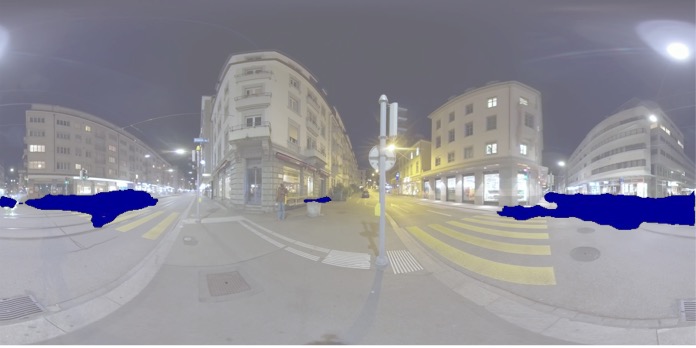}
    \hfil
    \includegraphics[width=0.24\textwidth]{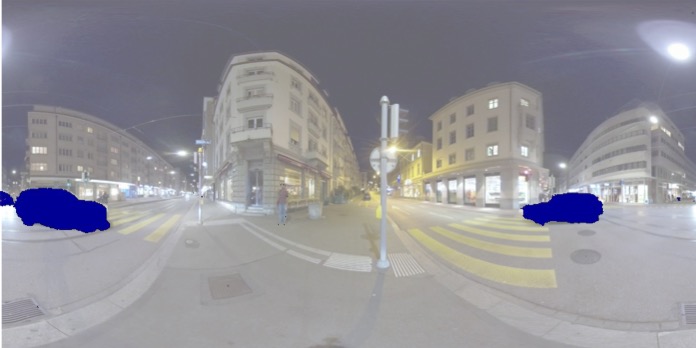}
    \\
     Visual Scene \hspace{18mm} Detected Background  \hspace{15mm} Semantic prediction \hspace{17mm}Semantic GT\\ 
        \resizebox{0.35\linewidth}{!}{
\begin{tikzpicture}[tight background, scale=0.75, every node/.style={font=\small}]
	\draw[white, fill=car, draw=white] (0,0) rectangle (1* 4, 1) node[pos=0.5] {Car};
	\draw[white, fill=train, draw=white] (1 * 4,0) rectangle (2 * 4, 1) node[pos=0.5] {Train};
	\draw[white, fill=motorcycle, draw=white] (2 * 4,0) rectangle (3 * 4, 1) node[pos=0.5] {Motorcycle};
\end{tikzpicture}}
    \caption{Qualitative results of auditory semantic prediction by our approach. First column shows the visual scene, the second for the computed background image, the third for the object masks predicted by our approach, and the fourth for the ground truth. 
    }
    \label{fig:result1}
\end{figure*}

\begin{figure}
\begin{tabular}{cccc}
  \centering 
  \includegraphics[width=0.3\textwidth,height=0.185\textwidth]{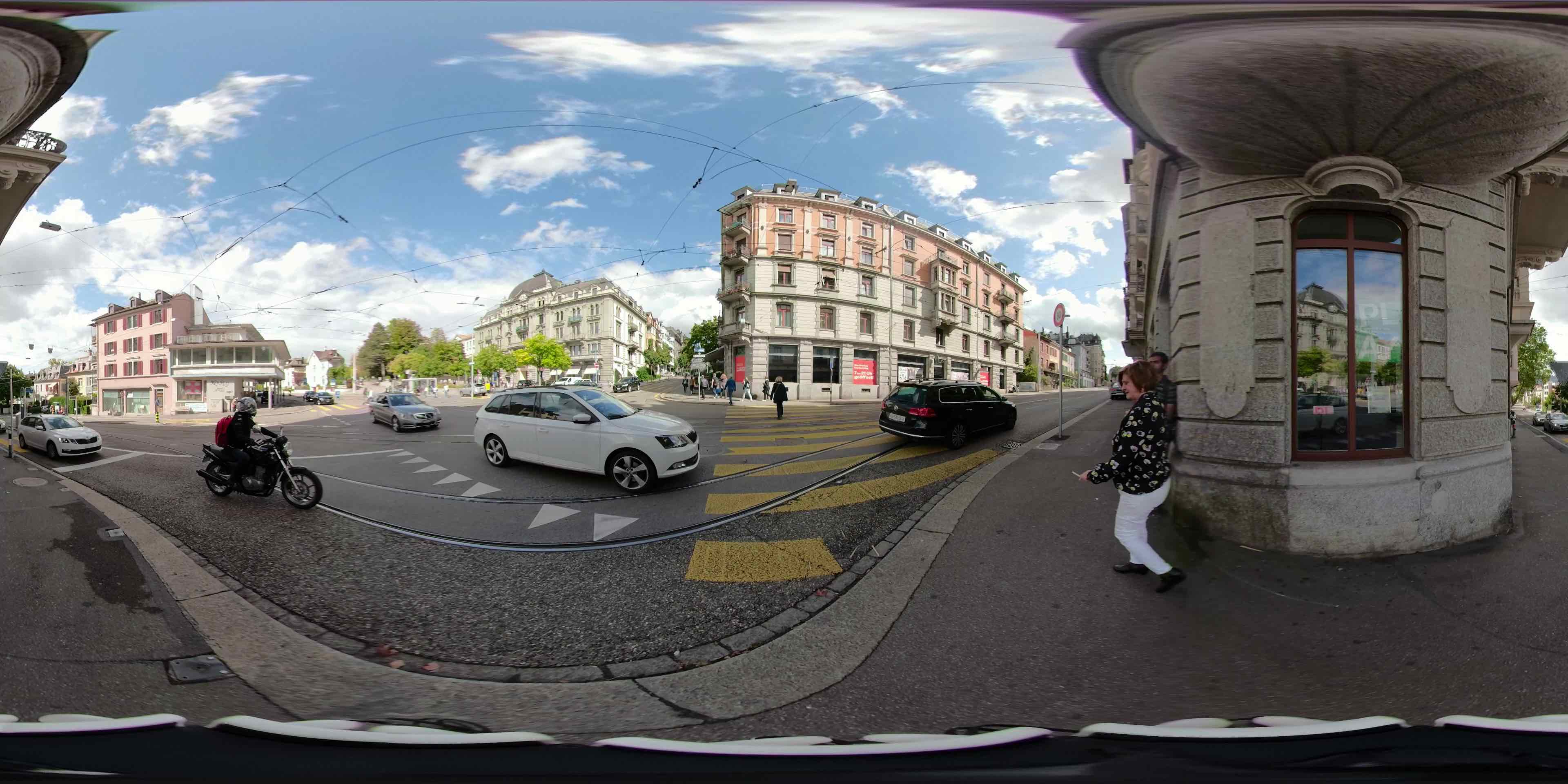} & 
  \includegraphics[width=0.3\textwidth,height=0.185\textwidth]{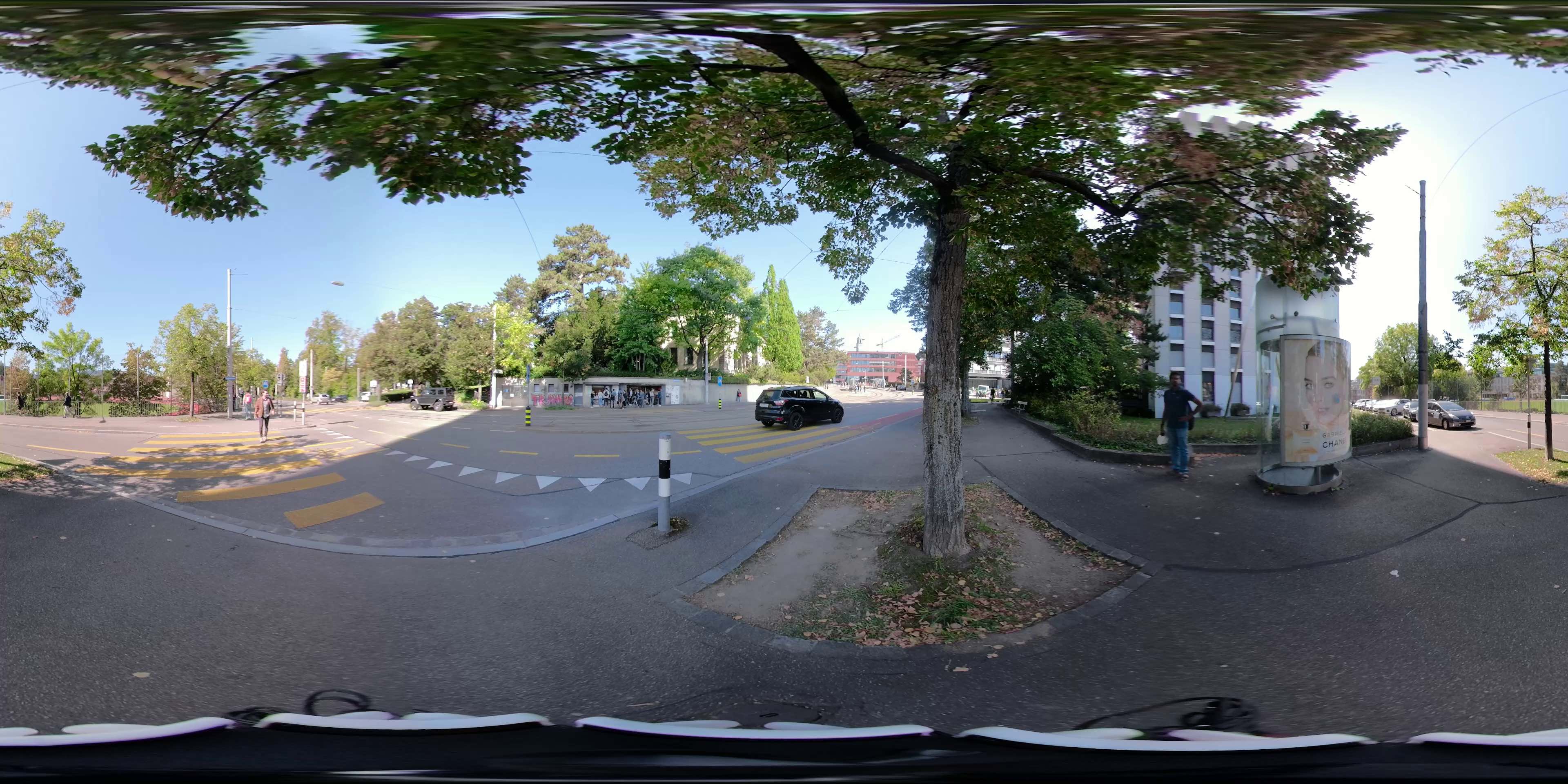} & 
  \includegraphics[width=0.3\textwidth,height=0.185\textwidth]{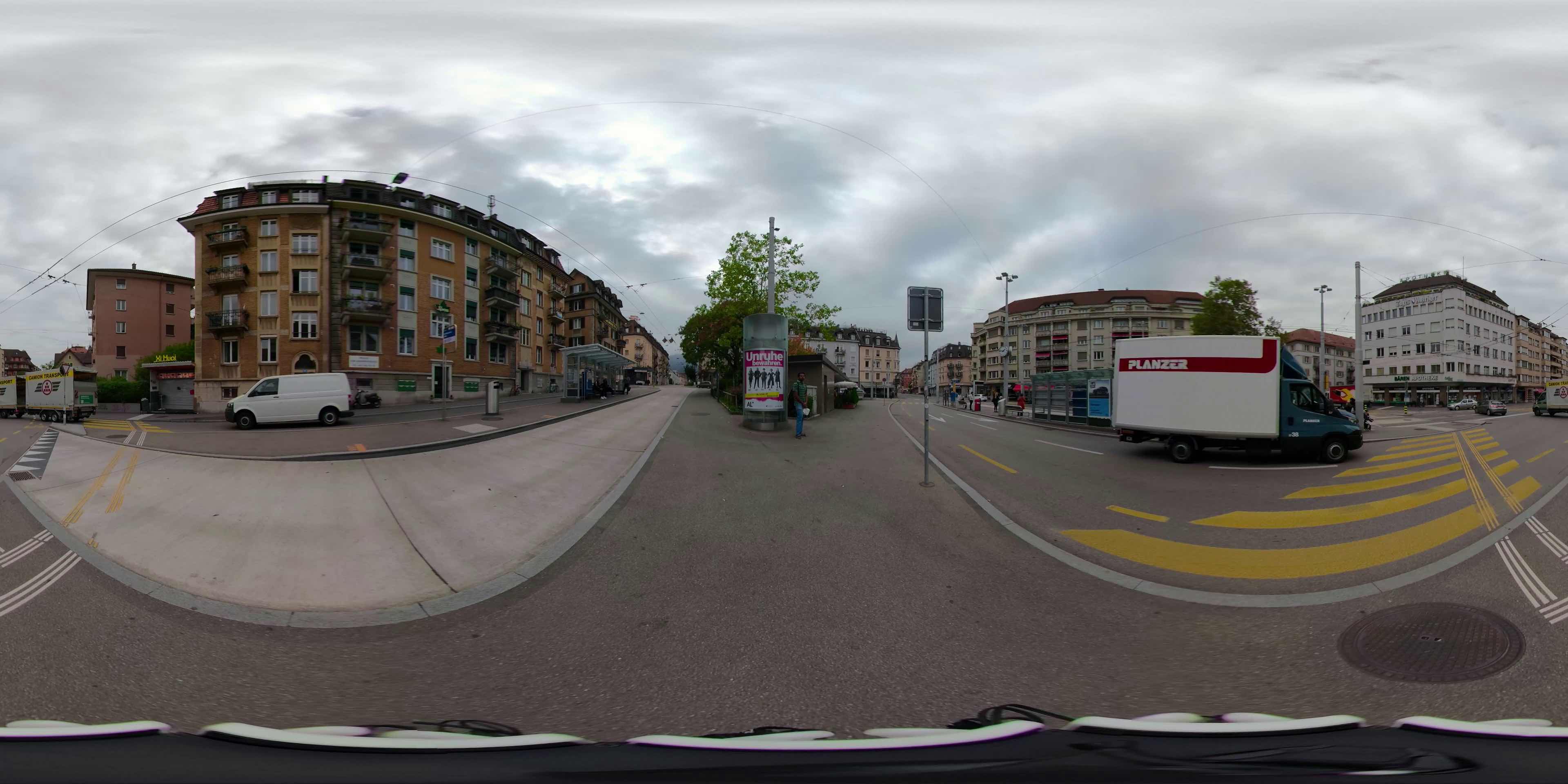} 
  \\
  \includegraphics[width=0.3\textwidth,height=0.185\textwidth]{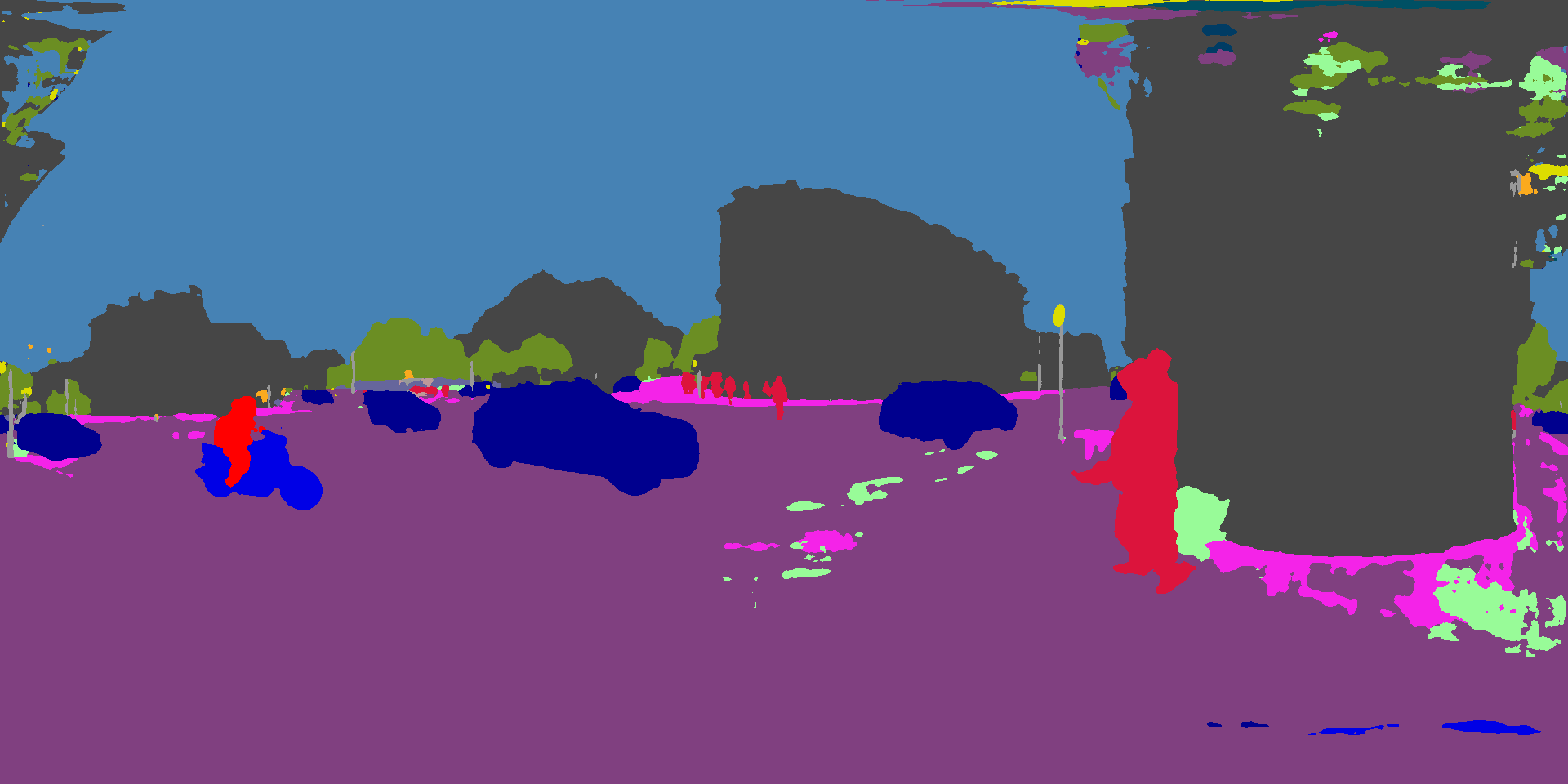} & 
  \includegraphics[width=0.3\textwidth,height=0.185\textwidth]{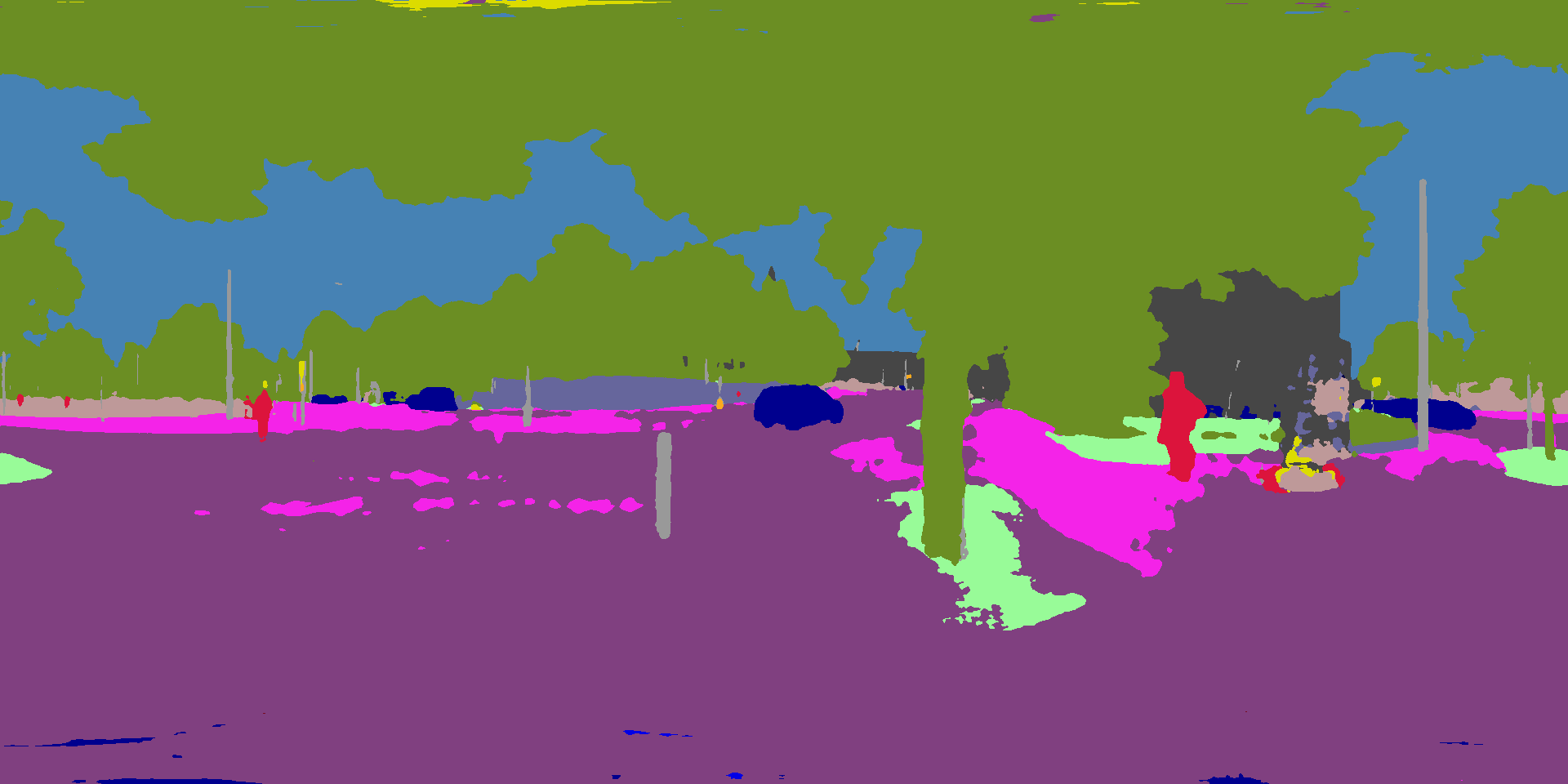} & 
  \includegraphics[width=0.3\textwidth,height=0.185\textwidth]{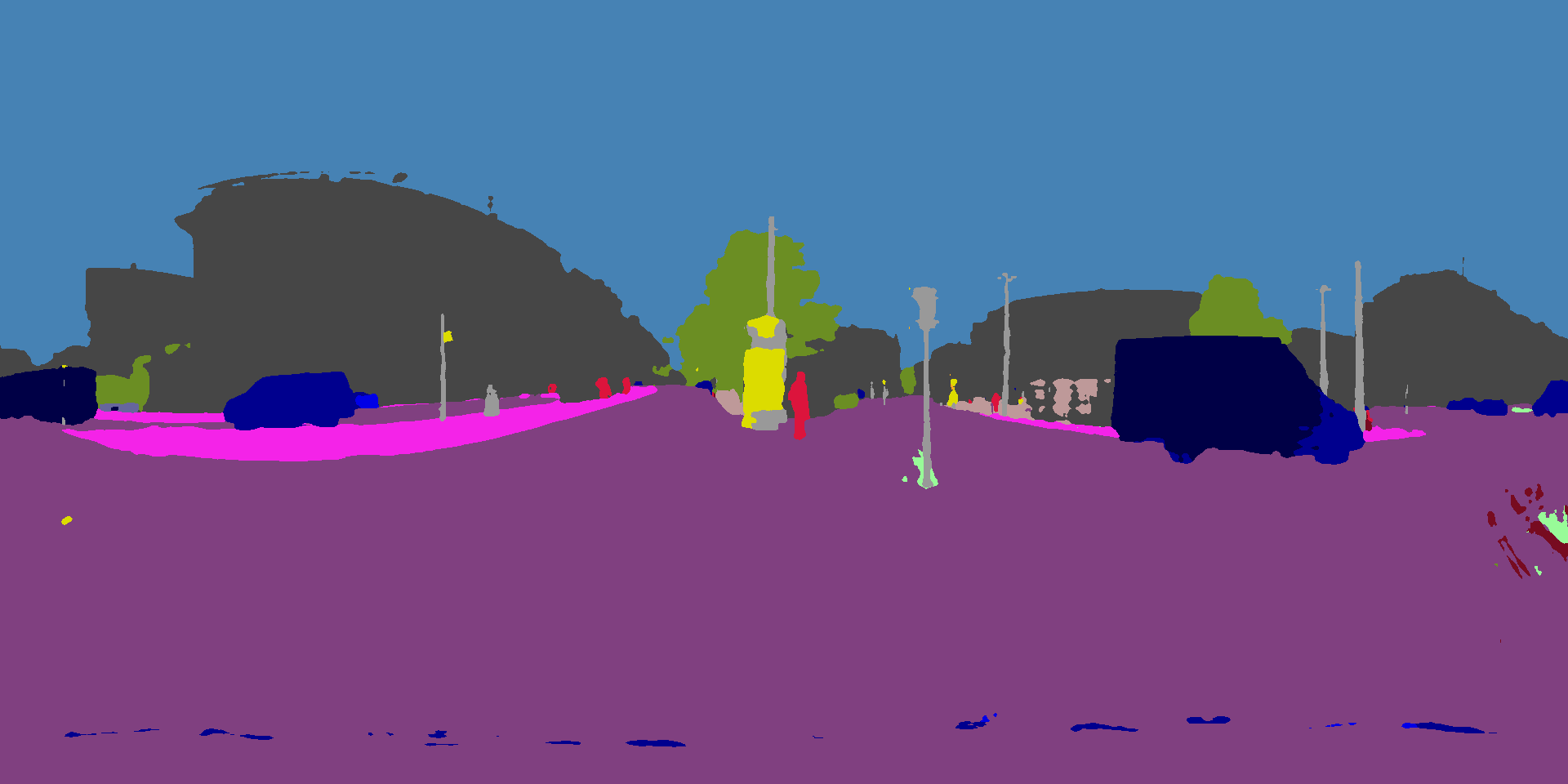} 
  \\
  \includegraphics[width=0.3\textwidth,height=0.185\textwidth]{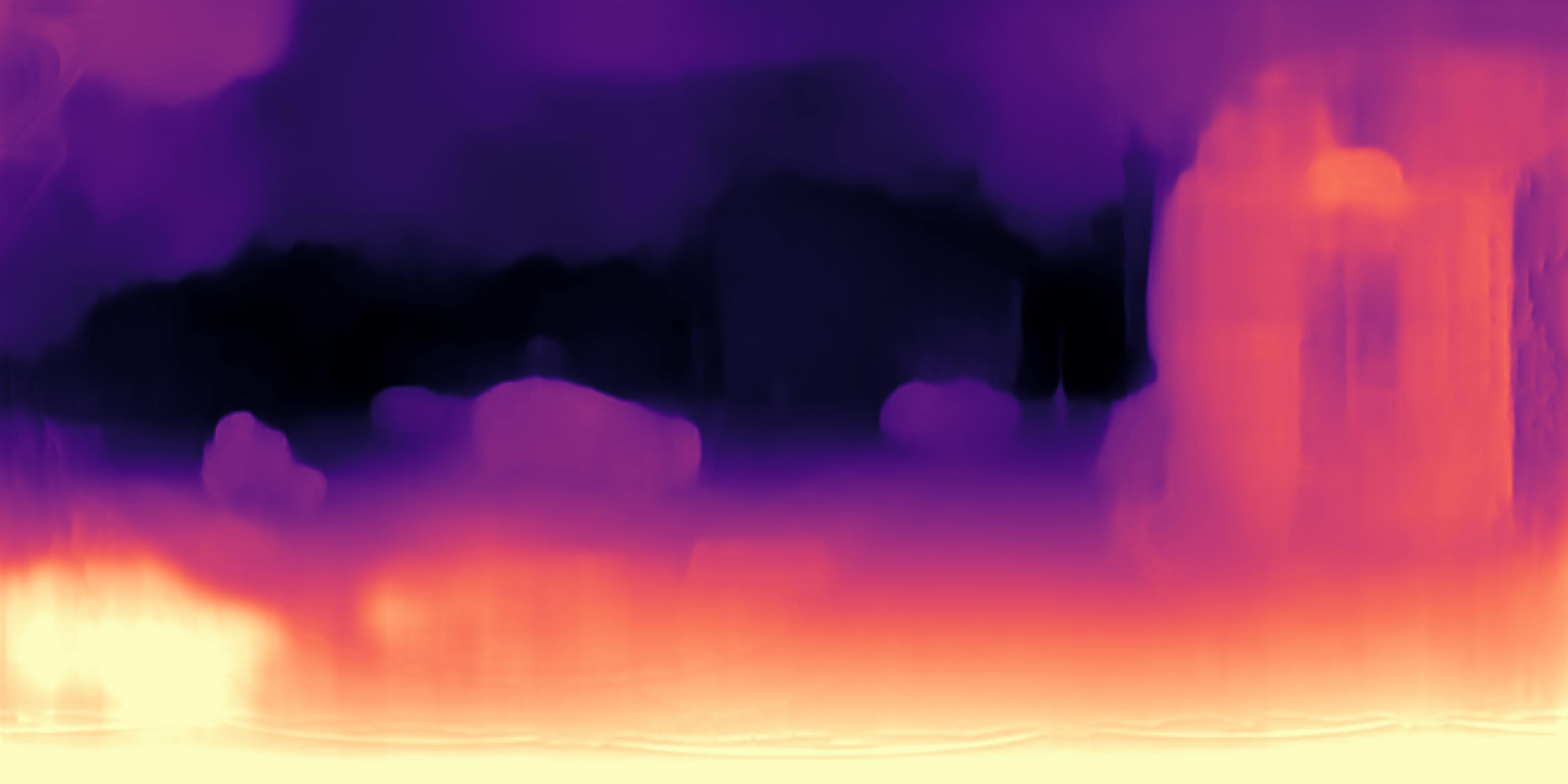} & 
  \includegraphics[width=0.3\textwidth,height=0.185\textwidth]{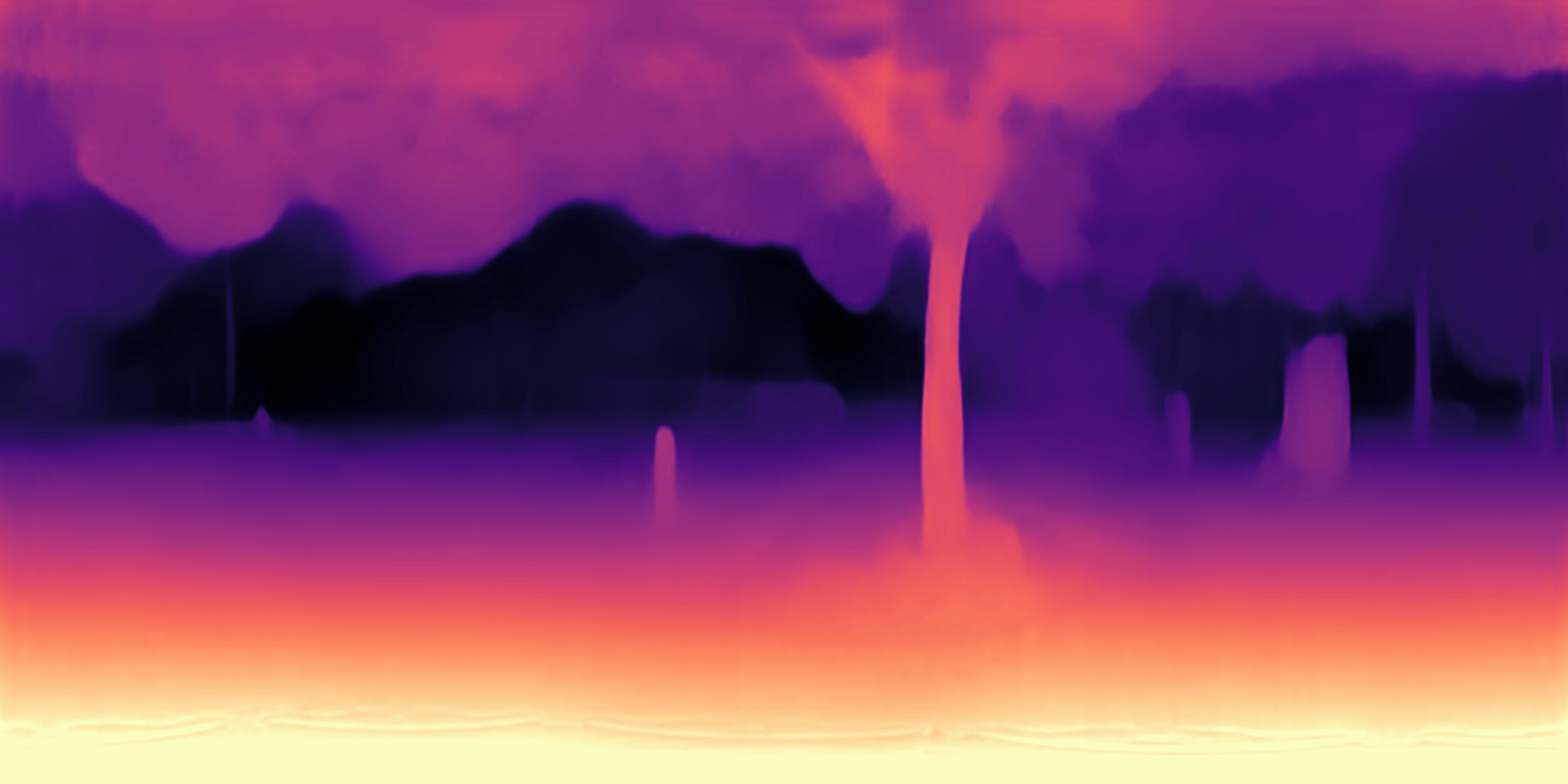} & 
  \includegraphics[width=0.3\textwidth,height=0.185\textwidth]{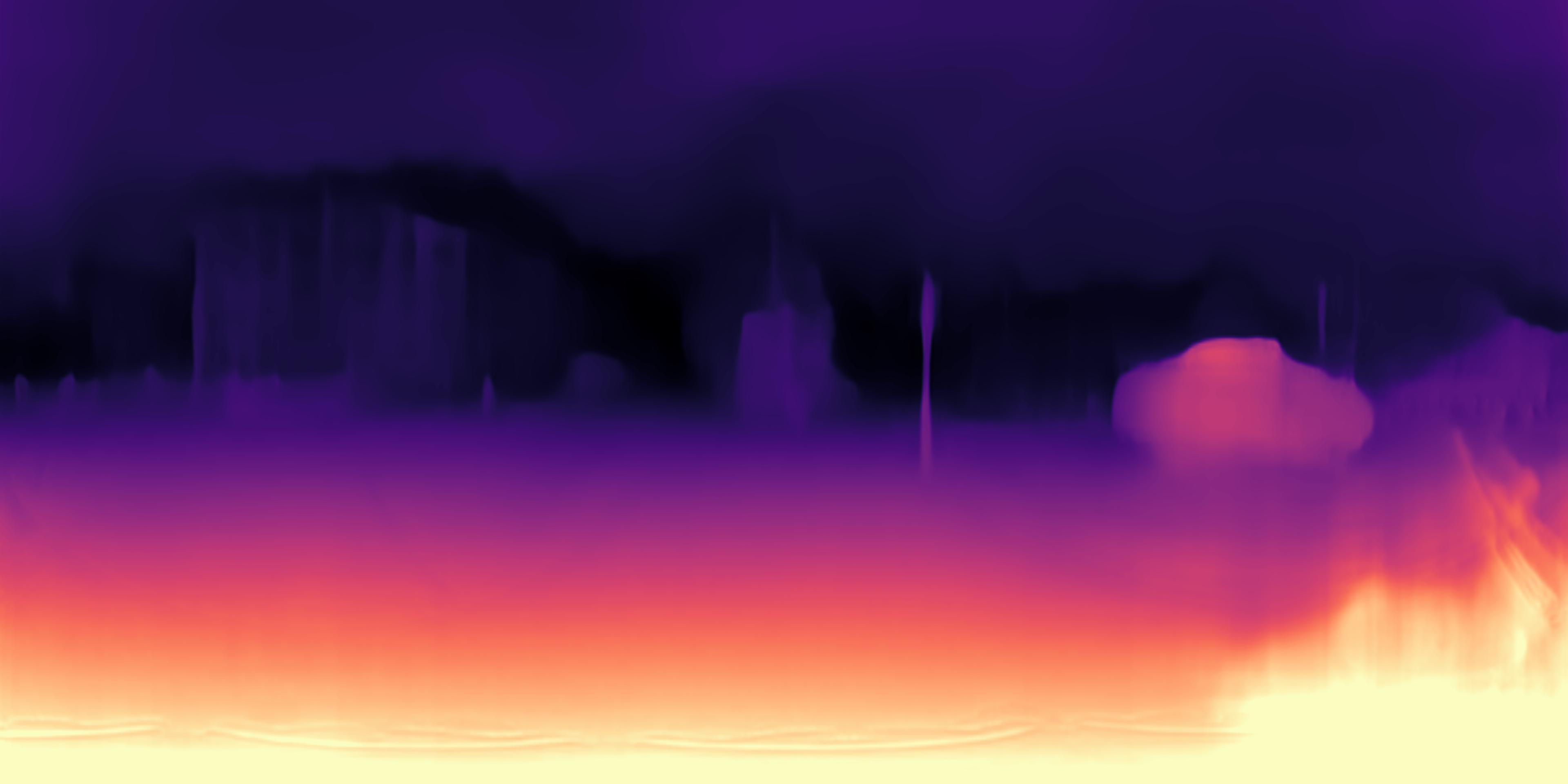} 
  \\
  \includegraphics[width=0.3\textwidth,height=0.185\textwidth]{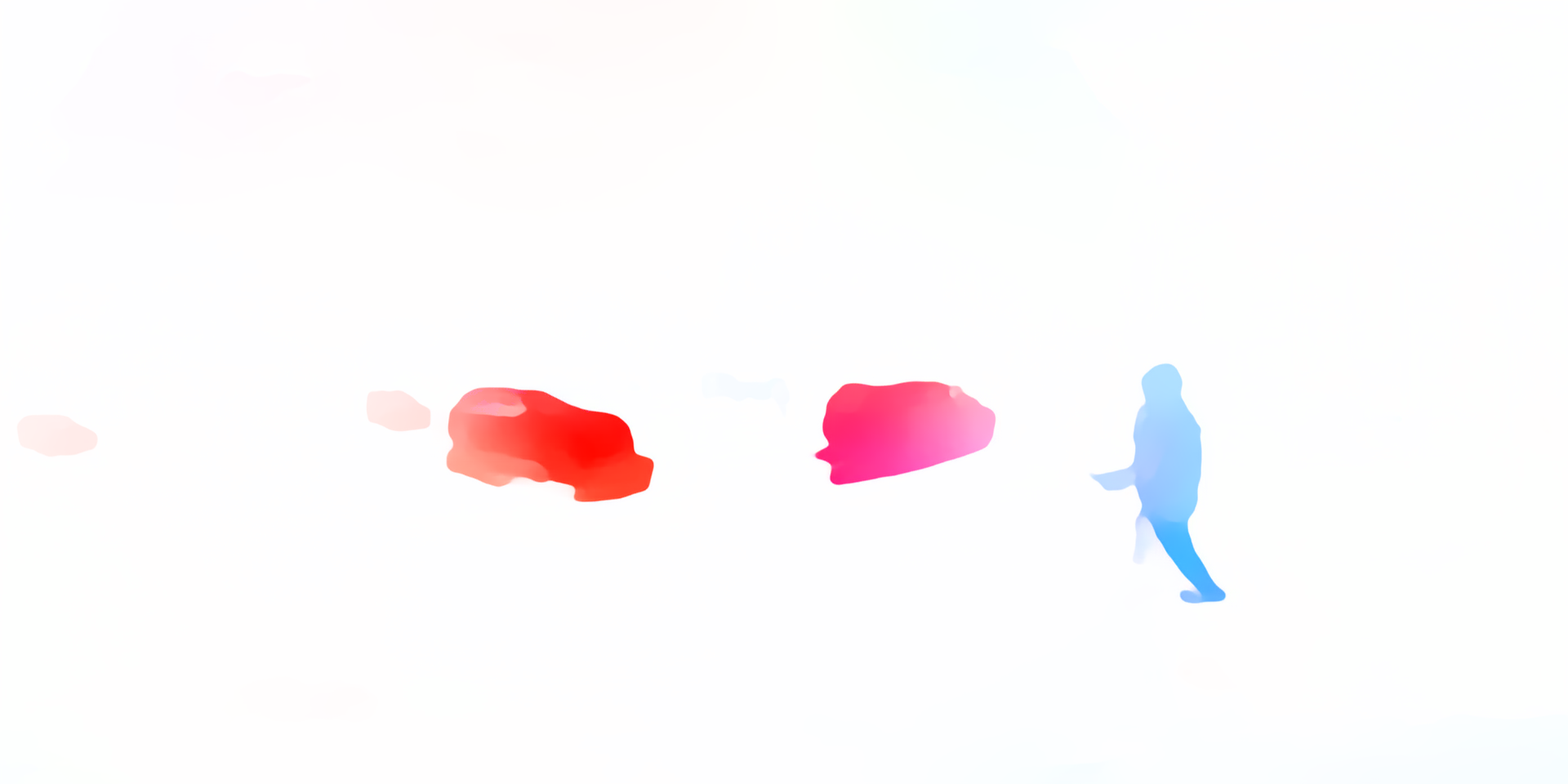} & 
  \includegraphics[width=0.3\textwidth,height=0.185\textwidth]{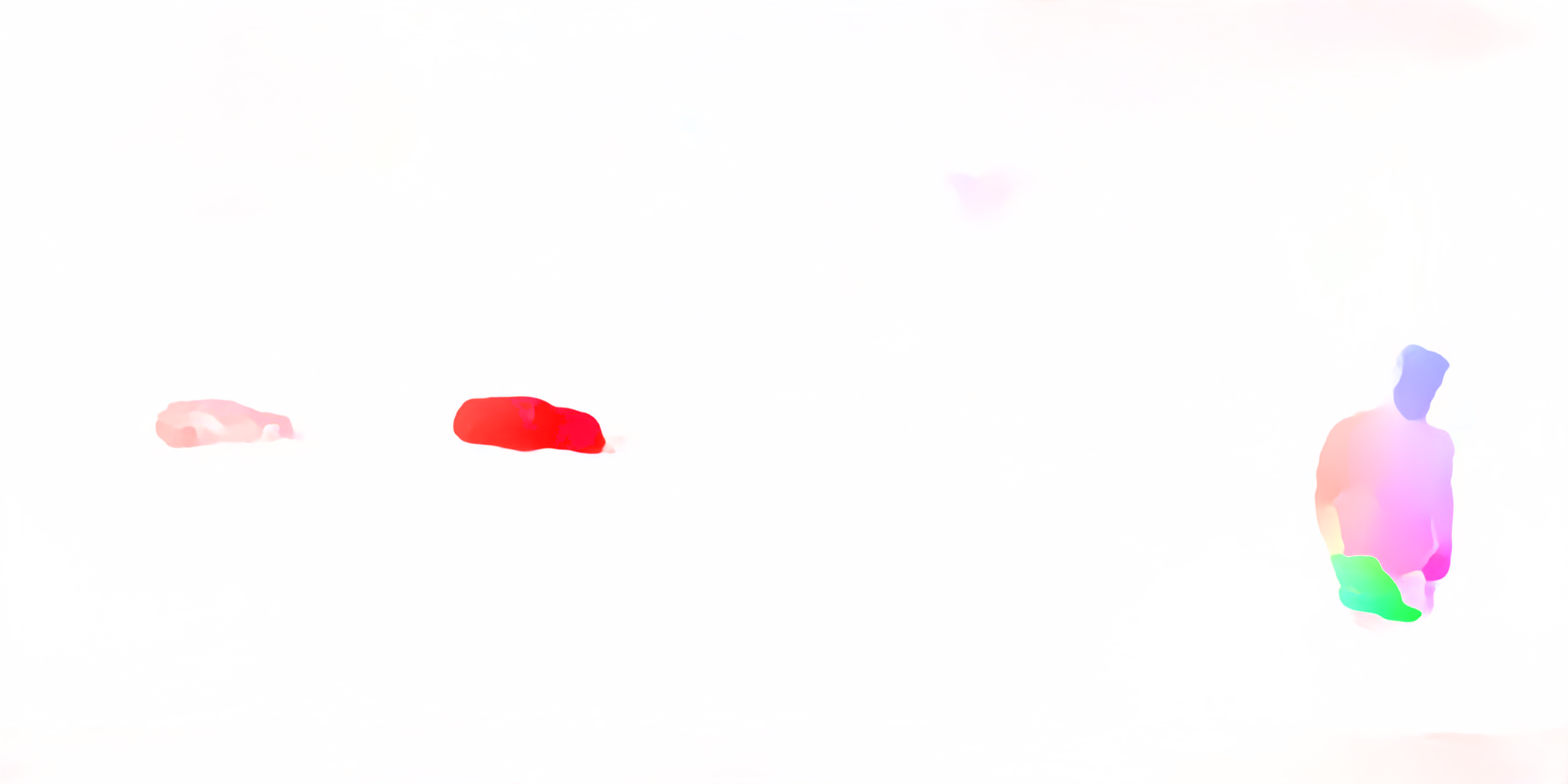} & 
  \includegraphics[width=0.3\textwidth,height=0.185\textwidth]{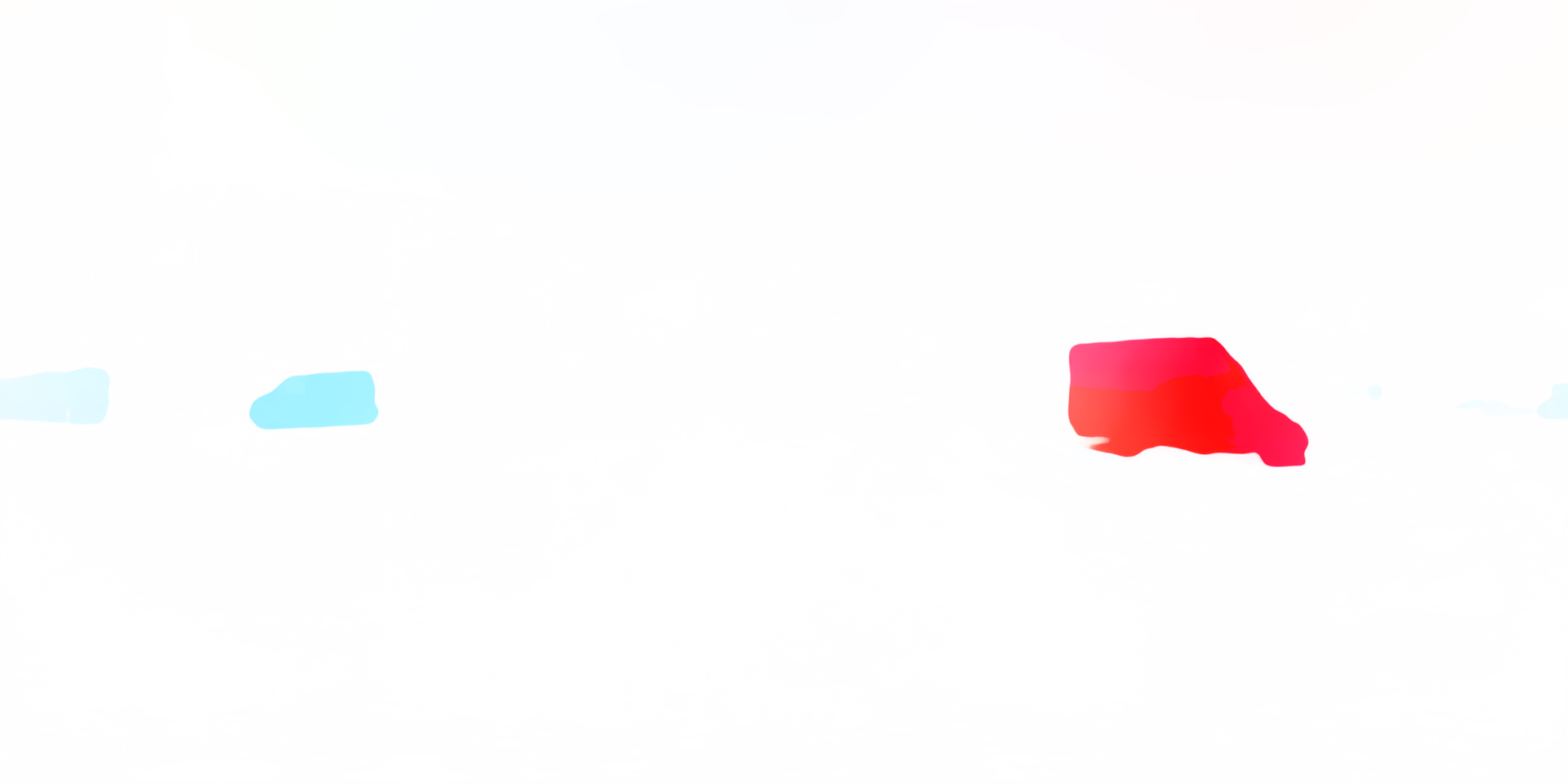} \\
  \end{tabular} 
  \caption{An illustration of samples of semantic segmentation masks, depth predictions and motion estimates generated from our dataset using the pre-trained state of the art vision models. }
  \label{fig:sample_flow}
\end{figure} 

\subsection{Spatial Sound Super-resolution} 
Tab.~\ref{tab:spatialresolutionresults} shows the results of S$^3$R as a stand-alone task (first row) and under the multi-task setting. To keep it simple, we only estimate the sound signals of microphone pair $(1,6)$  from the microphone pair $(3,8)$. We can see from Fig.~\ref{fig:setup} that these two pairs are perpendicular in orientation, so the prediction is quite challenging. 
We can observe from Tab.~\ref{tab:spatialresolutionresults} that the multi-task learning with the semantic prediction, depth prediction and motion prediction outperforms the accuracy of the stand-alone S$^{3}$R model. Hence, the multi-task learning also helps S$^{3}$R -- the same conclusion as for the other three tasks. This trend is also observed for all the three different types of encoders.

We also conduct a user study for the subjective assessment of the generated binaural sounds.
The participants listen to ground-truth binaural sounds, binaural sounds generated by the \emph{Mono} approach (Tab.~\ref{tab:maintable}) and the ones generated by our approach with binaural sounds and under the multi-task learning framework. We present two (out of three) randomly picked sounds and ask the user to select a preferred one in terms of binaural sound quality. Fig.~\ref{fig:subjective:sounds} shows the percentage of times each method is chosen as the preferred one. We see that \textit{Ours} is close to the ground truth selection. This shows that our predicted binaural sounds for a different orientation are of great quality.

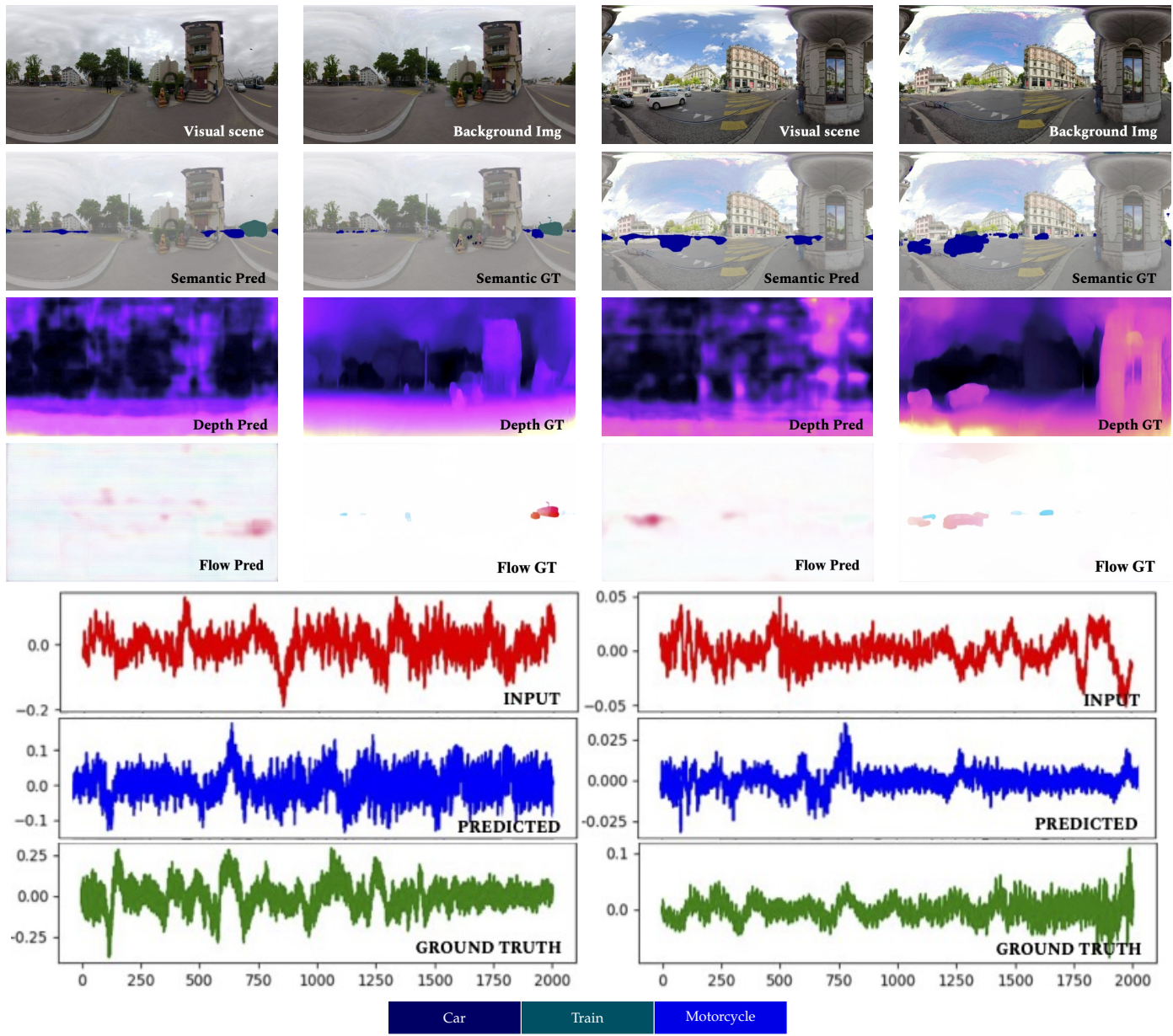
\begin{figure*}[tb]
\begin{tabular}{cccc}
  \centering 
  \includegraphics[trim=0 0 0 0,clip,width=0.235\textwidth,height=0.12\textwidth]{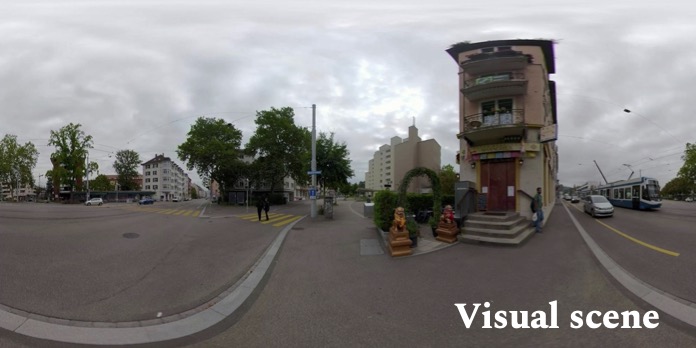} & 
  \includegraphics[trim=0 0 0 0,clip,width=0.235\textwidth,height=0.12\textwidth]{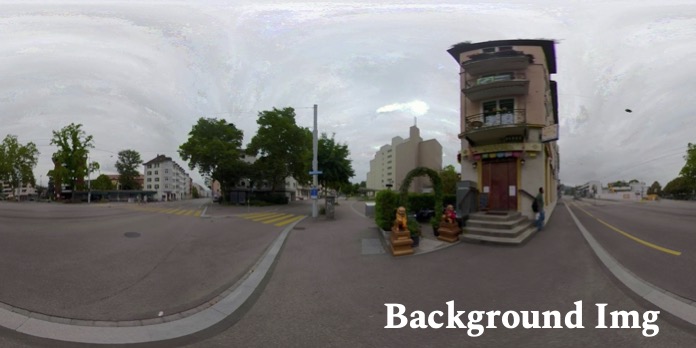} & 
  \includegraphics[trim=0 0 0 0,clip,width=0.235\textwidth,height=0.12\textwidth]{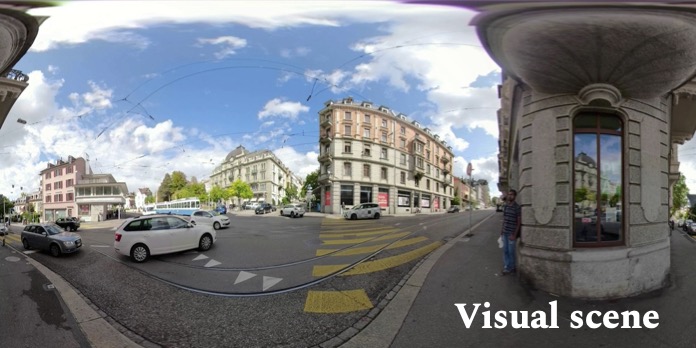} & 
  \includegraphics[trim=0 0 0 0,clip,width=0.235\textwidth,height=0.12\textwidth]{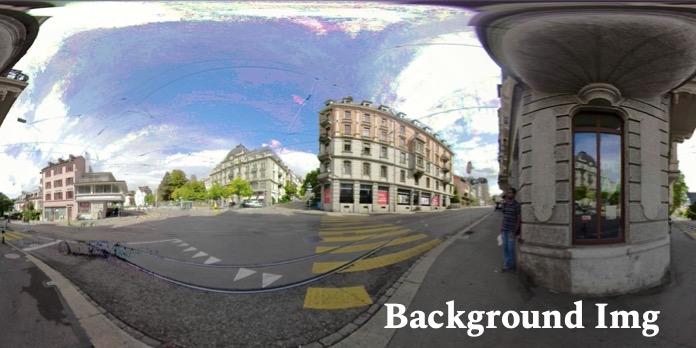}
  \\
  \includegraphics[trim=0 0 0 0,clip,width=0.235\textwidth,height=0.12\textwidth]{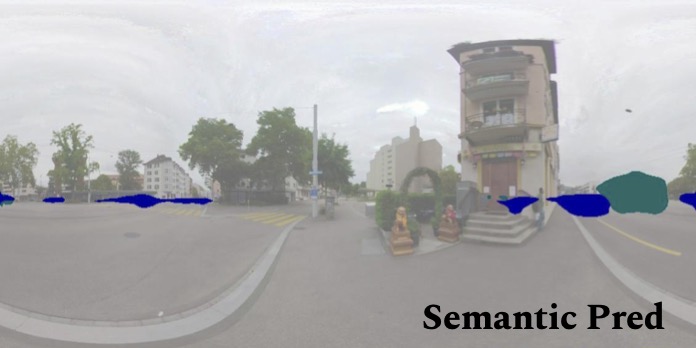} & 
  \includegraphics[trim=0 0 0 0,clip,width=0.235\textwidth,height=0.12\textwidth]{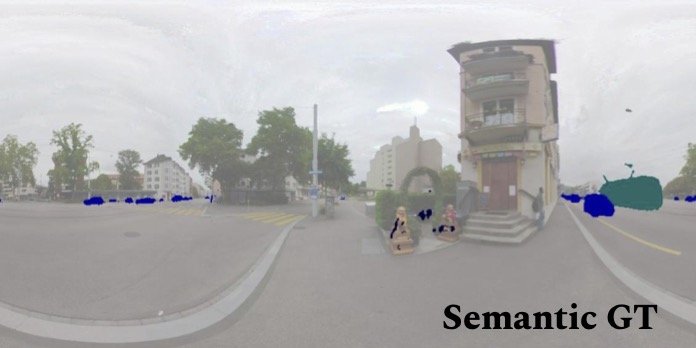} &
  \includegraphics[trim=0 0 0 0,clip,width=0.235\textwidth,height=0.12\textwidth]{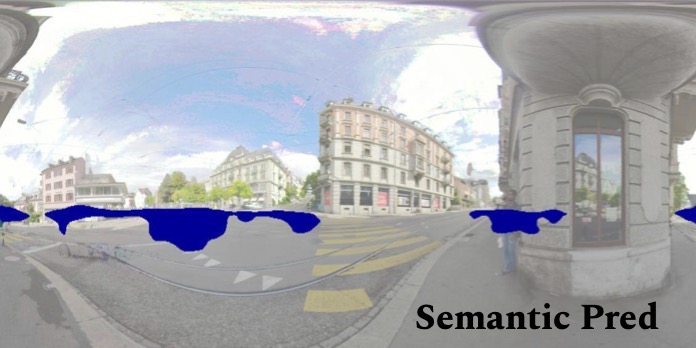} &
  \includegraphics[trim=0 0 0 0,clip,width=0.235\textwidth,height=0.12\textwidth]{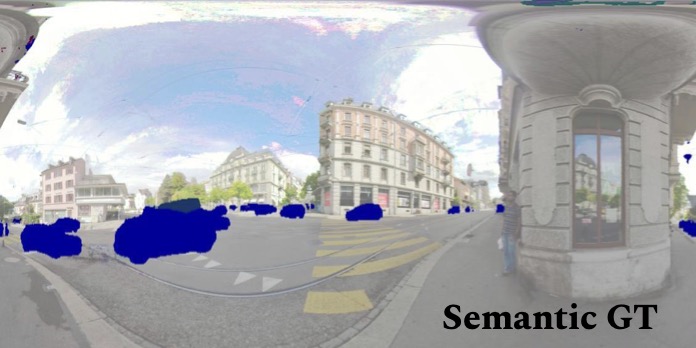} 
  \\
  \includegraphics[trim=0 00 0 0,clip,width=0.235\textwidth,height=0.12\textwidth]{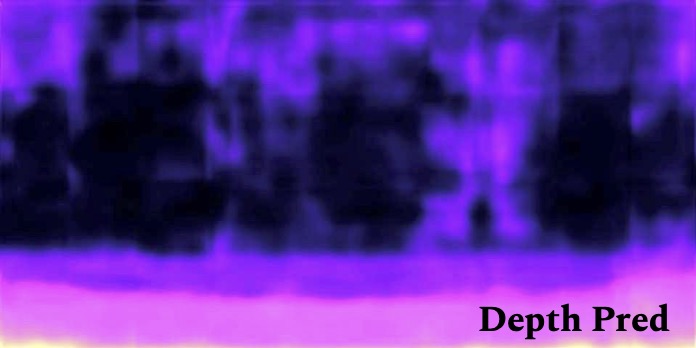} & 
  \includegraphics[trim=0 0 0 0,clip,width=0.235\textwidth,height=0.12\textwidth]{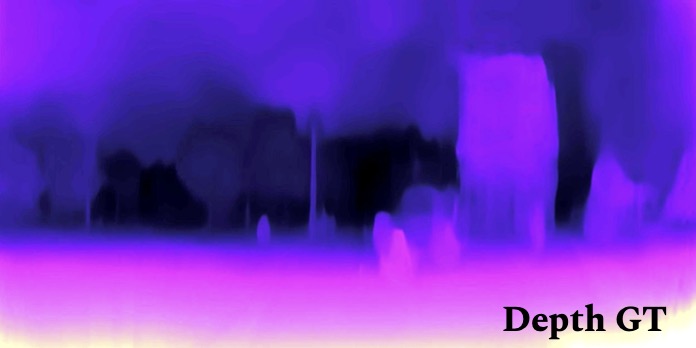} & 
  \includegraphics[trim=0 0 0 0,clip,width=0.235\textwidth,height=0.12\textwidth]{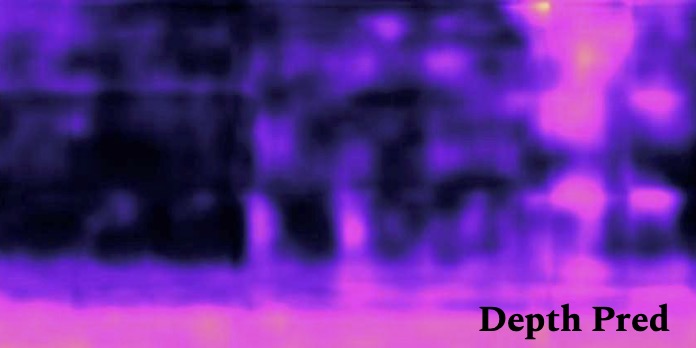} & 
  \includegraphics[trim=0 0 0 0,clip,width=0.235\textwidth,height=0.12\textwidth]{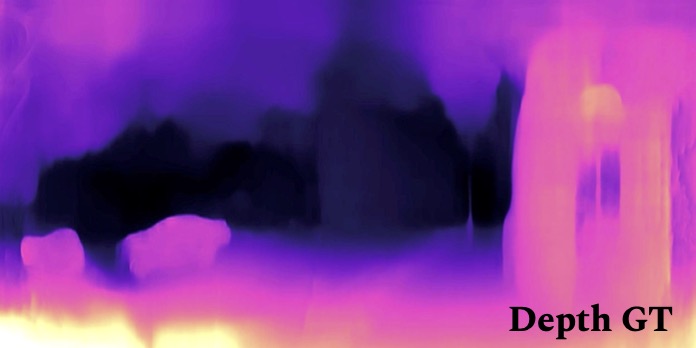}
  \\
  \includegraphics[trim=0 0 0 0,clip,width=0.235\textwidth,height=0.12\textwidth]{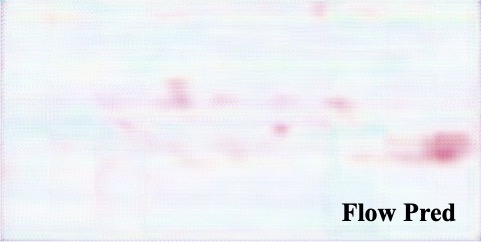} & 
  \includegraphics[trim=0 0 0 0,clip,width=0.235\textwidth,height=0.12\textwidth]{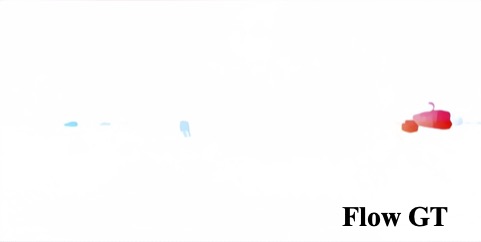} &
  \includegraphics[trim=0 0 0 0,clip,width=0.235\textwidth,height=0.12\textwidth]{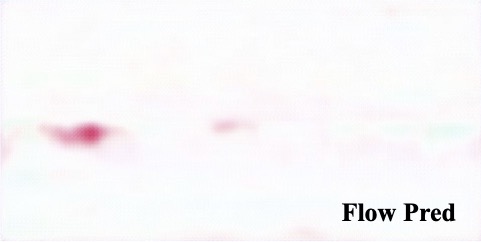} & 
  \includegraphics[trim=0 0 0 0,clip,width=0.235\textwidth,height=0.12\textwidth]{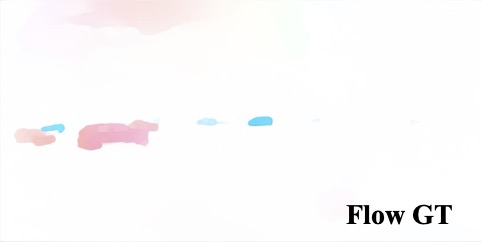}
  \\
  \multicolumn{4}{c}{\includegraphics[width=\textwidth,height=0.35\textwidth]{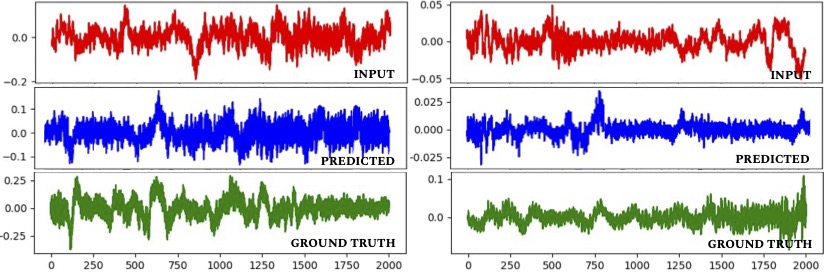}} 
  \\
 \multicolumn{4}{c}{    \resizebox{0.35\linewidth}{!}{
\begin{tikzpicture}[tight background, scale=0.75, every node/.style={font=\small}]
	\draw[white, fill=car, draw=white] (0,0) rectangle (1* 4, 1) node[pos=0.5] {Car};
	\draw[white, fill=train, draw=white] (1 * 4,0) rectangle (2 * 4, 1) node[pos=0.5] {Train};
	\draw[white, fill=motorcycle, draw=white] (2 * 4,0) rectangle (3 * 4, 1) node[pos=0.5] {Motorcycle};
\end{tikzpicture}} }
  \caption{Qualitative results of all the four tasks. 
  Better view in color. }
  \label{fig:multitask}
  \end{tabular}
\end{figure*}

\subsection{Spectrogram vs. DDSP}
In this section, we compare spectrogram and DDSP for our tasks. The results can be seen in Tab. \ref{tab:maintable}, Tab. \ref{tab:depth:motion} and Tab. \ref{tab:spatialresolutionresults}.
The tables also show the results obtained by combining the audio features from DDSP and from spectrogram based encoding. It can be found that the semantic prediction performance (mIoU) of \emph{Ours(B)} using DDSP is $23.01$\% on the AuditoryTestPseudo dataset, in contrast to $32.01$\% when using spectrogram. This trend is also observed for the other three tasks, and also observed for all tasks under the multi-task setting. This shows that for our considered tasks spectrogram-based encoding is more useful compared to DDSP-based encoding. This probably can be attributed to the great capability of convolutional neural networks in extracting features out of spectrogram images.   

Since DDSP and spectrogram can be complementary as they encode features in a very different manner. We also combine the two and evaluate the performance. The results are also tabulated in the three tables. One can observe that the combination of the two gives the best results. For instance, on the AuditoryTestPseudo test set, \emph{Ours(B)} under this setting yields $45.54\%$ mIoU compared to $25.22\%$ and $43.40\%$ mIoU when using DDSP and spectrogram, respectively. This confirms  that auditory features from ddsp-based methods and spectrogram-based methods are indeed complementary and can be combined to further boost the performance of auditory perception tasks.

\subsection{Qualitative results}
We show qualitative results in Fig.~\ref{fig:result1} for the task of auditory semantic prediction. We also show the detected background image and the ground truth segmentation mask.
The last three rows are devoted to the results in 
rainy, foggy and night conditions. We observe that our model remains robust in those adverse visual conditions.  If rain is too big, it might become an adverse auditory condition as well due to the generated noise. Unfortunately, we do not have very big rain in our dataset to verify this. 

In Fig.~\ref{fig:multitask}, we show two results and ground truths for all the tasks. 
It is quite clear that the major sound-making objects can be properly detected. We see that the depth results reflect the general layout of the scene and the motion results captures the motion of the major sound-making objects, though they are still much coarser than the results of a vision system under clear conditions. However, this is still very valuable given the fact that binaural sounds are of very low-resolution -- only two channels in total.

\subsection{Limitations and future work}
Our vision teacher models are pretrained on perspective images and we apply them to panoramic images. This is due to the limited amount of datasets and annotations for panoramic images. We would like to note that most of the interesting objects appear in the equatorial region. For that region, the distortion of the panoramic images is less severe, and hence the results are less affected. The results by the vision teachers can be seen in Fig. \ref{fig:sample_flow}. 

In future, we plan to incorporate a 3D LiDAR to our sensor setup to get accurate ground-truth for depth prediction. For the semantic prediction, our method is trained and evaluated only for 3 object classes. This is because some classes such as police cars are very rare in a middle-sized dataset. Human is an important class that should be considered in this application. However, not all humans in a scene are sound-making objects. This will raise challenges for collecting the ground truth. Therefore, human class is not considered in this work. We would like to point out that due to these challenges, many of the recent works~\cite{vehicle:tracking:sound:iccv19,irie2019seeing} only deals with one class in real world environment. Another potential future direction is to investigate the usefulness of Harmonic Convolutional Networks (HCN) \cite{dap} for our auditory tasks. HCN is recently proposed to model priors in audio signals by explicitly utilizing the harmonic structure, and has shown great success in unsupervised audio restoration and supervised musical source separation. 

\section{Conclusion} 
\label{sec:conclusion} 
The work develops an approach to predict the semantic labels of sound-making objects, scene depth and object motion in a panoramic image frame, given binaural sounds of the scene alone. To enhance this task, a novel auxiliary task named spatial sound super-resolution is also proposed. All the four tasks are  formulated as multi-task learning and are trained in an end-to-end fashion.
This work has also proposed a novel visual-audio dataset. Extensive experiments have shown that 1) the proposed method achieves good results for all the four tasks; 2) the four tasks are mutually beneficial; 3) the number and orientations of microphones are both important for auditory perception; 4) DDSP and spectrogram representations are complementary and their combination yields the best performance for all the considered auditory tasks.

%
%

\ifCLASSOPTIONcompsoc
  \section*{Acknowledgments}
\else
  \section*{Acknowledgment}
\fi

This work is funded by Toyota Motor Europe via the research project TRACE-Z\"urich. 



\bibliographystyle{IEEEtran}
\bibliography{IEEEabrv,refs}

\begin{thebibliography}{10}
\providecommand{\url}[1]{#1}
\csname url@samestyle\endcsname
\providecommand{\newblock}{\relax}
\providecommand{\bibinfo}[2]{#2}
\providecommand{\BIBentrySTDinterwordspacing}{\spaceskip=0pt\relax}
\providecommand{\BIBentryALTinterwordstretchfactor}{4}
\providecommand{\BIBentryALTinterwordspacing}{\spaceskip=\fontdimen2\font plus
\BIBentryALTinterwordstretchfactor\fontdimen3\font minus
  \fontdimen4\font\relax}
\providecommand{\BIBforeignlanguage}[2]{{%
\expandafter\ifx\csname l@#1\endcsname\relax
\typeout{** WARNING: IEEEtran.bst: No hyphenation pattern has been}%
\typeout{** loaded for the language `#1'. Using the pattern for}%
\typeout{** the default language instead.}%
\else
\language=\csname l@#1\endcsname
\fi
#2}}
\providecommand{\BIBdecl}{\relax}
\BIBdecl

\bibitem{everingham2010pascal}
M.~Everingham, L.~Van~Gool, C.~K. Williams, J.~Winn, and A.~Zisserman, ``The
  pascal visual object classes (voc) challenge,'' \emph{IJCV}, vol.~88, no.~2,
  pp. 303--338, 2010.

\bibitem{Cordts2016Cityscapes}
M.~Cordts, M.~Omran, S.~Ramos, T.~Rehfeld, M.~Enzweiler, R.~Benenson,
  U.~Franke, S.~Roth, and B.~Schiele, ``The cityscapes dataset for semantic
  urban scene understanding,'' in \emph{Proc. of the IEEE Conference on
  Computer Vision and Pattern Recognition (CVPR)}, 2016.

\bibitem{SDV21_arXiv}
C.~Sakaridis, D.~Dai, and L.~Van~Gool, ``{ACDC}: The adverse conditions dataset
  with correspondences for semantic driving scene understanding,'' in
  \emph{International Conference on Computer Vision (ICCV)}, 2021.

\bibitem{soundscape:icra17}
L.~{Marchegiani} and I.~{Posner}, ``Leveraging the urban soundscape: Auditory
  perception for smart vehicles,'' in \emph{IEEE International Conference on
  Robotics and Automation (ICRA)}, 2017.

\bibitem{hear:around:corners:21}
Y.~Schulz, A.~K. Mattar, T.~M. Hehn, and J.~F.~P. Kooij, ``Hearing what you
  cannot see: Acoustic vehicle detection around corners,'' \emph{IEEE Robotics
  and Automation Letters}, vol.~6, pp. 2587--2594, 2021.

\bibitem{vehicle:tracking:sound:iccv19}
C.~Gan, H.~Zhao, P.~Chen, D.~Cox, and A.~Torralba, ``Self-supervised moving
  vehicle tracking with stereo sound,'' in \emph{The IEEE International
  Conference on Computer Vision (ICCV)}, October 2019.

\bibitem{echolocating:listeners:00}
L.~D. Rosenblum, M.~S. Gordon, and L.~Jarquin, ``Echolocating distance by
  moving and stationary listeners,'' \emph{Ecological Psychology}, vol.~12,
  no.~3, pp. 181--206, 2000.

\bibitem{gaver1993world}
W.~W. Gaver, ``What in the world do we hear?: An ecological approach to
  auditory event perception,'' \emph{Ecological psychology}, vol.~5, no.~1, pp.
  1--29, 1993.

\bibitem{ORSpoken}
A.~{Balajee Vasudevan}, D.~{Dai}, and L.~{Van Gool}, ``Object referring in
  visual scene with spoken language,'' in \emph{Winter Conference on
  Applications of Computer Vision (WACV)}, 2018.

\bibitem{deruyttere2019talk2car}
T.~Deruyttere, S.~Vandenhende, D.~Grujicic, L.~Van~Gool, and M.~F. Moens,
  ``Talk2car: Taking control of your self-driving car,'' in \emph{Conference on
  Empirical Methods in Natural Language Processing}, 2019.

\bibitem{sound:pixels:eccv18}
H.~Zhao, C.~Gan, A.~Rouditchenko, C.~Vondrick, J.~McDermott, and A.~Torralba,
  ``The sound of pixels,'' in \emph{The European Conference on Computer Vision
  (ECCV)}, September 2018.

\bibitem{scene-aware-audio}
D.~Li, T.~R. Langlois, and C.~Zheng, ``Scene-aware audio for 360\textdegree{}
  videos,'' \emph{ACM Trans. Graph.}, vol.~37, no.~4, 2018.

\bibitem{room:geometry:acoustic:response:12}
F.~{Antonacci}, J.~{Filos}, M.~R.~P. {Thomas}, E.~A.~P. {Habets}, A.~{Sarti},
  P.~A. {Naylor}, and S.~{Tubaro}, ``Inference of room geometry from acoustic
  impulse responses,'' \emph{IEEE Transactions on Audio, Speech, and Language
  Processing}, vol.~20, no.~10, pp. 2683--2695, 2012.

\bibitem{separate:sound:watching:video:18}
R.~Gao, R.~Feris, and K.~Grauman, ``Learning to separate object sounds by
  watching unlabeled video,'' in \emph{European Conference on Computer Vision
  (ECCV)}, 2018.

\bibitem{aytar2016soundnet}
Y.~Aytar, C.~Vondrick, and A.~Torralba, ``Soundnet: Learning sound
  representations from unlabeled video,'' in \emph{Advances in Neural
  Information Processing Systems (NeurIPS)}, 2016.

\bibitem{sounds:motion:iccv19}
H.~Zhao, C.~Gan, W.-C. Ma, and A.~Torralba, ``The sound of motions,'' in
  \emph{The IEEE International Conference on Computer Vision (ICCV)}, October
  2019.

\bibitem{sound:localization:head:movement:14}
K.~I. McAnally and R.~L. Martin, ``Sound localization with head movement:
  implications for 3-d audio displays,'' \emph{Frontiers in Neuroscience}, vol.
  210, no.~8, 2014.

\bibitem{sound:localization:head:movement:67}
W.~R. Thurlow, J.~W. Mangels, and P.~S. Runge, ``Head movements during sound
  localizationtd,'' \emph{The Journal of the Acoustical Society of America},
  vol. 489, no.~42, 1967.

\bibitem{huang2012human}
W.~Huang, L.~Alem, and M.~A. Livingston, \emph{Human factors in augmented
  reality environments}.\hskip 1em plus 0.5em minus 0.4em\relax Springer
  Science and Business Media, 2012.

\bibitem{mousavian2016joint}
A.~Mousavian, H.~Pirsiavash, and J.~Ko{\v{s}}eck{\'a}, ``Joint semantic
  segmentation and depth estimation with deep convolutional networks,'' in
  \emph{International Conference on 3D Vision (3DV)}, 2016, pp. 611--619.

\bibitem{multitasking:review:20}
S.~Vandenhende, S.~Georgoulis, W.~Van~Gansbeke, M.~Proesmans, D.~Dai, and
  L.~Van~Gool, ``Multi-task learning for dense prediction tasks: A survey,''
  \emph{IEEE Transactions on Pattern Analysis and Machine Intelligence}, 2021.

\bibitem{engel2019ddsp}
J.~Engel, C.~Gu, A.~Roberts \emph{et~al.}, ``Ddsp: Differentiable digital
  signal processing,'' in \emph{International Conference on Learning
  Representations (ICLR)}, 2020.

\bibitem{merging:of:sense}
R.~{Fendrich}, ``The merging of the senses,'' \emph{Journal of Cognitive
  Neuroscience}, vol.~5, no.~3, pp. 373--374, July 1993.

\bibitem{emotion:cross-model:transfer:18}
S.~Albanie, A.~Nagrani, A.~Vedaldi, and A.~Zisserman, ``Emotion recognition in
  speech using cross-modal transfer in the wild,'' in \emph{ACM Multimedia},
  2018.

\bibitem{ambient:sound:eccv16}
A.~Owens, J.~Wu, J.~H. McDermott, W.~T. Freeman, and A.~Torralba, ``Ambient
  sound provides supervision for visual learning,'' in \emph{European
  Conference on Computer Vision (ECCV)}, B.~Leibe, J.~Matas, N.~Sebe, and
  M.~Welling, Eds., 2016.

\bibitem{Owens_2018_ECCV}
A.~Owens and A.~A. Efros, ``Audio-visual scene analysis with self-supervised
  multisensory features,'' in \emph{The European Conference on Computer Vision
  (ECCV)}, 2018.

\bibitem{spatial:alignment:nips20}
P.~Morgado, Y.~Li, and N.~Nvasconcelos, ``Learning representations from
  audio-visual spatial alignment,'' \emph{Advances in Neural Information
  Processing Systems (NeurIPS)}, 2020.

\bibitem{left:right:flipping:audio:cvpr20}
K.~Yang, B.~Russell, and J.~Salamon, ``Telling left from right: Learning
  spatial correspondence of sight and sound,'' in \emph{IEEE/CVF Conference on
  Computer Vision and Pattern Recognition (CVPR)}, 2020.

\bibitem{VisualEchoes:eccv20}
R.~Gao, C.~Chen, Z.~Al-Halah, C.~Schissler, and K.~Grauman, ``Visualechoes:
  Spatial image representation learning through echolocation,'' in \emph{ECCV},
  2020.

\bibitem{audio-visual:instance:cvpr21}
P.~Morgado, I.~Misra, and N.~Vasconcelos, ``Robust audio-visual instance
  discrimination,'' in \emph{IEEE/CVF Conference on Computer Vision and Pattern
  Recognition (CVPR)}, June 2021, pp. 12\,934--12\,945.

\bibitem{distill:audio:visual:cvpr21}
Y.~Chen, Y.~Xian, A.~S. Koepke, and Z.~Akata, ``Distilling audio-visual
  knowledge by compositional contrastive learning,'' in \emph{IEEE Conference
  on Computer Vision and Pattern Recognition (CVPR)}, Jun. 2021.

\bibitem{alwassel_2020_xdc}
H.~Alwassel, D.~Mahajan, B.~Korbar, L.~Torresani, B.~Ghanem, and D.~Tran,
  ``Self-supervised learning by cross-modal audio-video clustering,'' in
  \emph{Advances in Neural Information Processing Systems (NeurIPS)}, 2020.

\bibitem{blauert1997spatial}
J.~Blauert, \emph{Spatial hearing: the psychophysics of human sound
  localization}.\hskip 1em plus 0.5em minus 0.4em\relax MIT press, 1997.

\bibitem{survey:sound:localization:15}
S.~Argentieri, P.~Danès, and P.~Souères, ``A survey on sound source
  localization in robotics: From binaural to array processing methods,''
  \emph{Computer Speech \& Language}, vol.~34, no.~1, pp. 87 -- 112, 2015.

\bibitem{localization:sound:source:review:17}
C.~Rascon and I.~Meza, ``Localization of sound sources in robotics: A review,''
  \emph{Robotics and Autonomous Systems}, vol.~96, pp. 184 -- 210, 2017.

\bibitem{SoundCompass:2014}
J.~Tiete, F.~Dom{\'i}nguez, B.~da~Silva, L.~Segers, K.~Steenhaut, and
  A.~Touhafi, ``Soundcompass: A distributed mems microphone array-based sensor
  for sound source localization,'' in \emph{Sensors}, 2014.

\bibitem{sound:location:09}
A.~{Saxena} and A.~Y. {Ng}, ``Learning sound location from a single
  microphone,'' in \emph{IEEE International Conference on Robotics and
  Automation}, 2009.

\bibitem{klee2006kalman}
U.~Klee, T.~Gehrig, and J.~McDonough, ``Kalman filters for time delay of
  arrival-based source localization,'' \emph{EURASIP Journal on Advances in
  Signal Processing}, vol. 2006, no.~1, p. 012378, 2006.

\bibitem{deepwave:nips19}
M.~M. J.-A. Simeoni, S.~Kashani, P.~Hurley, and M.~Vetterli, ``Deepwave: A
  recurrent neural-network for real-time acoustic imaging,'' 2019.

\bibitem{localize:sound:source:scene:cvpr18}
A.~Senocak, T.-H. Oh, J.~Kim, M.-H. Yang, and I.~So~Kweon, ``Learning to
  localize sound source in visual scenes,'' in \emph{The IEEE Conference on
  Computer Vision and Pattern Recognition (CVPR)}, 2018.

\bibitem{look:listen:learn:iccv17}
R.~Arandjelovic and A.~Zisserman, ``Look, listen and learn,'' in \emph{The IEEE
  International Conference on Computer Vision (ICCV)}, 2017.

\bibitem{objects:sound:eccv18}
R.~Arandjelovi{\'{c}} and A.~Zisserman, ``Objects that sound,'' in
  \emph{European Conference on Computer Vision (ECCV)}, 2018.

\bibitem{Harmony:in:Motion:cvpr07}
Z.~{Barzelay} and Y.~Y. {Schechner}, ``Harmony in motion,'' in \emph{IEEE
  Conference on Computer Vision and Pattern Recognition (CVPR)}, 2007.

\bibitem{cosegmenting:sounds:objects:iccv19}
R.~Gao and K.~Grauman, ``Co-separating sounds of visual objects,'' in \emph{The
  IEEE International Conference on Computer Vision (ICCV)}, October 2019.

\bibitem{tian2018audio}
Y.~Tian, J.~Shi, B.~Li, Z.~Duan, and C.~Xu, ``Audio-visual event localization
  in unconstrained videos,'' in \emph{Proceedings of the European Conference on
  Computer Vision (ECCV)}, 2018.

\bibitem{marchegiani2018listening}
L.~Marchegiani and P.~Newman, ``Listening for sirens: Locating and classifying
  acoustic alarms in city scenes,'' \emph{arXiv preprint arXiv:1810.04989},
  2018.

\bibitem{acoustic:based:safety:emergency:09}
B.~{Fazenda}, {Hidajat Atmoko}, {Fengshou Gu}, {Luyang Guan}, and A.~{Ball},
  ``Acoustic based safety emergency vehicle detection for intelligent transport
  systems,'' in \emph{ICCAS-SICE}, 2009.

\bibitem{dataset:taxonomy:14}
J.~Salamon, C.~Jacoby, and J.~P. Bello, ``A dataset and taxonomy for urban
  sound research,'' in \emph{ACM International Conference on Multimedia}, 2014.

\bibitem{zuren2019tro}
J.~Zürn, W.~Burgard, and A.~Valada, ``Self-supervised visual terrain
  classification from unsupervised acoustic feature learning,'' \emph{IEEE
  Transactions on Robotics}, vol.~37, no.~2, pp. 466--481, 2021.

\bibitem{wet:audio}
I.~Abdić, L.~Fridman, D.~E. Brown, W.~Angell, B.~Reimer, E.~Marchi, and
  B.~Schuller, ``Detecting road surface wetness from audio: A deep learning
  approach,'' in \emph{International Conference on Pattern Recognition (ICPR)},
  2016.

\bibitem{more:than:meets:eye}
F.~R. Valverde, J.~V. Hurtado, and A.~Valada, ``There is more than meets the
  eye: Self-supervised multi-object detection and tracking with sound by
  distilling multimodal knowledge,'' in \emph{The IEEE/CVF Conference on
  Computer Vision and Pattern Recognition (CVPR)}, June 2021.

\bibitem{irie2019seeing}
G.~Irie, M.~Ostrek, H.~Wang, H.~Kameoka, A.~Kimura, T.~Kawanishi, and
  K.~Kashino, ``Seeing through sounds: Predicting visual semantic segmentation
  results from multichannel audio signals,'' in \emph{IEEE International
  Conference on Acoustics, Speech and Signal Processing (ICASSP)}, 2019, pp.
  3961--3964.

\bibitem{3d:room:geometry:audio-visual:sensors:17}
H.~{Kim}, L.~{Remaggi}, P.~J. {Jackson}, F.~M. {Fazi}, and A.~{Hilton}, ``3d
  room geometry reconstruction using audio-visual sensors,'' in
  \emph{International Conference on 3D Vision (3DV)}, 2017, pp. 621--629.

\bibitem{glass:reconstruction:acoustic:15}
M.~{Ye}, {Yu Zhang}, R.~{Yang}, and D.~{Manocha}, ``3d reconstruction in the
  presence of glasses by acoustic and stereo fusion,'' in \emph{IEEE Conference
  on Computer Vision and Pattern Recognition (CVPR)}, 2015.

\bibitem{accoustic:echoes:reveal:room:shape:13}
I.~Dokmanic, R.~Parhizkar, A.~Walther, Y.~M. Lu, and M.~Vetterli, ``Acoustic
  echoes reveal room shape,'' \emph{Proceedings of the National Academy of
  Sciences}, vol. 110, no.~30, pp. 6. 12\,186--12\,191, 2013.

\bibitem{BatVision:icra:20}
J.~H. Christensen, S.~Hornauer, and S.~X. Yu, ``Batvision: Learning to see 3d
  spatial layout with two ears,'' \emph{IEEE International Conference on
  Robotics and Automation (ICRA)}, 2020.

\bibitem{depthwithechoes:cvpr21}
K.~Parida, S.~Srivastava, and G.~Sharma, ``Beyond image to depth: Improving
  depth prediction using echoes,'' in \emph{IEEE/CVF conference on computer
  vision and pattern recognition (CVPR)}, 2021.

\bibitem{Gan2020LookLA}
C.~Gan, Y.~Zhang, J.~Wu, B.~Gong, and J.~Tenenbaum, ``Look, listen, and act:
  Towards audio-visual embodied navigation,'' \emph{2020 IEEE International
  Conference on Robotics and Automation (ICRA)}, pp. 9701--9707, 2020.

\bibitem{soundspace:eccv20}
C.~Chen, U.~Jain, C.~Schissler, S.~V.~A. Gari, Z.~Al-Halah, V.~K. Ithapu,
  P.~Robinson, and K.~Grauman, ``Soundspaces: Audio-visual navigation in 3d
  environments,'' in \emph{European Conference on Computer Vision (ECCV)},
  2020.

\bibitem{audio-visual:nav:cvpr21}
C.~Chen, Z.~Al-Halah, and K.~Grauman, ``Semantic audio-visual navigation,'' in
  \emph{IEEE/CVF Conference on Computer Vision and Pattern Recognition (CVPR)},
  June 2021.

\bibitem{waypoint:audio-visual:nav}
C.~Chen, S.~Majumder, Z.~Al-Halah, R.~Gao, S.~K. Ramakrishnan, and K.~Grauman,
  ``Learning to set waypoints for audio-visual navigation,'' in
  \emph{International Conference on Learning Representations (ICLR)}, 2021.

\bibitem{eliakim2018fully}
I.~Eliakim, Z.~Cohen, G.~Kosa, and Y.~Yovel, ``A fully autonomous terrestrial
  bat-like acoustic robot,'' \emph{PLoS computational biology}, vol.~14, no.~9,
  p. e1006406, 2018.

\bibitem{3d:sound}
D.~R. Begault and L.~J. Trejo, \emph{3-D sound for virtual reality and
  multimedia}.\hskip 1em plus 0.5em minus 0.4em\relax Academic Press
  Professional, Inc., 2000.

\bibitem{spatial:soundscapes:2020}
C.~Rajguru, M.~Obrist, and G.~Memoli, ``Spatial soundscapes and virtual worlds:
  Challenges and opportunities,'' \emph{Frontiers in Psychology}, vol.~11,
  2020.

\bibitem{Owens_2016_CVPR}
A.~Owens, P.~Isola, J.~McDermott, A.~Torralba, E.~H. Adelson, and W.~T.
  Freeman, ``Visually indicated sounds,'' in \emph{The IEEE Conference on
  Computer Vision and Pattern Recognition (CVPR)}, 2016.

\bibitem{360:sound:generation:nips18}
T.~L. Pedro~Morgado, Nuno~Vasconcelos and O.~Wang, ``Self-supervised generation
  of spatial audio for 360 deg video,'' in \emph{Neural Information Processing
  Systems (NIPS)}, 2018.

\bibitem{gao20192}
R.~Gao and K.~Grauman, ``2.5 d visual sound,'' in \emph{IEEE Conference on
  Computer Vision and Pattern Recognition (CVPR)}, 2019.

\bibitem{sep-stereo:eccv20}
H.~Zhou, X.~Xu, D.~Lin, X.~Wang, and Z.~Liu, ``Sep-stereo: Visually guided
  stereophonic audio generation by associating source separation,'' in
  \emph{European Conference on Computer Vision (ECCV)}, 2020.

\bibitem{visually:informed:binaural:audio:21}
X.~Xu, H.~Zhou, Z.~Liu, B.~Dai, X.~Wang, and D.~Lin, ``Visually informed
  binaural audio generation without binaural audios,'' in \emph{IEEE conference
  on computer vision and pattern recognition (CVPR)}, 2021.

\bibitem{binauralspeechsynthesis}
A.~Richard, D.~Markovic, I.~D. Gebru, S.~Krenn, G.~Butler, F.~de~la Torre, and
  Y.~Sheikh, ``Neural synthesis of binaural speech from mono audio,'' in
  \emph{International Conference on Learning Representations (ICLR)}, 2021.

\bibitem{chen2018encoder}
L.-C. Chen, Y.~Zhu, G.~Papandreou, F.~Schroff, and H.~Adam, ``Encoder-decoder
  with atrous separable convolution for semantic image segmentation,'' in
  \emph{Proceedings of the European conference on computer vision (ECCV)},
  2018.

\bibitem{background:substraction:11}
S.~{Brutzer}, B.~{Höferlin}, and G.~{Heidemann}, ``Evaluation of background
  subtraction techniques for video surveillance,'' in \emph{IEEE Conference on
  Computer Vision and Pattern Recognition (CVPR)}, 2011.

\bibitem{monodepth2:iccv19}
C.~Godard, O.~M. Aodha, M.~Firman, and G.~J. Brostow, ``Digging into
  self-supervised monocular depth estimation,'' in \emph{Proceedings of the
  IEEE International Conference on Computer Vision}, 2019.

\bibitem{Geiger2013IJRR}
A.~Geiger, P.~Lenz, C.~Stiller, and R.~Urtasun, ``Vision meets robotics: The
  kitti dataset,'' \emph{International Journal of Robotics Research (IJRR)},
  2013.

\bibitem{ilg2017flownet}
E.~Ilg, N.~Mayer, T.~Saikia, M.~Keuper, A.~Dosovitskiy, and T.~Brox, ``Flownet
  2.0: Evolution of optical flow estimation with deep networks,'' in \emph{the
  IEEE conference on computer vision and pattern recognition (CVPR)}, 2017.

\bibitem{role:head:movements:sound:localization:40}
H.~Wallach, ``The role of head movements and vestibular and visual cues in
  sound localization,'' \emph{Journal of Experimental Psychology}, vol.~27,
  no.~4, p. 339, 1940.

\bibitem{invert:fourier:transform:83}
D.~{Griffin} and {Jae Lim}, ``Signal estimation from modified short-time
  fourier transform,'' in \emph{ICASSP '83. IEEE International Conference on
  Acoustics, Speech, and Signal Processing}, vol.~8, April 1983, pp. 804--807.

\bibitem{chen2017deeplab}
L.-C. Chen, G.~Papandreou, I.~Kokkinos, K.~Murphy, and A.~L. Yuille, ``Deeplab:
  Semantic image segmentation with deep convolutional nets, atrous convolution,
  and fully connected crfs,'' \emph{IEEE transactions on pattern analysis and
  machine intelligence}, vol.~40, no.~4, pp. 834--848, 2017.

\bibitem{kim2018crepe}
J.~W. Kim, J.~Salamon, P.~Li, and J.~P. Bello, ``Crepe: A convolutional
  representation for pitch estimation,'' in \emph{2018 IEEE International
  Conference on Acoustics, Speech and Signal Processing (ICASSP)}.\hskip 1em
  plus 0.5em minus 0.4em\relax IEEE, 2018, pp. 161--165.

\bibitem{russell2008labelme}
B.~C. Russell, A.~Torralba, K.~P. Murphy, and W.~T. Freeman, ``Labelme: a
  database and web-based tool for image annotation,'' \emph{International
  journal of computer vision}, vol.~77, no. 1-3, pp. 157--173, 2008.

\bibitem{kingma2014adam}
D.~P. Kingma and J.~Ba, ``Adam: A method for stochastic optimization,''
  \emph{arXiv preprint arXiv:1412.6980}, 2014.

\bibitem{dap}
Z.~Zhang, Y.~Wang, C.~Gan, J.~Wu, J.~B. Tenenbaum, A.~Torralba, and W.~T.
  Freeman, ``Deep audio priors emerge from harmonic convolutional networks,''
  in \emph{International Conference on Learning Representations (ICLR)}, 2020.

\end{thebibliography}

%

\begin{IEEEbiography}[{\includegraphics[width=1in,clip,keepaspectratio]{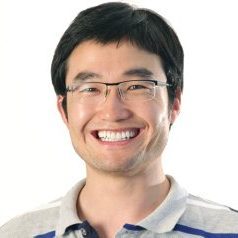}}]{Dengxin Dai}
 is a Senior Researcher and head of the Vision for Autonomous Systems Group at MPI for Informatics. He is also a Lecturer at ETH Zurich. In 2016, he obtained his Ph.D. in Computer Vision at ETH Zurich and has been a group leader since then working on autonomous driving.  His research interests lie in robust perception in adverse weather and illumination conditions, sensor fusion, multi-task learning, object recognition under limited supervision, and auditory perception. He is a member of the ELLIS Society. He has been a program committee member of several major computer vision conferences and has organized multiple well-received workshops. He was guest editor of IJCV, and was/is area chair for WACV 2020, CVPR 2021, CVPR 2022, and ECCV 2022. In 2021, he has received the Golden Owl Award at ETH Zurich for his excellent teaching. 
 
\end{IEEEbiography}

\begin{IEEEbiography}[{\includegraphics[width=1in,clip,keepaspectratio]{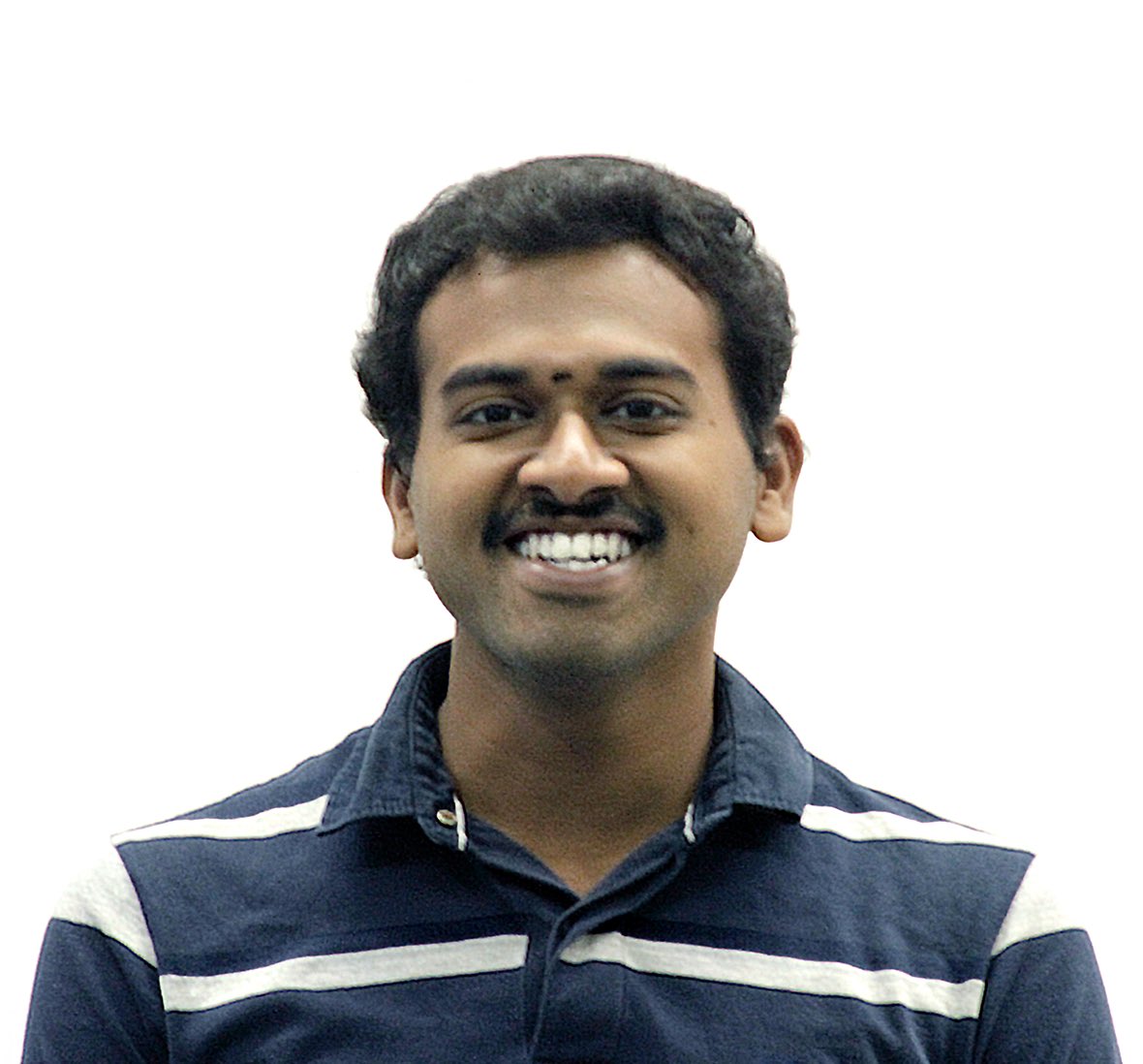}}]{Arun Balajee Vasudevan}
is is a Ph.D student at the CVL group, ETH Zurich. Before joining in the lab, he received his MSc in Computer Science from EPFL and Bachlors in Electrical Engineering in Indian Institute of Technology Jodhpur.
He is part of the Toyota TRACE project focusing on the driver-car interaction. His research interests are in Natural language understanding, visual-audio perception and navigation.
\end{IEEEbiography}

\begin{IEEEbiography}[{\includegraphics[width=1in,clip,keepaspectratio]{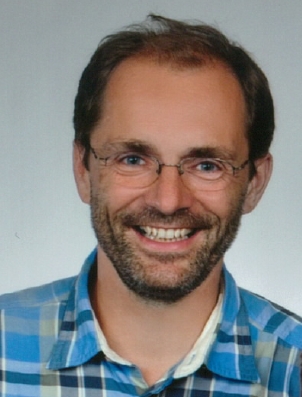}}]{Jiri Matas}
is a full professor of Computer Vision in Prague. He has served in various roles at major computer vision conferences (e.g. ICCV, CVPR, ICPR, NIPS, ECCV), co-chaired ECCV 2004, 2016, CVPR 2007 and a good number of workshop series on topics such as Object Tracking, Vision for All Seasons and Autonomous Driving. He is on the editorial board of IJCV and was the Associate Editor-in-Chief of IEEE T. PAMI. He served on the computer science panel of ERC. 
\end{IEEEbiography}


\begin{IEEEbiography}[{\includegraphics[width=1in,clip,keepaspectratio]{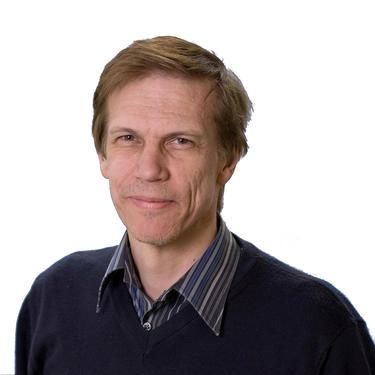}}]{Luc Van Gool}
is a full professor for Computer Vision at ETH Zurich and the KU Leuven. He leads research and teaches at both places. He has authored over 300 papers. He has been a program committee member of several, major computer vision conferences (e.g.\ Program Chair ICCV’05, Beijing, General Chair of ICCV’11, Barcelona, and of ECCV’14, Zurich). His main interests include 3D reconstruction and modeling, object recognition, and autonomous driving. He received several Best Paper awards (e.g.\ David Marr Prize ’98, Best Paper CVPR’07). He received the Koenderink Award in 2016 and the ‘Distinguished Researcher’ nomination by the IEEE Computer Society in 2017. In 2015 he also received the 5-yearly Excellence Prize by the Flemish Fund for Scientific Research. He was the holder of an ERC Advanced Grant (VarCity). Currently, he leads computer vision research for autonomous driving in the context of the Toyota TRACE labs in Leuven and at ETH, and has an extensive collaboration with Huawei on the topic of image and video enhancement.
\end{IEEEbiography}




\end{document}